\documentclass[11pt]{article}
\pdfoutput=1
\usepackage{amssymb,jheppub,slashed}
\usepackage{braket,commath,nicefrac}
\usepackage{subcaption,siunitx,amsmath}
\usepackage{slashed,float}
\usepackage{xcolor}
\usepackage{braket}
\usepackage{scalerel}

\newcommand{\smallA}{\scaleto{A}{4pt}}
\newcommand{\smallB}{\scaleto{B}{4pt}}

\title{Measuring Quantum Discord at the LHC}
\author[a]{Tao Han,}
\author[a]{Matthew Low,}
\author[b]{Navin McGinnis,}
\author[b]{and Shufang Su}

\affiliation[a]{PITT PACC, Department of Physics and Astronomy,\\ University of Pittsburgh, 3941 O’Hara St., Pittsburgh, PA 15260, USA}
\affiliation[b]{Department of Physics, University of Arizona, Tucson, Arizona 85721, USA}

\emailAdd{than@pitt.edu}
\emailAdd{matthew.w.low@pitt.edu}
\emailAdd{nmcginnis@arizona.edu}
\emailAdd{shufang@arizona.edu}

\preprint{
\begin{flushright}
PITT-PACC-2316
\end{flushright}}

\abstract{
There has been an increasing interest in exploring quantities associated with quantum information at colliders.  We perform a detailed analysis describing how to measure the quantum discord in the top anti-top quantum state at the Large Hadron Collider (LHC).  While for pure states, quantum discord, entanglement, and Bell nonlocality all probe the same correlations, for mixed states they probe different aspects of quantum correlations.  The quantum discord, in particular, is interesting because it aims to encapsulate all correlations between systems that cannot have a classical origin.  We employ two complementary approaches for the study of the top anti-top system, namely the decay method and the kinematic method.  We highlight subtleties associated with measuring discord for reconstructed quantum states at colliders.  Usually quantum discord is difficult to compute due to an extremization that must be performed.  We show, however, that for the $t\bar{t}$ system this extremization can be performed analytically and we provide closed-form formulas for the quantum discord.  We demonstrate that with current LHC datasets, quantum discord can be observed at $3.6-5.7\sigma$, depending on the signal region, with the decay method and can be measured at a precision of $0.1-0.2\%$ with the kinematic method.  At the high luminosity LHC, the observation of quantum discord is expected to be $>5\sigma$ using both the decay and kinematic methods and can be measured with a precision of $5\%$ with the decay method and $0.05\%$ with the kinematic method.
Additionally, we identify the kinematic cuts at the LHC to isolate the $t\bar{t}$ state that is separable but has non-zero discord.  By systematically investigating quantum discord for the first time through a detailed collider analysis, this work expands the toolkit for quantum information studies in particle physics and lays the groundwork for deeper insights into the quantum properties in high-energy collisions.}

\begin{document}

\maketitle
\newpage

\section{Introduction}
\label{sec:intro}

Experiments at the CERN Large Hadron Collider (LHC) have been leading the exploration of fundamental physics at the energy frontier for more than a decade, with the milestone discovery of the Higgs boson, precision determinations of the Standard Model (SM) properties in the QCD and electroweak sectors, and the active search for new physics beyond the SM.  Recently, there has been an increasing body of work on exploring quantities associated with quantum entanglement, Bell nonlocality, and quantum information at colliders~\cite{Afik:2020onf,Djouadi:2024lyv, White:2024bjp, Wildridge:2024yeg, Altomonte:2024upf, Gao:2024leu, Fang:2024ple, Wu:2024ovc, Sullivan:2024wzl, Cheng:2024rxi, Ravina:2024ard, Du:2024sly, CMS:2024zkc, Ye:2024dqx, Ruzi:2024cbt, Gabrielli:2024kbz, LoChiatto:2024dmx, ParticleDataGroup:2024cfk, Dong:2024xsb, Larkoski:2024uoc, Dong:2024xsg, Cheng:2024btk, Wu:2024asu, deNova:2024qeo, White:2024nuc, Maltoni:2024wyh, CMS:2024pts, Afik:2024uif, Kowalska:2024kbs, ATLAS:2024kxj, Maltoni:2024csn, Subba:2024mnl, Morales:2024jhj, Duch:2024pwm, Aguilar-Saavedra:2024whi, Aoude:2024xpx, Wu:2024mtj, Jung:2024ans, Aguilar-Saavedra:2024vpd, Barr:2024djo, Sarkar:2024dxe, Belhaj:2024bqk, Blasone:2024dud, Aguilar-Saavedra:2024hwd, Maltoni:2024tul, Aguilar-Saavedra:2024fig, Li:2024luk, CMS:2024hgo, Simpson:2024hbr, CMS:2024vqh, Blasone:2024bmo, Khor:2023xar, DeFabritiis:2023llu, Liu:2023bnr, Ehataht:2023zzt, Cheng:2023qmz, Kats:2023zxb, ATLAS:2023fsd, Altomonte:2023mug, Han:2023fci, Bernal:2023jba, Jung:2023fpv, Sakurai:2023nsc, FerreiradaSilva:2023mhf, Ma:2023yvd, DeFabritiis:2023tkh, Bi:2023uop, Bernal:2023ruk, Aoude:2023hxv, Aguilar-Saavedra:2023hss, Dudal:2023mij, Morales:2023gow, Dong:2023xiw, Bittencourt:2023asd, Dudal:2023pbc, DeFabritiis:2023ubj, Ghosh:2023rpj, Fabbrichesi:2023cev, ATLAS:2023jzs, Gross:2022hyw, Altakach:2022ywa, Mantani:2022dao, Severi:2022qjy, Aguilar-Saavedra:2022mpg, Ashby-Pickering:2022umy, Aguilar-Saavedra:2022wam, Ehlers:2022oke, Ehlers:2022oal, Afik:2022dgh, Hung:2022azf, Fabbrichesi:2022ovb, Kurashvili:2022ybg, Hentschinski:2022rsa, Aguilar-Saavedra:2022uye, Ramos:2022gia, Aoude:2022imd, Afik:2022kwm, Larkoski:2022lmv, Severi:2021cnj, Bittencourt:2021xcx, Karlberg:2021kwr, Eckstein:2021pgm, Fabbrichesi:2021npl, Vollmann:2021moq, Feal:2020myr, Iskander:2020rkb, Aoude:2020mlg, Ramos:2020kaj, Grossi:2024jae, Han:2025ewp, Fabbrichesi:2025ywl}.  The large amount of data accumulated at the LHC offers a diverse plethora of quantum states to study, thus providing a great testing ground for quantum tomography and for the high-energy exploration of quantum information. 

It is encouraging that the ATLAS~\cite{ATLAS:2023fsd} and CMS~\cite{CMS:2024vqh} experiments have both observed quantum entanglement in the $t\bar{t}$ system. Beyond that, substantial work has been devoted to outlining the measurements of quantum entanglement and Bell nonlocality in other channels at the LHC and examining their quantum properties~\cite{Fabbrichesi:2022ovb,Ashby-Pickering:2022umy,Aguilar-Saavedra:2022mpg,Aguilar-Saavedra:2023hss,Aoude:2023hxv,Aguilar-Saavedra:2024hwd,Barr:2024djo,Aguilar-Saavedra:2024whi,Morales:2024jhj,Afik:2024uif,Gao:2024leu,Fu:2024bki}.  Entanglement and Bell nonlocality have also been shown to exhibit sensitivty to the effects of new physics beyond the SM in a variety of final states~\cite{Aoude:2022imd,Fabbrichesi:2022ovb,Aoude:2023hxv}. 

The space of quantum correlations is much richer and more diverse than just quantum entanglement and Bell nonlocality.  These range from quantum discord, which is non-zero in more quantum systems than entanglement, to negative conditional entropy, which is present in fewer quantum systems than Bell nonlocality.  Among the many different quantum correlations, in this work we focus on quantum discord~\cite{Zurek_2000,Ollivier:2001fdq}.

Quantum discord is a different form of correlation than quantum entanglement.  While entanglement identifies quantum states in which one subsystem cannot be fully described without knowledge of the other subsystem, quantum discord more directly characterizes states which have correlations of a quantum origin.  Heuristically, quantum discord quantifies correlations within a quantum state that are not invariant under measurements.  Such a property is a hallmark of quantum systems and completely absent in classical systems.

One interesting feature of quantum discord is that it can be non-zero for some separable states.  This can be counterintuitive for those who are accustomed to orienting their view of quantum systems around entanglement.  In this sense, quantum discord is often viewed as a more robust metric for evaluating systems that display quantum behavior~\cite{Bera_2017}.  Another interesting feature of quantum discord is that it is not necessarily symmetric between subsystems.  In the $t\bar{t}$ system, a difference between the discord of $t$ and the discord of $\bar{t}$ would be a measure of CP violation in $t\bar{t}$ production~\cite{Afik:2022dgh}.

Recently, it was recognized that quantum discord may be measurable at high energy colliders~\cite{Afik:2022dgh}.  The study importantly showed that over phase space the quantum discord of the $t\bar{t}$ quantum state spans the full range of possible discord values making it an excellent system in which to measure quantum discord.  In this work, we move from the idea of quantum discord at the LHC to a concrete description of how to measure it at the LHC, including simple formulae, and projections for the sensitivity and precision.  Additionally, we import some fundamental concepts from quantum information, like entropy and mutual information, to colliders to place quantum discord in a fuller context.  Finally, we resolve one of the major theoretical challenges in quantum discord at colliders, namely the impact of fictitious states~\cite{Afik:2022kwm,Cheng:2023qmz,Cheng:2024btk}.

Our comprehensive collider study is performed using events that are showered, hadronized, and run through a detector simulation, with distributions that are unfolded.  We perform the analysis using two complementary approaches, namely the decay method and the kinematic method \cite{Cheng:2024btk}.  In the decay method, our analysis closely follows existing experimental strategies for measuring entanglement at the LHC~\cite{ATLAS:2023fsd,CMS:2024vqh}: reconstructing the $t\bar{t}$ quantum states by unfolding the angular distribution of decay products of the $t\bar t$ system.  We find sensitivity of $\approx 10\%$ for a measurement of discord with current data at the LHC, and project that by the end of the high luminosity LHC (HL-LHC) this could improve to less than $5\%$.  The kinematic method is simpler in the treatment of the $t\bar t$ system and the results are more encouraging, with the sensitivity improved by more than an order of magnitude. Furthermore, we also outline the theoretical subtleties of the discord measurement which would need to be adapted for other collider systems. Our study offers the first fine-grained look at quantum behavior for LHC physics. 

The remainder of the paper is as follows. After properly defining the top anti-top system as a bipartite quantum state at the LHC in Sec.~\ref{sec:tt}, we introduce the concepts of quantum information relevant to the collider observables, such as various forms of entropy, and provide closed-form formulae for evaluating the quantum discord of the $t\bar{t}$ system in Sec.~\ref{sec:qi}.  We carry out a detailed phenomenological study for discord at the LHC for top quark production and decay in Sec.~\ref{sec:collider}.  We provide some further discussion and draw our conclusions in Sec.~\ref{sec:conclusions}.  Some technical details are provided in a few appendices.

\section{The Top Anti-Top Quantum State}
\label{sec:tt}

In this work we study the $t\bar{t}$ system which can be described as a bipartite qubit system.  The quantum state $\rho_{AB}$ which is comprised of the spin of $t$ and the spin of $\bar{t}$ can be written using the Fano-Bloch decomposition as~\cite{Fano:1983zz}
\begin{equation}
\label{eq:Fano_decomp}
\rho_{AB} = \frac{1}{4} \left(
\mathbb{I}_2 \otimes \mathbb{I}_2
+ \sum_i B^+_i \sigma_i \otimes \mathbb{I}_2
+ \sum_j B^-_j \mathbb{I}_2 \otimes \sigma_j
+ \sum_{ij} C_{ij} \sigma_i \otimes \sigma_j
\right),
\end{equation}
where both $i,j$ run from 1 to 3, $\sigma_i$ are the Pauli matrices, $\mathbb{I}_2$ is the $2 \times 2$ identity matrix, $B^+_i$ is a vector describing the net polarization of the top, $B^-_j$ is a vector describing the net polarization of the anti-top, and $C_{ij}$ is the spin correlation matrix.  The subscript $AB$ of $\rho_{AB}$ indicates this is the joint system of subsystem $A$, which is the qubit of the top spin, and subsystem $B$, which is the qubit of the anti-top spin.

To measure the spins, one needs to specify a set of quantization axes for the spin called the basis.  Common basis choices include the helicity basis $\{ k, r, n \}$, the beam basis $\{ x, y, z \}$, and the diagonal basis~\cite{Cheng:2023qmz,Cheng:2024btk}, which are discussed in more detail in Sec.~\ref{sec:decay}. From CP invariance, $B^+_{i} = B^-_{i}$ and $C_{ij} = C_{ji}$.  In the $t\bar{t}$ system at leading order, $B^+_{i} = B^-_{i} = 0$ which means that the spin correlation matrix fully characterizes the quantum state.  States with zero polarization are called Bell-diagonal states and can be fully described by the three eigenvalues of the spin correlation matrix.

For the subset of phase space where the top quark scattering angle $\theta \approx \pi/2$ in the partonic center-of-mass (COM) frame, we can write a decomposition of $\rho_{t\bar{t}}$ that provides a useful heuristic to understand the behavior of the quantum informational quantities we study~\cite{Barr:2024djo}.  At $\theta = \pi/2$, the $t$ and $\bar{t}$ travel perpendicular to the beam which means the beam basis and helicity basis coincide.  

Bell states are pure states that have the maximal amount of entanglement and maximally violate Bell's inequality. First, consider the pure Bell states $\rho^{(+)}$ and $\rho^{(-)}$, defined along the $z$ direction,
\begin{equation} \label{eq:BellState}
\rho^{(\pm)} =  \ket{\psi^{(\pm)}} \bra{\psi^{(\pm)}},
\qquad\qquad
\psi^{(\pm)} = \frac{1}{\sqrt{2}}( \ket{\downarrow \uparrow} \pm \ket {\uparrow \downarrow} ),
\end{equation}
where $\ket{\uparrow}$ and $\ket{\downarrow}$ are the eigenstates of $\sigma_z$.  Next, consider the separable mixed states
\begin{align}
\label{eq:rhoMixX}
\rho^{(X)}_{\rm mix} &=
\frac{1}{2} (\ket{++}\bra{++} + \ket{--}\bra{--}), \\
\label{eq:rhoMixY}
\rho^{(Y)}_{\rm mix} &=
\frac{1}{2} (\ket{\leftarrow\rightarrow}\bra{\leftarrow\rightarrow} + \ket{\rightarrow\leftarrow}\bra{\rightarrow\leftarrow}), \\
\label{eq:rhoMixZ}
\rho^{(Z)}_{\rm mix} &=
\frac{1}{2} (\ket{\downarrow \uparrow}\bra{\downarrow \uparrow} + \ket{\uparrow \downarrow}\bra{\uparrow\downarrow}),
\end{align}
where $\ket{+}$ and $\ket{-}$ are the eigenstates of $\sigma_x$ and $\ket{\leftarrow}$ and $\ket{\rightarrow}$ are the eigenstates of $\sigma_y$.

The quantum state $\rho_{q\bar{q}}$ that would be formed from $q\bar{q}$ is
\begin{equation}
\label{eq:rhoqq}
\rho_{q\bar{q}} = a \rho^{(+)} + (1-a) \rho^{(X)}_{\rm mix},
\qquad\qquad
a = \frac{\beta^2}{2-\beta^2},
\end{equation}
where $\beta=\sqrt{1-4m_{t}^{2}/M_{t\bar{t}}^{2}}$~ is the top quark velocity, and $M_{t\bar{t}}$ is the invariant mass of the top anti-top system. Near threshold $\beta \to 0$, the state approaches the mixed separable state $\rho^{(X)}_{\rm mix}$, while at high-$p_T$, $\beta \to 1$, the state approaches the pure entangled state $\rho^{(+)}$.

The quantum state $\rho_{gg}$ that would be formed from $gg$ is
\begin{equation}
\label{eq:rhogg}
\rho_{gg} = a_1 \rho^{(+)} + a_2 \rho^{(-)}
+ a_3 \rho^{(X)}_{\rm mix} + a_4 \rho^{(Y)}_{\rm mix},
\end{equation}
where the coefficients are
\begin{equation}
a_1 = \frac{\beta^4}{1+2\beta^2-2\beta^4},
\qquad
a_2 = \frac{(1-\beta^2)^2}{1+2\beta^2-2\beta^4},
\qquad
a_3 = a_4 = \frac{2\beta^2(1-\beta^2)^2}{1+2\beta^2-2\beta^4},
\end{equation}
and are constrained as $a_1 + a_2 + a_3 + a_4 = 1$~\cite{Fabbrichesi:2022ovb,Barr:2024djo}.  Near threshold the state approaches the pure entangled state $\rho^{(-)}$.  At high-$p_T$ the state approaches the pure entangled state $\rho^{(+)}$.  In between these limits there is a significant mixed component of the quantum state.

The $t\bar{t}$ quantum state is a mixed state comprised of the density matrices of the quark annihilation and gluon fusion partonic $t\bar{t}$ production channels, $\rho_{q\bar{q}}$ and $\rho_{gg}$.   At threshold, while the $q\bar{q}$ fraction is smaller than the $gg$ fraction, it is sufficiently large that the $t\bar{t}$ state should be treated as mixed.  On the other hand, at high-$p_T$, the $t\bar{t}$ state approaches the pure state $\rho^{(+)}$.

\section{Quantum Information}
\label{sec:qi}

In this section, we introduce the relevant quantum information quantities that can be measured at the LHC.  In each case, we will show the predicted distribution as a function of the top quark velocity $\beta$ and scattering angle $\theta$.

The description of subsystem $A$ alone, $\rho_A$, is obtained from total quantum state $\rho_{AB}$ by performing the partial trace over the $B$ subsystem.
\begin{equation} \label{eq:partialTraceA}
\rho_A = \text{tr}_{\smallB} \rho_{AB}
= \sum_b (\mathbb{I}_2 \otimes \bra{b}) \rho_{AB} (\mathbb{I}_2 \otimes \ket{b}),
\end{equation}
where $b$ indexes an orthonomal basis for subsystem $B$.  Similarly, the description of subsystem $B$ alone is
\begin{equation} \label{eq:partialTraceB}
\rho_B = \text{tr}_{\smallA} \rho_{AB}
= \sum_a (\bra{a} \otimes \mathbb{I}_2 ) \rho_{AB} (\ket{a} \otimes \mathbb{I}_2 ).
\end{equation}
Both $\rho_A$ and $\rho_B$ are called reduced density matrices.

\subsection{Von Neumann Entropy}
\label{sec:entropy}

In classical information theory the Shannon entropy $H$ of a random variable characterizes the amount of uncertainty about the value of that variable.  For example, a single bit $X$ that is always known to be $0$ has the minimal entropy of $H(X)=0$.  On the other hand, a single bit that is equally likely to be $0$ or $1$ has the largest possible uncertainty about its value and, consequently, has the maximal entropy of $H(X)=1$.  For a system of $N$ bits the maximal entropy is $H(X_1, \ldots, X_N)=N$.

For a quantum state $\rho$, the quantum generalization of Shannon entropy is the Von Neumann entropy $S(\rho)$, which is defined as~\cite{Fano:1957zz}
\begin{equation}
\label{eq:entropy}
S(\rho) = - \text{tr} (\rho \log_2 \rho),
\end{equation}
where $\log_2$ is the base-2 logarithm.  The Von Neumann entropy is subadditive~\cite{Araki:1970ba}, which means
\begin{equation} 
\label{eq:entropy-subadd}
S(\rho_{AB}) \leq S(\rho_A) + S(\rho_B),
\end{equation}
where $\rho_A$ are $\rho_B$ are the reduced density matrices.  The equality in Eq.~\eqref{eq:entropy-subadd} only applies for product states.

\begin{figure}
\centering
  \includegraphics[scale=0.45]{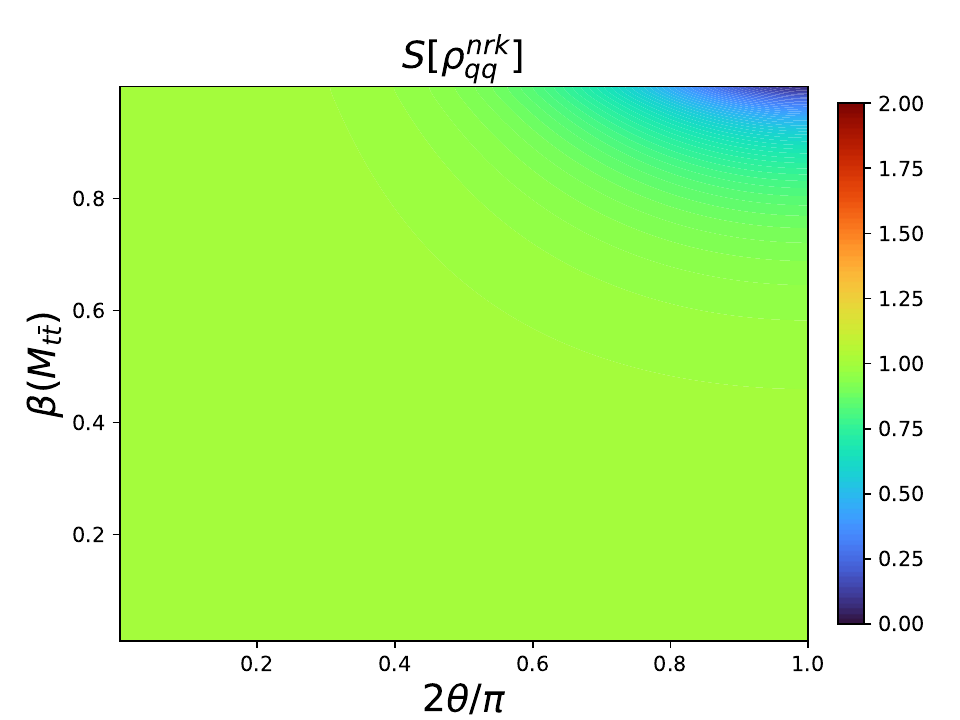}
  \includegraphics[scale=0.45]{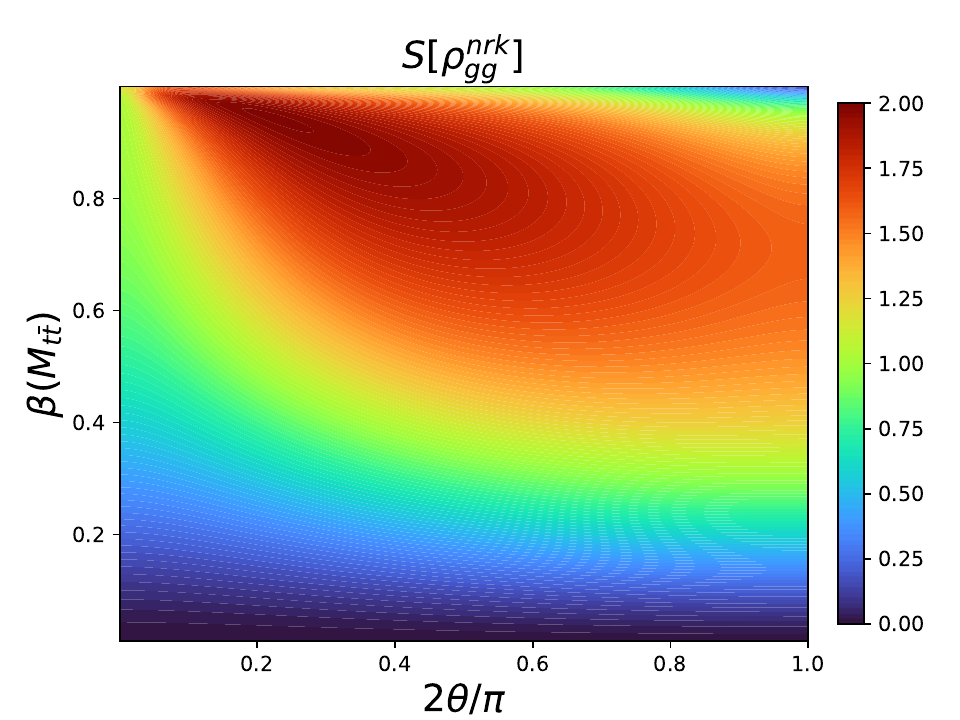}\\
  \includegraphics[scale=0.45]{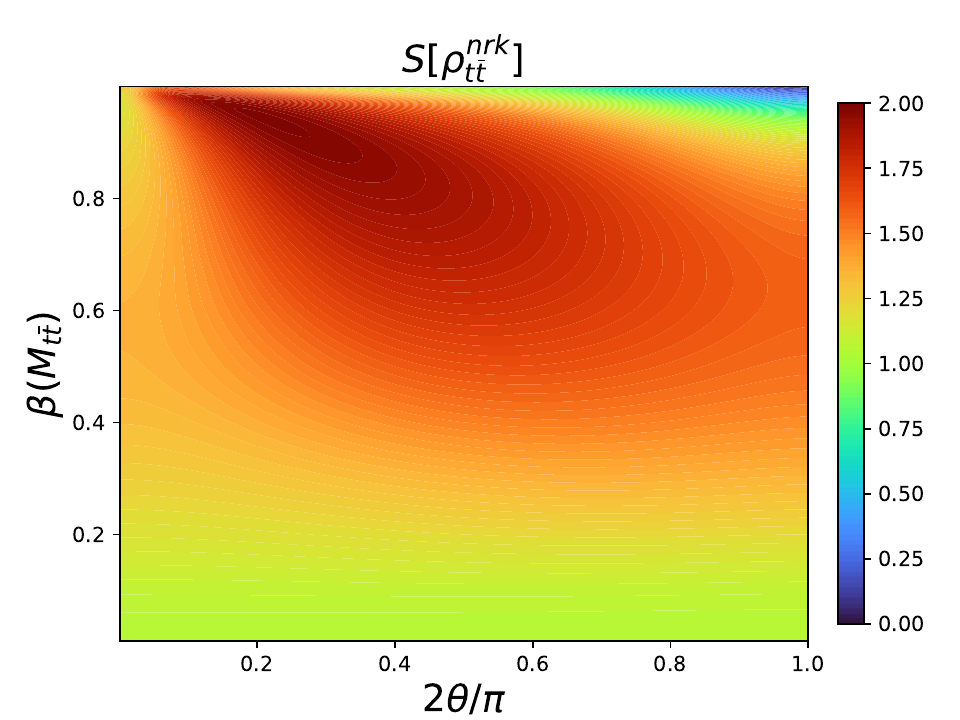}  
  \includegraphics[scale=0.45]{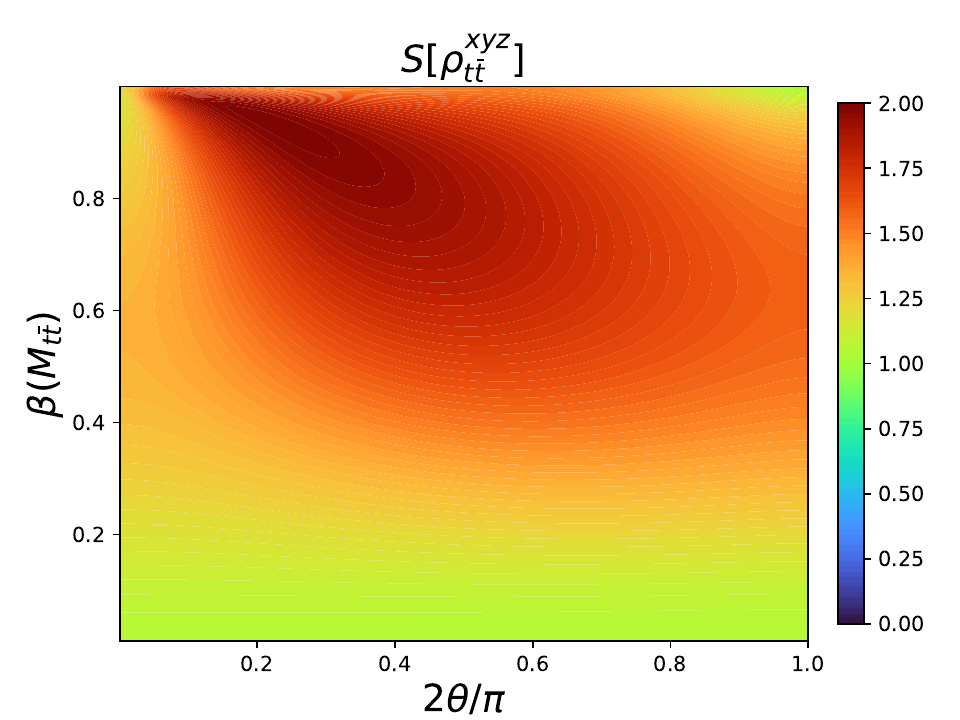}  
  \caption{The Von Neumann entropy for $q\bar{q} \to t\bar{t}$ (top left), $gg \to t\bar{t}$ (top right), $pp \to t\bar{t}$ (bottom left) in the helicity basis.  The process $pp \to t\bar{t}$ (bottom right) is also shown in the beam basis.  For a two qubit system the entropy ranges from 0 to 2.}
\label{fig:QI_entropy}
\end{figure}

The entropy is zero only for pure states. It becomes maximal for a maximally-mixed state, following the intuition of the classical case.  The entropy for the $t\bar{t}$ system is shown in Fig.~\ref{fig:QI_entropy} for $pp \to t\bar{t}$ and its subprocesses.  Since the $t\bar{t}$ system involves two qubits the entropy ranges from $S=0$ to $S=2$. 

From $q\bar{q}$ (Fig.~\ref{fig:QI_entropy}, top left), the $t\bar{t}$ produced over most of phase space corresponds to $\rho^{(X)}_{\rm mix}$ which has a moderate entropy value of $S=1$.  In the high-$p_T$ region the state approaches $\rho^{(+)}$ which has the minimal entopy of $S=0$.  The $gg$-initiated quantum state (Fig.~\ref{fig:QI_entropy}, top right), on the other hand, varies between the states $\rho^{(-)}$ and $\rho^{(+)}$ with minimal entropies of $S=0$, at threshold and in the high-$p_T$ regions, respectively, and a nearly full mixed state with entropies approaching the maximal value of $S=2$.  

The total quantum state from $pp\to t\bar{t}$ (Fig.~\ref{fig:QI_entropy}, bottom left) has non-zero entropy over all of phase space except in the very high-$p_T$ region where the total state approaches $\rho^{(+)}$.  In the beam basis (Fig.~\ref{fig:QI_entropy}, bottom right), this region of zero entropy disappears due to angular averaging.

\subsection{Mutual Information}
\label{sec:mutualinfo}

In classical information theory, the mutual information $I$ of two bits characterizes how much information is obtained about one bit by observing the other bit.  Classically, it is symmetric with respect to the two bits.

For example, consider four coins $c_1$, $c_2$, $c_3$, and $c_4$ to be flipped.  If subsystem $A$ contains $c_1$ and $c_2$ and subsystem $B$ contains $c_3$ and $c_4$, the mutual information of the total system is $I=0$ since an observation of the results of $c_1$ and $c_2$ do not provide any information as to the results of $c_3$ and $c_4$.  On the other hand, if subsystem $A$ contains $c_1$ and $c_2$ and subsystem $B$ contains $c_2$ and $c_3$ then the mutual information of the total system is $I=1$.  An observation of either subsystem reveals the value of $c_2$ which is 1 bit of information about the other subsystem.

The quantum analog of the mutual information has two forms.  The first is the ``total'' mutual information $I(\rho)$ defined as
\begin{equation}
\label{eq:total_mutual_information}
I(\rho_{AB}) = S(\rho_A) + S(\rho_B) - S(\rho_{AB}).
\end{equation}
By subadditivity, Eq.~\eqref{eq:entropy-subadd}, the total mutual information is always $\geq 0$.  $I(\rho_{AB})$ is symmetric with respect to the two subsystems.  Since entropies themselves are bounded, the maximum value for a system with $N$ qubits is $N$.  In the $t\bar{t}$ system $I(\rho_{t\bar{t}}) \leq 2$.

\begin{figure}
\centering
  \includegraphics[scale=0.45]{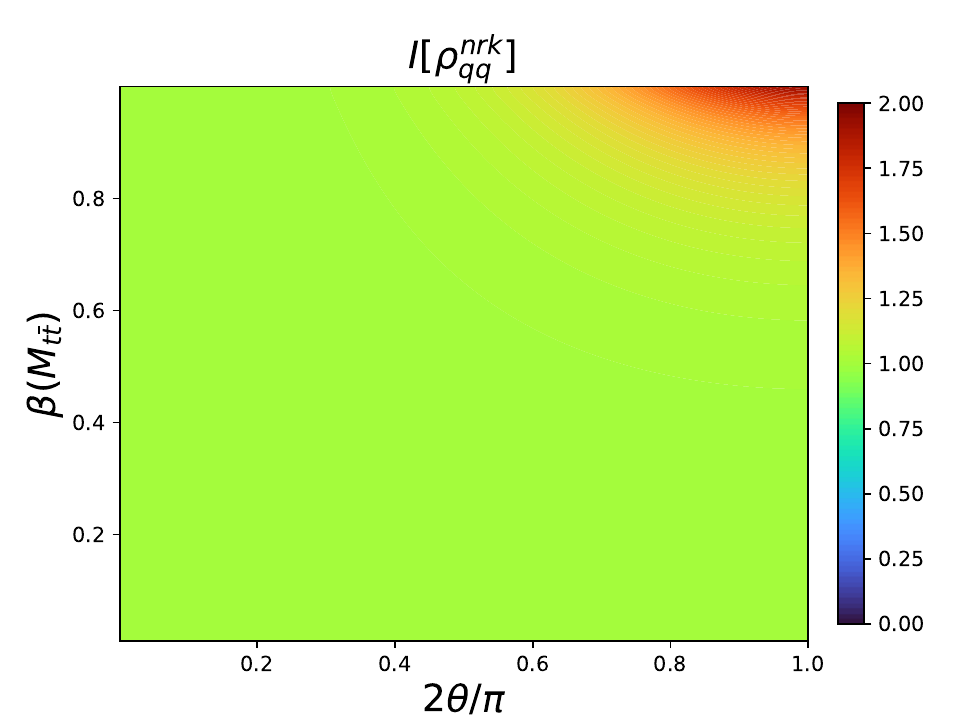}
  \includegraphics[scale=0.45]{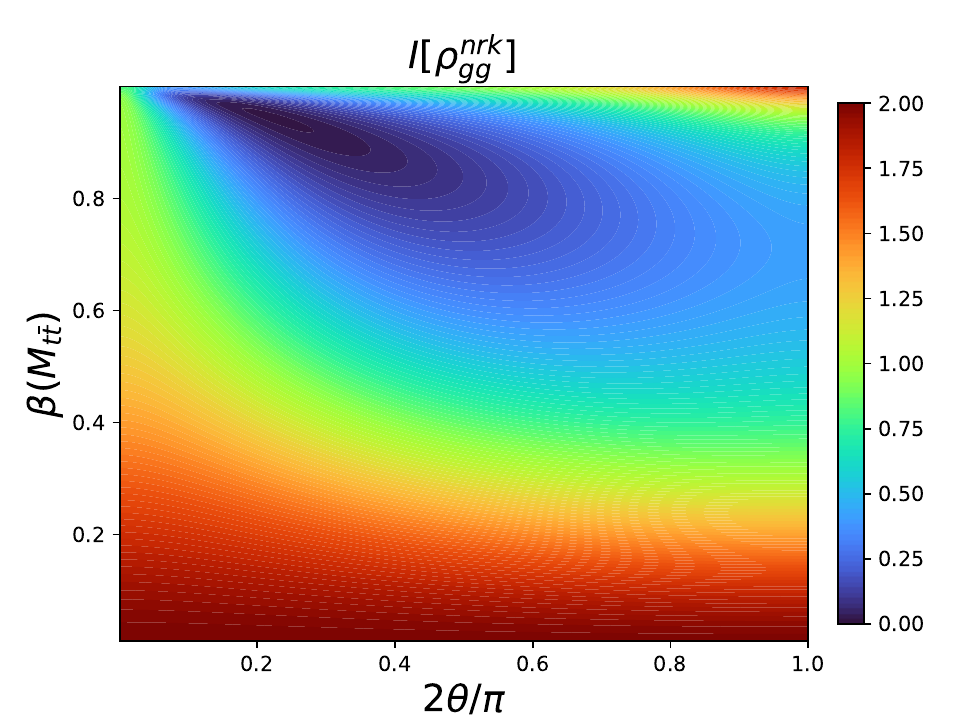}\\
  \includegraphics[scale=0.45]{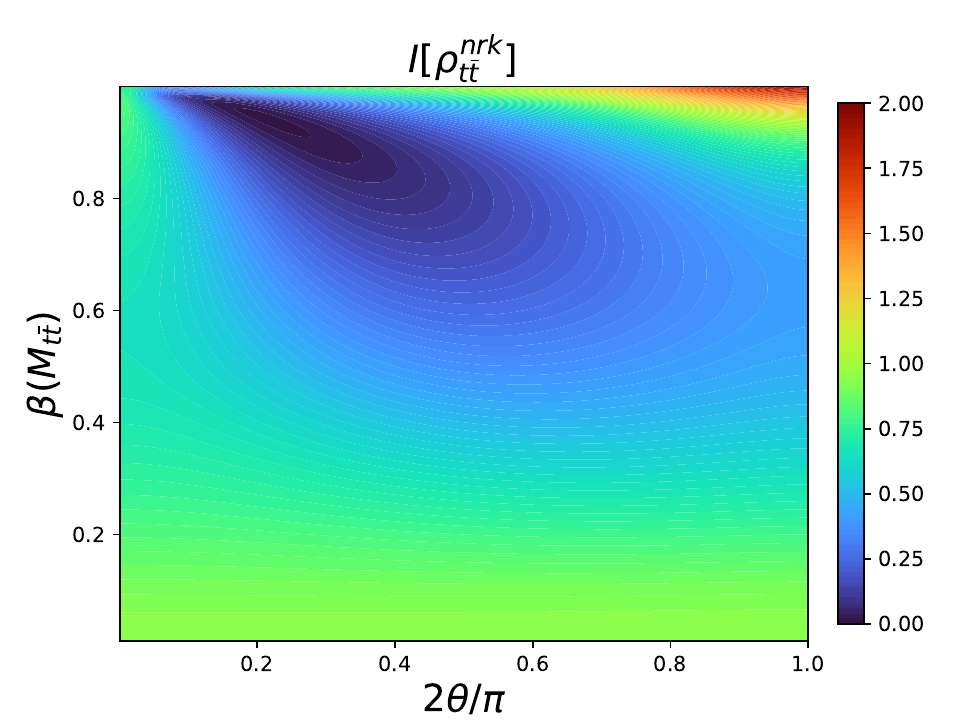}  
  \includegraphics[scale=0.45]{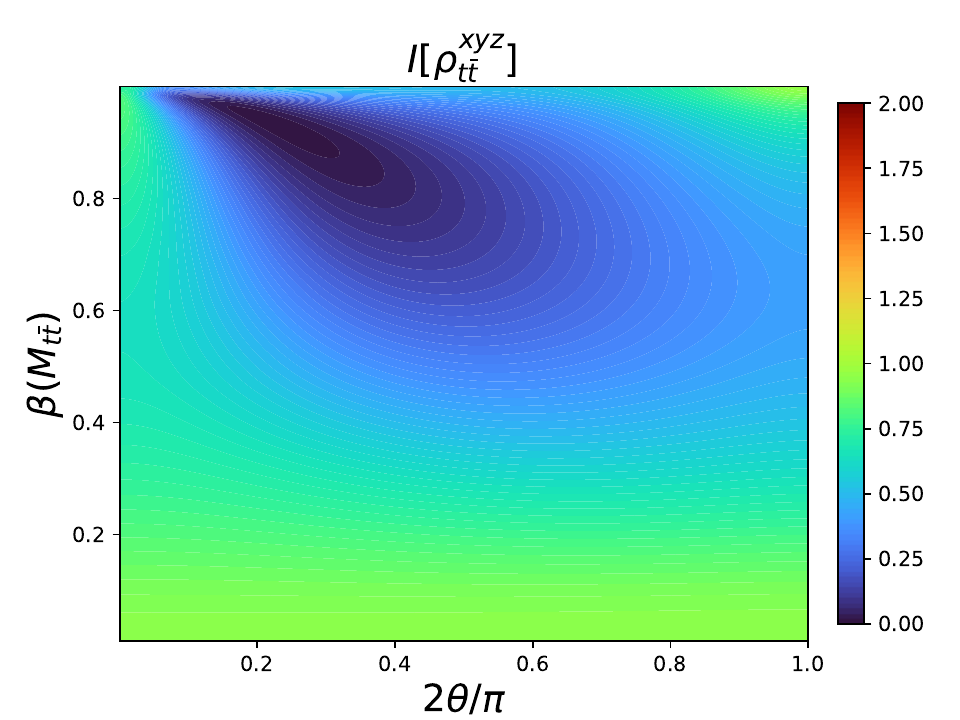}  
  \caption{The total mutual information for $q\bar{q} \to t\bar{t}$ (top left), $gg \to t\bar{t}$ (top right), $pp \to t\bar{t}$ (bottom left) in the helicity basis.  The process $pp \to t\bar{t}$ (bottom right) is also shown in the beam basis.  For a two qubit system the total mutual information ranges from 0 to 2.}
\label{fig:QI_totalmutualinformation}
\end{figure}

The total mutual information of the $t\bar{t}$ system is shown in Fig.~\ref{fig:QI_totalmutualinformation}.  In $q\bar{q}$ production (Fig.~\ref{fig:QI_totalmutualinformation}, top left), the $t\bar{t}$ quantum state has a single bit of mutual information over most of phase space because $I(\rho^{(X)}_{\rm mix}) = 1$.  In the high-$p_T$ region the total mutual information becomes maximal $I(\rho^{(+)}) = 2$.  In $gg$ production (Fig.~\ref{fig:QI_totalmutualinformation}, top right), the total mutual information reaches near maximal values at threshold, $I(\rho^{(-)}) = 2$, and at high-$p_T$, $I(\rho^{(+)}) = 2$.  In between these cases, the mixed state takes a range of values, even approaching $I=0$.  In $pp \to t\bar{t}$ (Fig.~\ref{fig:QI_totalmutualinformation}, bottom left) there is some total mutual information at threshold, which shrinks to near zero, before reaching nearly $I(\rho^{(+)})=2$ at high-$p_T$.

There is a second definition of mutual information which is known as the ``classical'' mutual information $J_A(\rho)$.  We first define the axis-dependent quantity $J_A(\rho;\hat{n})$ as
\begin{equation}
\label{eq:classical_mutual_information_aux}
J_A(\rho_{AB} ; \hat{n}) = S(\rho_A) - S(\rho_A | \rho_B ; \hat{n}).
\end{equation}
The term $S(\rho_A | \rho_B ; \hat{n})$ is the entropy of subsystem $A$ given a measurement of subsystem $B$.  A measurement is defined by a set of one-dimensional projectors $\Pi_{\hat{n}}$ where $\hat{n}$ corresponds to the axis along which the spin of $B$ is measured.  A complete measurement has the set of $\Pi_{+\hat{n}}$ and $\Pi_{-\hat{n}}$ which are
\begin{equation}
\Pi_{\pm \hat{n}} = \mathbb{I}_2 \otimes \ket{\pm n} \bra{\pm n}.
\end{equation}
The eigenstate $\ket{\pm n}$ is defined by $\vec{\sigma} \cdot \hat{n} \ket{\pm n} = \pm  \ket{\pm n}$.  The term $S(\rho_A | \rho_B ; \hat{n})$ is then written as
\begin{equation}
S(\rho_A | \rho_B ; \hat{n}) = p_{+\hat{n}} S(\rho_{+\hat{n}})
+ p_{-\hat{n}} S(\rho_{-\hat{n}}).
\end{equation}
where the measurement probabilities are
\begin{equation}
p_{\pm \hat{n}} = \text{tr}( \Pi_{\pm \hat{n}} \rho_{AB} \Pi_{\pm \hat{n}} ),
\end{equation}
and are used to define the normalized state of $A$ after the measurement of $B$
\begin{equation}
\rho_{\pm \hat{n}} = \frac{1}{p_{\pm \hat{n}}}
\text{tr}_{\smallB}( \Pi_{\pm \hat{n}} \rho_{AB} \Pi_{\pm \hat{n}} ),
\end{equation}
The quantum property that measurements on a state alter the state is reflected in the dependence of Eq.~\eqref{eq:classical_mutual_information_aux} on the measurement projectors $\Pi_{\pm \hat{n}}$, or equivalently, the measurement axis $\hat{n}$. 

The classical mutual information removes the $\hat{n}$-dependence by maximizing over all possible $\hat{n}$ values
\begin{equation}
\label{eq:classical_mutual_information}
J_A(\rho_{AB}) = \underset{\hat{n}}{\text{max}}  \;
J_A(\rho_{AB} ; \hat{n}).
\end{equation}
This corresponds to the choice of measurement that least disturbs subsystem $A$.  The classical mutual information is not necessarily symmetric between $A$ and $B$.  We can also write
\begin{equation} 
\label{eq:classical_mutual_information2}
J_A(\rho_{AB}) = S(\rho_A) - \underset{\hat{n}}{\text{min}}\left( p_{+\hat{n}} S(\rho_{+\hat{n}})
+ p_{-\hat{n}} S(\rho_{-\hat{n}}) \right).
\end{equation}
The classical moniker is appropriate because it represents the correlations in the system that are invariant under subsystem measurements.  Correlations that are fundamentally quantum mechanical will change as one subsystem is measured and do not contribute to Eq.~\eqref{eq:classical_mutual_information2}.  The remaining correlations are classical in nature.  Note that $J_A(\rho) \leq I(\rho)$ for any $\rho$ and for two qubits $J_A(\rho)$ ranges from 0 to 1~\cite{PhysRevA.84.042124}.

\begin{figure}
\centering
  \includegraphics[scale=0.45]{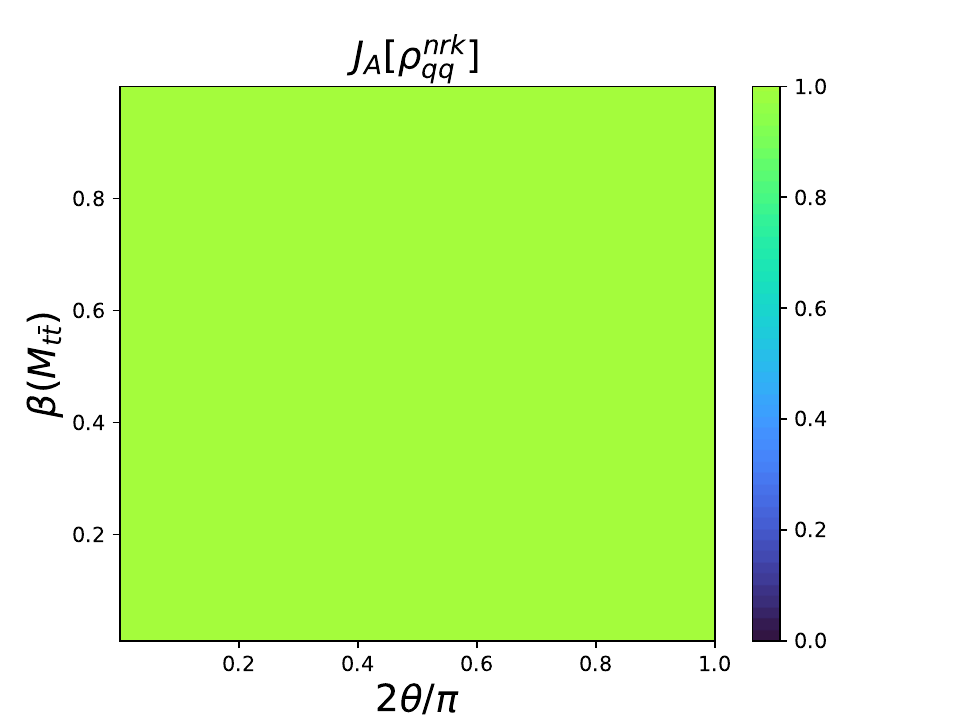}
  \includegraphics[scale=0.45]{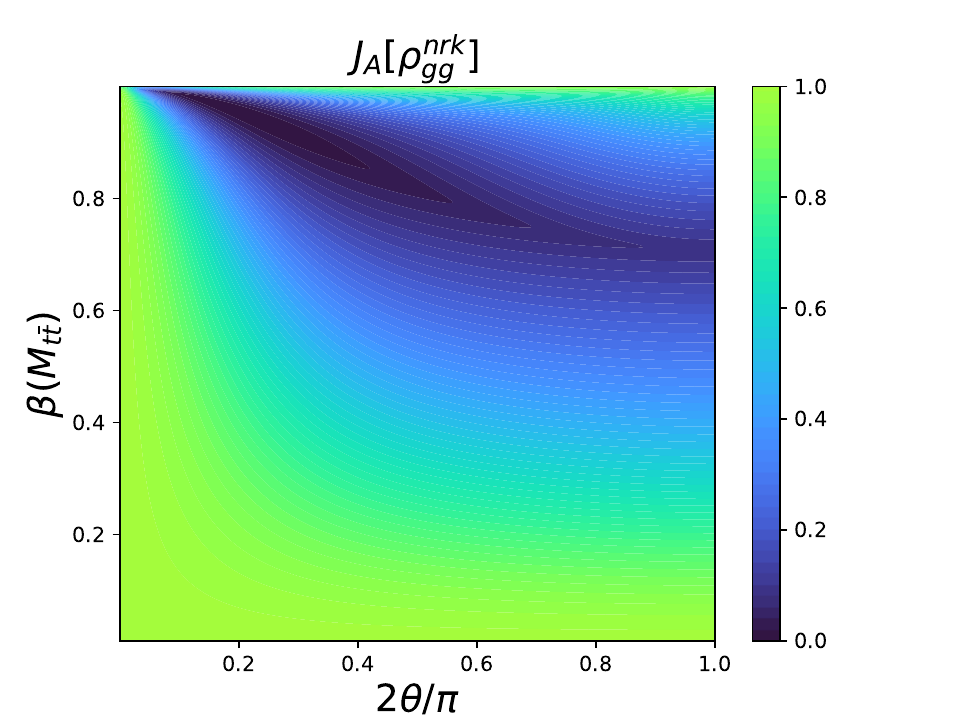}\\
  \includegraphics[scale=0.45]{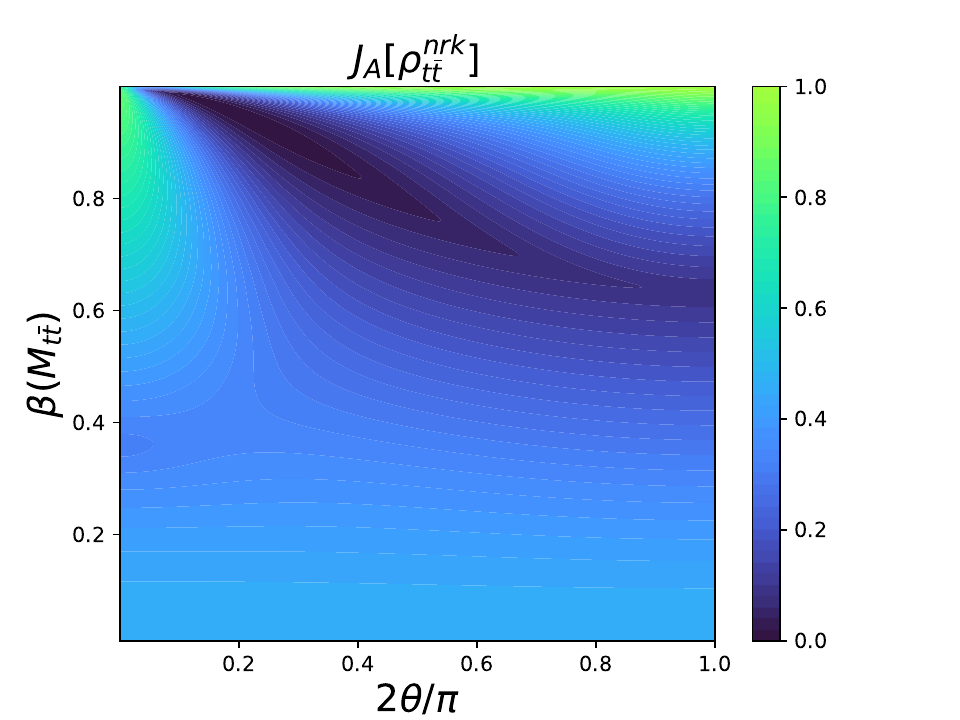}  
  \includegraphics[scale=0.45]{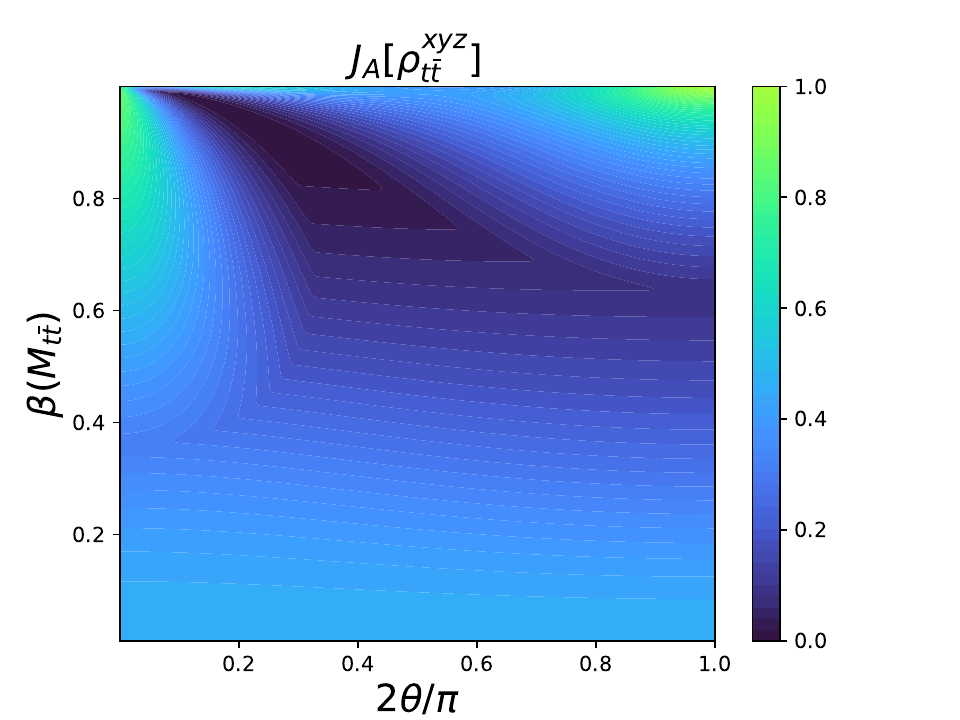}  
  \caption{The classical mutual information for $q\bar{q} \to t\bar{t}$ (top left), $gg \to t\bar{t}$ (top right), $pp \to t\bar{t}$ (bottom left) in the helicity basis.  The process $pp \to t\bar{t}$ (bottom right) is also shown in the beam basis.  For a two qubit system the classical mutual information ranges from 0 to 1.}
\label{fig:QI_classicalmutualinformation}
\end{figure}

Figure~\ref{fig:QI_classicalmutualinformation} shows the classical mutual information in the $t\bar{t}$ system.  For the $q\bar{q}$ subprocess (Fig.~\ref{fig:QI_classicalmutualinformation}, top left) $J_A(\rho^{(X)}_{\rm mix})=1$ and $J_A(\rho^{(+)})=1$ leading to a uniform value over the full phase space.  Comparing this to Fig.~\ref{fig:QI_totalmutualinformation}, we see that while the state produced in $q\bar{q}$ is correlated over the entire phase, only in the high-$p_T$ region does the state have quantum correlations.  The pattern is similar in $gg$ production (Fig.~\ref{fig:QI_classicalmutualinformation}, top right) where both $J_A(\rho^{(-)})=1$ and $J_A(\rho^{(+)})=1$ which is the maximal amount of classical correlation.  The $pp \to t\bar{t}$ case (Fig.~\ref{fig:QI_classicalmutualinformation}, bottom left) indicates a comparatively low amount of classical mutual information except where the state approaches $\rho^{(+)}$.

\subsection{Quantum Discord}
\label{sec:discord}

The quantum discord $D_A(\rho)$ is the difference between the total mutual information and the classical mutual information~\cite{Zurek:2003zz,Ollivier:2001fdq}.  Classically, this quantity vanishes. However in quantum mechanical systems it can be non-zero.  Heuristically, it captures the correlations in a system that are quantum mechanical in nature and cannot be attributed to classical correlations.

For pure states, the quantum discord coincides with the concurrence,\footnote{The concurrence is a measure of entanglement and ranges from $0$ to $1$.  A value of 0 indicates a separable state while a non-zero value indicates non-zero entanglement.} however, for mixed states they probe different quantum mechanical aspects of a system.  In particular, separable states can have either zero or non-zero discord.  In order words, the set of zero discord states is a subset of the set of separable states.

The quantum discord of $A$ is defined as $D_A(\rho_{AB}) = I(\rho_{AB}) - J_A(\rho_{AB})$ which leads to
\begin{equation} \label{eq:DA}
D_A(\rho_{AB})
= S(\rho_B) - S(\rho_{AB}) + \underset{\hat{n}}{\text{min}}\left( p_{+\hat{n}} S(\rho_{+\hat{n}})
+ p_{-\hat{n}} S(\rho_{-\hat{n}}) \right).
\end{equation}
Like the classical mutual information, the quantum discord does not need to be symmetric with respect to the two subsystems.  In the $t\bar{t}$ system, $D_A(\rho_{AB}) = D_B(\rho_{AB})$ is enforced at leading order by CP-invariance.  Therefore, the presence of CP violation in $t\bar{t}$ production is manifested by $D_A(\rho_{AB}) \neq D_B(\rho_{AB})$.

Since $I(\rho_{AB}) \geq J_A(\rho_{AB})$, quantum discord is always $\geq 0$ and in two qubit systems it is bounded above by 1~\cite{PhysRevA.82.052122,Xi_2011}.

For the parametrization of Eq.~\eqref{eq:Fano_decomp} we have~\cite{Afik:2022dgh}
\begin{equation}
p_{\pm \hat{n}} = \frac{1 \pm \mathbf{\hat{n}} \cdot \mathbf{B^-}}{2},
\qquad\qquad
\rho_{\pm \hat{n}} = 
 \frac{\mathbb{I}_2 + \mathbf{B^+_{\pm\hat{n}}} \cdot\mathbf{\sigma} }{2},
\qquad\qquad
\mathbf{B^+_{\pm\hat{n}}} =
\frac{\mathbf{B^+} \pm \mathbf{C} \cdot \mathbf{\hat{n}}}{1 \pm \mathbf{\hat{n}} \cdot \mathbf{B^-}}.
\end{equation}
Due to the minimization in Eq.~\eqref{eq:DA}, computing quantum discord exactly is generally considered to be difficult.  In fact, a number of proxies have been suggested to avoid this minimization~\cite{Bera_2017}.  Fortunately, the $t\bar{t}$ state falls into the set of states for which an analytic solution to Eq.~\eqref{eq:DA} is known~\cite{Luo:2008ecu}.  This solution allows us to compute and measure quantum discord exactly.

In the beam basis, where the quantum state is fully characterized by the spin correlation matrix
\begin{equation} 
C_{ij} = \left(\begin{array}{ccc}
C_\perp & 0 & 0 \\
0 & C_\perp & 0  \\
0 & 0 & C_z 
\end{array}\right),
\end{equation}
the quantum discord of $A$ is given by
\begin{equation} \label{eq:DA-beam}
\begin{aligned}
D_A(\rho_{t\bar{t}}) & =
 1 + \frac{1}{2}(1+C_z) \log_2\left(\frac{1+C_z}{4}\right)
+ \frac{1}{4}(1+2C_\perp-C_z) \log_2\left(\frac{1+2C_\perp-C_z}{4}\right) \\
& \quad + \frac{1}{4}(1-2C_\perp-C_z) \log_2\left(\frac{1-2C_\perp-C_z}{4}\right) \\
& \quad - \frac{1}{2}(1+C_{\rm max}) \log_2 \left(\frac{1+C_{\rm max}}{2}\right) - \frac{1}{2}(1-C_{\rm max}) \log_2 \left(\frac{1-C_{\rm max}}{2}\right),
\end{aligned}\end{equation}
where we have defined 
\begin{equation}
C_{\rm max} = \text{max}\{ |C_{\perp}|, |C_{z}| \}.
\end{equation}
Eq.~\eqref{eq:DA-beam} is the exact calculation of the quantum discord including the minimization calculation.

In the helicity basis, where the spin correlation is
\begin{equation} 
C_{ij} = \left(\begin{array}{ccc}
C_k & C_{kr} & 0 \\
C_{kr} & C_r & 0  \\
0 & 0 & C_n 
\end{array}\right),
\end{equation}
the quantum discord of $A$ is given exactly by
\begin{align}
D_{A}(\rho_{t\bar{t}})=1 &+ \frac{1}{4}(1-C_{k}-C_{n}-C_{r})\log_{2}\left(\frac{1-C_{k}-C_{n}-C_{r}}{4}\right) \nonumber  \\
&+ \frac{1}{4}(1+C_{k}-C_{n}+C_{r})\log_{2}\left(\frac{1+C_{k}-C_{n}+C_{r}}{4}\right) \nonumber \\
&+ \frac{1}{4}(1+C_{n} - \Delta)\log_{2}\left(\frac{1+C_{n} - \Delta}{4}\right) + \frac{1}{4}(1+C_{n} + \Delta)\log_{2}\left(\frac{1+C_{n} + \Delta}{4}\right) \nonumber \\
&-\frac{1}{2}(1+\lambda)\log_{2}\left(\frac{1+\lambda}{2}\right)-\frac{1}{2}(1-\lambda)\log_{2}\left(\frac{1-\lambda}{2}\right)
\end{align}
where we have defined
\begin{equation}
\Delta = \sqrt{C_{k}^{2} + 4C_{kr}^{2} +C_{r}^{2} - 2C_{k}C_{r}},
\qquad
\lambda = \text{max}\{|C_{n}|,\frac{1}{2}\left|C_{k}+C_{r}-\Delta\right|,\frac{1}{2}\left|C_{k}+C_{r}+\Delta\right|  \}.
\end{equation}
A full basis-independent derivation can be found in Appendix~\ref{sec:App_A}.

\begin{figure}
\centering
  \includegraphics[scale=0.45]{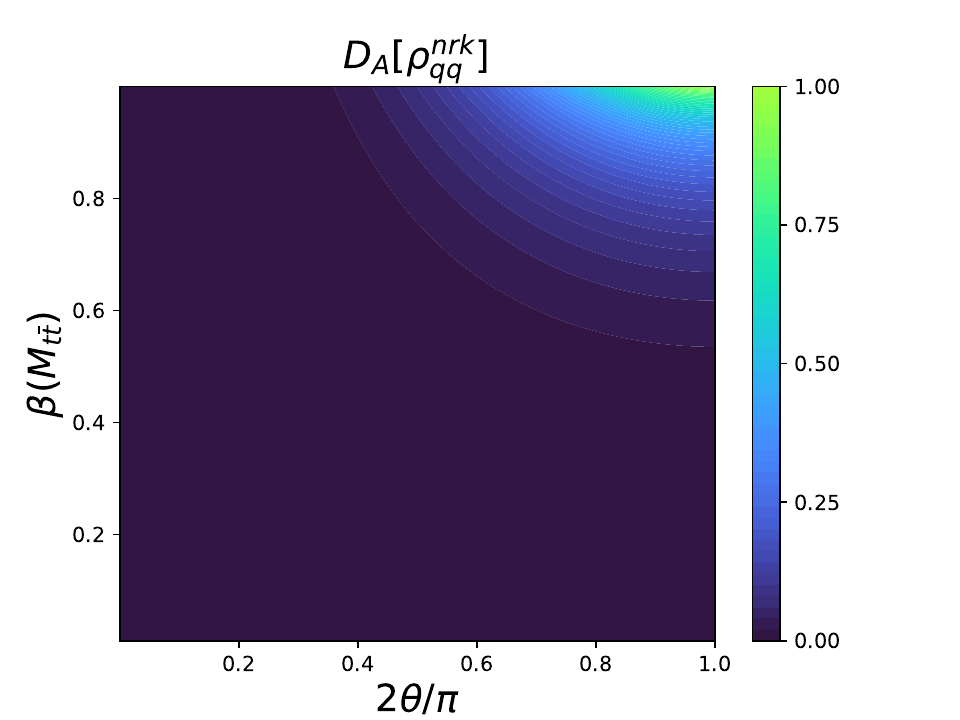}
  \includegraphics[scale=0.45]{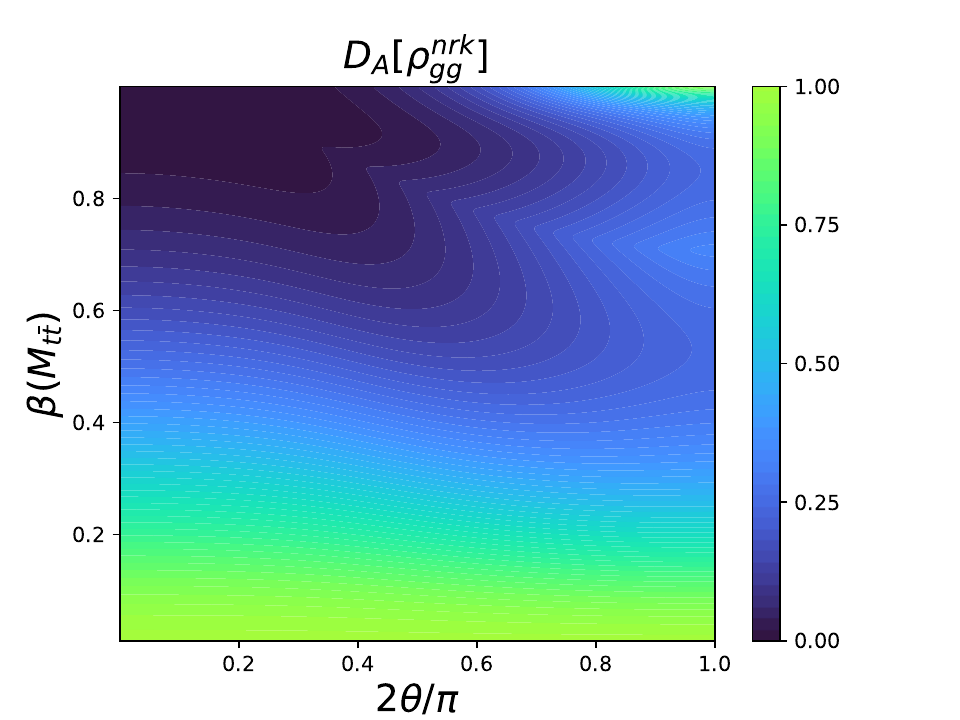}\\
  \includegraphics[scale=0.45]{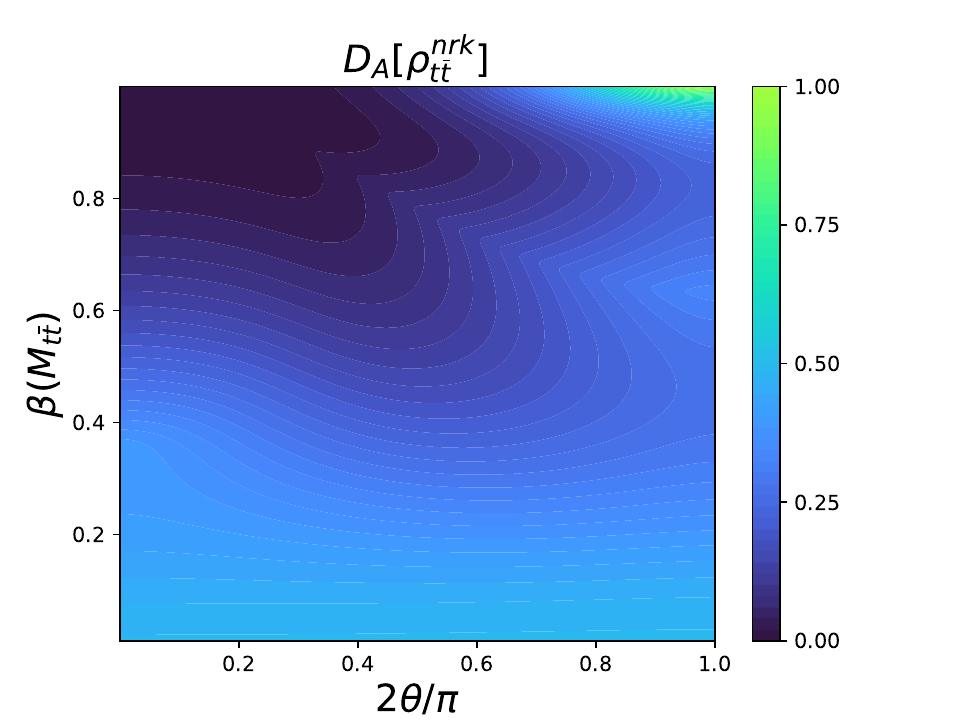}  
  \includegraphics[scale=0.45]{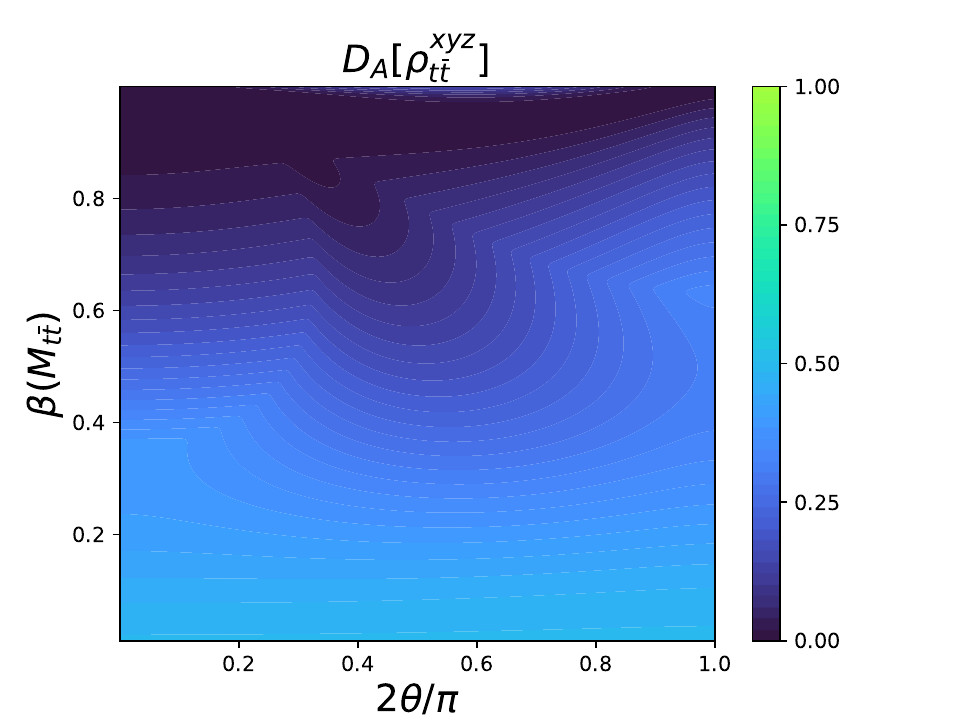}  
  \caption{The quantum discord for $q\bar{q} \to t\bar{t}$ (top left), $gg \to t\bar{t}$ (top right), $pp \to t\bar{t}$ (bottom left) in the helicity basis.  The process $pp \to t\bar{t}$ (bottom right) is also shown in the beam basis.  For a two qubit system the quantum discord ranges from 0 to 1.}
\label{fig:QI_quantumdiscord}
\end{figure}

Quantum discord for the $t\bar{t}$ system is shown in Fig.~\ref{fig:QI_quantumdiscord}. For the $q\bar{q}$ channel (Fig.~\ref{fig:QI_quantumdiscord}, top left), the discord is $D_A=0$ over the most of the phase space because $D_A(\rho^{(X)}_{\rm mix})=0$.  At high-$p_T$ we have $D_A(\rho^{(+)})=1$.  While the $t\bar{t}$ state originating from $q\bar{q}$ exhibits correlations, only when the state starts to approach $\rho^{(+)}$ are those quantum correlations.

In the $gg$ channel (Fig.~\ref{fig:QI_quantumdiscord}, top right), the two cases near the ends of phase space, corresponding to $\rho^{(\pm)}$ have $D_A(\rho^{(\pm)})=1$.  In between these regions, the discord is small, which is expected in any case since this region did not even exhibit substantial total correlations.  Finally, in $pp \to t\bar{t}$ (Fig.~\ref{fig:QI_quantumdiscord}, bottom left) the discord is small over the majority of the phase space where the state is quite mixed and approaches $D_A(\rho^{(+)})=1$ at high-$p_T$.  Note that while the quantum discord is small over much of phase space, it is still measurable as we will show in Sec.~\ref{sec:collider}.

\subsection{Conditional Entropy}
\label{sec:condentropy}

An additional quantity that is interesting in a quantum mechanical system is the quantum conditional entropy which is given by
\begin{equation}
S(\rho_A | \rho_B) = S(\rho_{AB}) - S(\rho_B),
\end{equation}
This is a direct analog of the classical conditional entropy.\footnote{Note that this is different than the quantity introduced in Eq.~\eqref{eq:classical_mutual_information_aux}.}  Physically, the conditional entropy measures the number of bits that must be shared with subsystem $A$ in order to reconstruct subsystem $B$.  Thus a larger conditional entropy indicates that subsystem $A$ needs more information to infer subsystem $B$.  By subadditivity, the maximal conditional entropy is $S=1$ for the two qubit system.  A conditional entropy of $S=0$ indicates that subsystem $A$ does not need any additional communication to fully reconstruct subsystem $B$.  Classically, $S(\rho_A | \rho_B) \geq 0$.

In a quantum system, however, the conditional entropy can be negative.  The physical interpretation of negative conditional entropy is that bits are available for potential future quantum communication between the two subsystems~\cite{Horodecki_2005}.  This is a stronger correlation than Bell nonlocality, meaning that the set of negative conditional entropy states is smaller than the set of Bell nonlocal states~\cite{Friis_2017}.

\begin{figure}
\centering
  \includegraphics[scale=0.45]{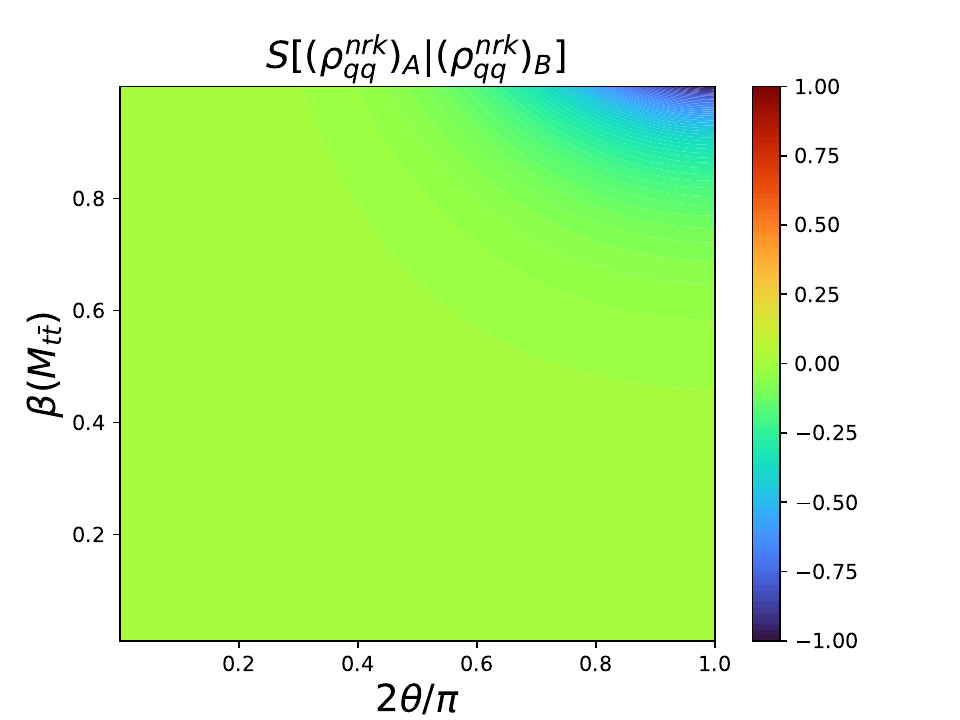}
  \includegraphics[scale=0.45]{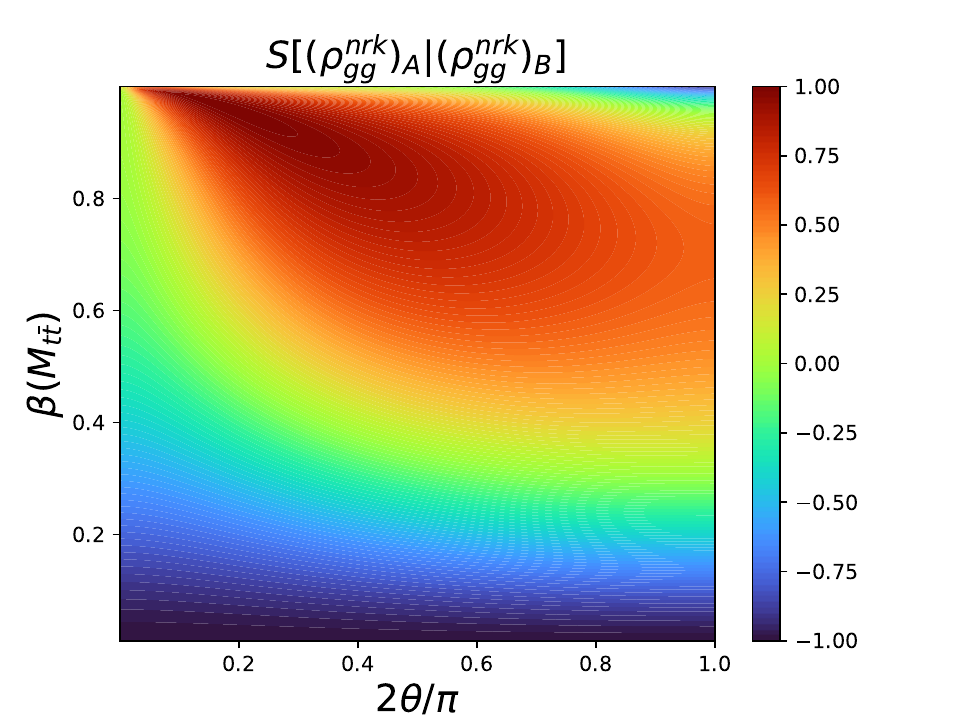}\\
  \includegraphics[scale=0.45]{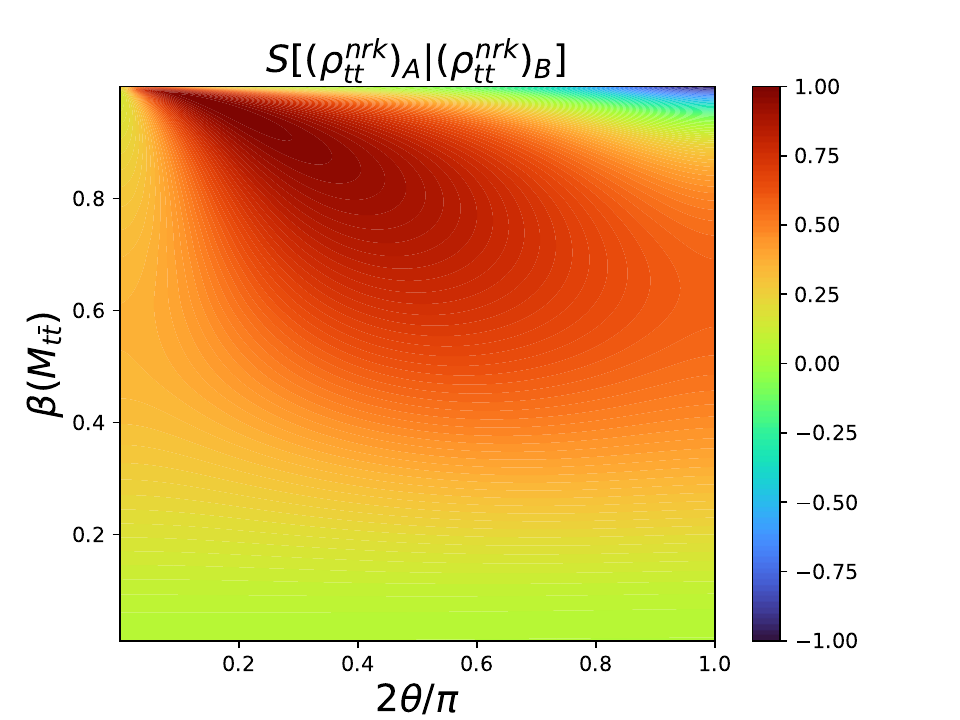}  
  \includegraphics[scale=0.45]{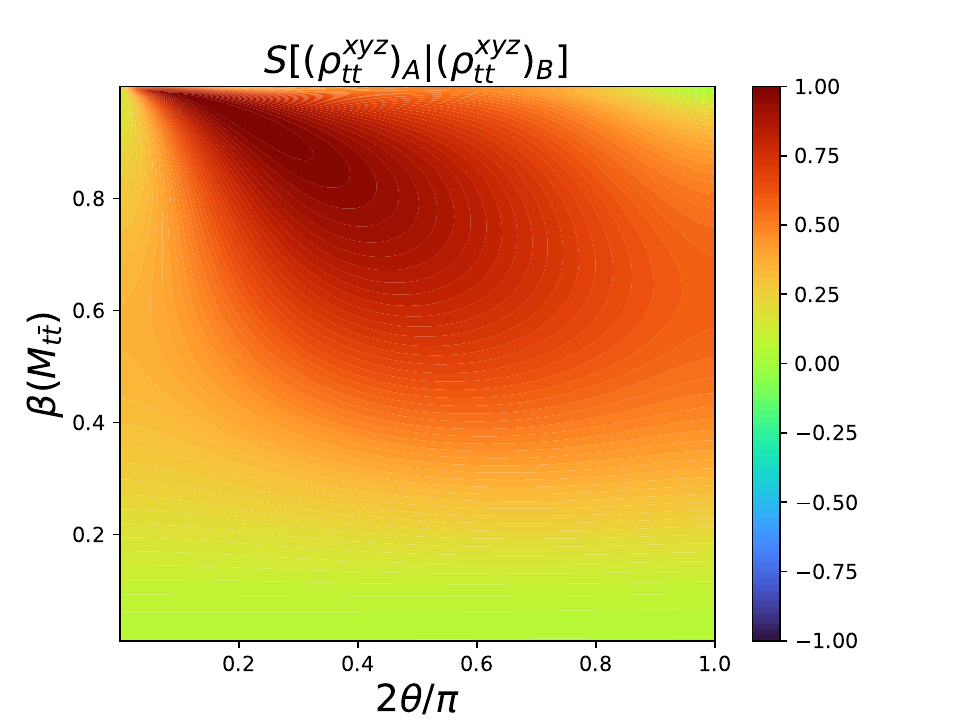}  
  \caption{The conditional entropy for $q\bar{q} \to t\bar{t}$ (top left), $gg \to t\bar{t}$ (top right), $pp \to t\bar{t}$ (bottom left) in the helicity basis.  The process $pp \to t\bar{t}$ (bottom right) is also shown in the beam basis. 
 For a two qubit system the conditional entropy ranges from -1 to 1.}
\label{fig:QI_conditionalentropy}
\end{figure}

In Fig.~\ref{fig:QI_conditionalentropy} we show the conditional entropy.  From the $q\bar{q}$ channel (Fig.~\ref{fig:QI_conditionalentropy}, top left) the conditional entropy is zero over most of phase space.  For $\rho^{(X)}_{\rm mix}$ subsystem $A$ has sufficient information to reconstruct subsystem $B$.  In the high-$p_T$ region the $t\bar{t}$ states approaches $\rho^{(+)}$ leading to a conditional entropy of $-1$.  The $gg$ channel (Fig.~\ref{fig:QI_conditionalentropy}, top right) covers the full range of conditional entropy values ranging from $-1$ near threshold and at high-$pT$ and reaching values near $1$ in between.  Similar to the other quantities studied, in the $pp \to t\bar{t}$ (Fig.~\ref{fig:QI_conditionalentropy}, bottom left) the contamination of $\rho^{(-)}$ state near threshold from $q\bar{q}$ dilute the quantum effects leading to negative conditional entropy only being possible at high-$p_T$.  A full study of negative conditional entropy at the LHC is beyond the scope of this study and left for future work.

\section{Collider Analysis}
\label{sec:collider}

Evaluating quantum information measures of the $t\bar{t}$ system at the LHC involves reconstructing the spin correlation matrix from collider observables. Previous works on entanglement and quantum tomography of the $t\bar{t}$ system have focused on leveraging the angular distributions of the top quark decay products to extract spin correlation coefficients~\cite{Fabbrichesi:2021npl,Severi:2021cnj,Dong:2023xiw,Han:2023fci}. We will refer to this approach to quantum tomography as the \textit{decay method}. 

Recently, a new method has emerged, referred to as the \textit{kinematic method}, which utilizes the relationship in a given system between the kinematics and the spin correlation matrix values to measure the spin correlation matrix directly from the reconstructed $t\bar{t}$ kinematics~\cite{Cheng:2024rxi}.  In this section, we outline these methods for reconstructing the $t\bar{t}$ density matrix and discuss their use in measuring the quantum discord of the $t\bar{t}$ quantum state at the LHC. 

\subsection{Decay Method}
\label{sec:decay}

Spin correlations of the $t\bar{t}$ system at hadron colliders have been studied for some time as a probe of fundamental properties of the top quark~\cite{Barger:1988jj}. Quantum tomography of top quarks using the decay method relies on the fact that the short top quark lifetime, $\tau_{t}\sim1/(\text{GeV})$, prevents the effects of QCD hadronization from diluting the transfer of spin information to decay products of the top. We may parameterize the decay of one of the tops as 
\begin{equation}
\bar{\Gamma}(\Omega)_{\alpha\beta} \propto \delta_{\alpha\beta} + \kappa\Omega_{i}\sigma^{i}_{\alpha\beta},
\end{equation}
where in this expression the indices $\alpha$ and $\beta$ label the top-spin indices, and $\Omega_i\ (i=1,2,3)$ represents the direction of the decay product of interest~\cite{Han:2023fci}. In this expression, most of the phase space of the decay width has been integrated, except for a single integral over the solid angle of one of the decay products. The proportionality of the remaining angular dependence, $\kappa$, is known as the spin analyzing power of the decay product. In the SM, values of the spin analyzing power have been calculated for down-type fermions $\kappa_{\ell,d}=+1$, up-type fermions $\kappa_{\nu,u}=-0.3$, both of which originate from the decay of the $W$, and also for the $b$-quark $\kappa_{b}=-0.4$~\cite{Brandenburg:2002xr}. 

Including the decay of both top quarks, we may parameterize the differential $t\bar{t}$ cross section as 
\begin{equation}
\frac{1}{\sigma}\frac{d^{2}\sigma}{d\cos\theta_{i}^{a}d\cos\theta_{j}^{b}} = \frac{1}{4}\left(1+\kappa_{a}\kappa_{b}C_{ij}\cos\theta_{i}^{a}\cos\theta_{j}^{b}\right),
\end{equation}
where $a$ and $b$ label a given particle in the decay chain of the top anti-top system. We have ignored the individual polarization coefficients of each top quark as they are known to vanish in the leading order cross section.  The angles $\theta^a_i$ and $\theta^b_j$, are defined as the relative directions between the momentum of the decay product evaluated in the respective top or anti-top rest frame and a choice of spin quantization axes, $\{\hat{i}\}$,
\begin{equation}
\cos\theta_{i}^{a}=\hat{a}\cdot\hat{i} \hspace{1.5cm} \cos\theta_{j}^{b}=\hat{b}\cdot\hat{j},
\end{equation}
where $\hat{a}$ and $\hat{b}$ are the momenta of the decay products evaluated in the rest frame of their respective parent particle.

The spin correlation coefficients can be extracted from the differential cross section in a variety of ways. In this paper, we will focus on the angular asymmetry
\begin{equation}
C_{ij}=\frac{4}{\kappa_{a}\kappa_{b}}\frac{N(\cos\theta_{i}^{a}\cos\theta_{j}^{b}>0) - N(\cos\theta_{i}^{a}\cos\theta_{j}^{b}<0)}{N(\cos\theta_{i}^{a}\cos\theta_{j}^{b}>0) + N(\cos\theta_{i}^{a}\cos\theta_{j}^{b}<0)}.
\label{eq:asymm}
\end{equation}
For further details regarding the angular decomposition of the $t\bar{t}$ cross section and reference to other methods of extracting the spin correlation coefficients, see Refs.~\cite{Fabbrichesi:2021npl,Severi:2021cnj,Dong:2023xiw,Han:2023fci}.

We will evaluate quantum information measures with respect to three different choices of spin quantization axes:
\begin{itemize}

\item {\bf Beam basis}: starting from the COM frame of the $t\bar{t}$ system the beam basis sets $\hat{z}$ along the beam direction and lets $\hat{x}$ and $\hat{y}$ span the orthogonal plane.

\item {\bf Helicity basis}: starting from the COM frame we define $\hat{k}$ along the top quark three-momentum forming an angle $\theta$ with respect to the $\hat{z}$ direction, then $\hat{r}$ is $\hat{r} = (\hat{z} - \cos\theta \hat{k})/\sin\theta$, and $\hat{n} = \hat{r} \times \hat{k}$.

\item {\bf Lab frame basis}: starting from the lab frame we set $\hat{z}$ along the beam direction and let $\hat{x}$ and $\hat{y}$ span the orthogonal plane.  This basis would be better labeled as the beam basis in the lab frame, but for simplicity we call it the lab frame basis.  This basis is useful because it corresponds to an ensemble of similarily-prepared systems as is needed for a genuine quantum state~\cite{Cheng:2024btk}.

\end{itemize}
Further discussion can be found in Refs.~\cite{Han:2023fci,Cheng:2023qmz}.

\subsection{Kinematic Method}
\label{sec:kinematic}

The kinematic method for reconstructing spin correlations for $t\bar{t}$ is complementary to the decay method, but has a number of advantages~\cite{Cheng:2023qmz}. This method exploits the symmetries of the $t\bar{t}$ cross section to express the measured spin density matrix in terms of reconstructed kinematic variables, namely, the $t\bar{t}$ invariant mass and the COM scattering angle. This simplifies the number of variables needed to evaluate the underlying quantum states of the $t\bar{t}$ system. Once the top quark kinematics are reconstructed in an event the spin correlations are directly evaluated using an analytical template. For example, at leading order in the helicity basis we have
\begin{equation}
C^{\rm helicity}_{ij}=\begin{pmatrix}
C_{k}(\theta,M_{t\bar{t}})&C_{kr}(\theta,M_{t\bar{t}})&0\\C_{kr}(\theta,M_{t\bar{t}})&C_{r}(\theta,M_{t\bar{t}})&0\\0&0&C_{n}(\theta,M_{t\bar{t}})
\end{pmatrix},
\label{eq:ttbar_substate}
\end{equation}
where $\theta$ and $M_{t\bar{t}}$ are the COM scattering angle and $t\bar{t}$ invariant mass respectively. Analytic formulae for the leading order $t\bar{t}$ density matrix coefficients used for the kinematic method can be found in Refs.~\cite{Baumgart:2012ay,Afik:2020onf,Aoude:2022imd,Barr:2024djo}.

Spin correlations in the fixed COM beam basis and lab frame beam basis are obtained after an $SO(3)$ rotation and Wigner rotation, respectively~\cite{Cheng:2024btk}:
\begin{equation}
C^{{\rm beam}} = R(\theta,\phi)C^{\rm helicity}R(\theta,\phi)^{T},
\end{equation}
\begin{equation}
C^{\rm lab} = R_{t}(\Lambda_{\Omega})C^{\rm beam}R_{\bar{t}}(\Lambda_{\Omega})^{T}.
\end{equation}
The $t\bar{t}$ density matrix weighted by its production rate from $q\bar{q}$ annihilation or gluon fusion is then calculated directly from an event as 
\begin{equation}
\label{eq:rhoThetaMtt}
\rho(\theta,M_{t\bar{t}})=\frac{L_{q\bar{q}}|\mathcal{M}_{q\bar{q}\to t\bar{t}}|^{2}\rho_{q\bar{q}\to t\bar{t}}(\theta,M_{t\bar{t}}) + L_{gg}|\mathcal{M}_{gg\to t\bar{t}}|^{2}\rho_{gg\to t\bar{t}}(\theta,M_{t\bar{t}})}{L_{q\bar{q}}|\mathcal{M}_{q\bar{q}\to t\bar{t}}|^{2} + L_{gg}|\mathcal{M}_{gg\to t\bar{t}}|^{2}},
\end{equation}
where $L_{I}$ and $|\mathcal{M}_{I\to t\bar{t}}|^{2}$ are the partonic luminosity functions and spin-averaged matrix elements for the $I=q\bar{q},gg$ production channels, respectively. We use the \texttt{NNPDF4.0} NNLO PDF set with $\alpha_{s}(m_{Z})=0.118$~\cite{NNPDF:2021njg} included in the \texttt{LHAPDF6} package~\cite{Buckley:2014ana}. In evaluating the PDFs, we set the factorization scale at $Q=M_{t\bar{t}}$.

\subsection{Fictitious States}
\label{sec:fictitious}

Quantum states produced at colliders and reconstructed from event data using either the decay method or kinematic method are defined as a sum over phase space
\begin{equation}
\rho_{t\bar{t}}(\Pi)=\sum_{\vec{k},\vec{v}\in\Pi}\rho(\vec{k},\vec{v})=\sum_{\vec{k},\vec{v}\in\Pi}\sum_{i,j}\rho(\vec{k},\vec{v})_{i,j}|i\rangle\langle j|,
\label{eq:full_qstate}
\end{equation}
where $\vec{k}$ is the top quark momentum in the COM frame and $\vec{v}$ is the velocity of the $t\bar{t}$ system relative to the lab frame. The basis vectors in the spin Hilbert space are defined as $|i\rangle=|\text{spin of }t\rangle\otimes|\text{spin of }\bar{t}\rangle$~\cite{Han:2023fci}. In this notation, each point in phase space $\vec{k},\vec{v}\in\Pi$ defines a quantum substate
\begin{equation}
\label{eq:sub_qstate}
\rho(\vec{k},\vec{v})=\sum_{i,j}\rho(\vec{k},\vec{v})_{i,j}|i\rangle\langle j|.
\end{equation}
When all quantum substates are evaluated with respect to the same spin quantization axis, Eq.~\eqref{eq:full_qstate} is a genuine quantum state. However, as discussed at length in Refs.~\cite{Han:2023fci,Cheng:2023qmz}, when the spin quantization axes are event-dependent, $\{|i\rangle\} =  \{|i(\vec{k})\rangle\}$, there is no obvious physical interpretation of the total quantum state and it is then referred to as a \textit{fictitious} quantum state~\cite{Afik:2022kwm}.

The leading order $t\bar{t}$ quantum state is completely defined by the spin correlation matrix $C_{ij}$ in Eq.~\eqref{eq:Fano_decomp}.  When the spin quantization axes are event-dependent, rather than $C_{ij}$, what is reconstructed at colliders is the averaged spin correlation matrix $\overline{C}_{ij}$, defined as
\begin{equation}
\overline{C}_{ij} =
\frac{1}{\sigma_{\Pi}}\int_{\Pi} d\Omega\frac{d\sigma}{d\Omega}C_{ij}(\Omega),
\label{eq:fictitious_state}
\end{equation}
where $C_{ij}(\Omega)$ is the spin correlation matrix of an underlying quantum substate at the phase space point $\Omega\in\Pi$ and $\sigma_{\Pi}$ is the production cross section integrated over the phase space region, $\Pi$.  

For any choice of basis, the decay method reconstructs the fictitious state, defined by $\overline{C}_{ij}$.  The kinematic method, on the other hand, directly calculates the substates, $C_{ij}(\Omega)$, event-by-event.  Thus, when using the kinematic method to measure a quantity $\mathcal{O}$, one can either reconstruct $\overline{C}_{ij}$ and use that to compute $\mathcal{O}(\overline{C}_{ij})$ or one can compute 
$\mathcal{O}$ event-by-event and take the average leading to $\overline{\mathcal{O}(C_{ij}(\Omega))}$.  For quantities that are linear functions of the spin correlation matrix components, these two averaging procedures are equivalent:
\begin{equation}
\mathcal{O}(\overline{C}_{ij})=
\mathcal{O}\left(\frac{1}{\sigma_{\Pi}}\int_{\Pi} d\Omega\frac{d\sigma}{d\Omega}C_{ij}(\Omega)\right)
=\frac{1}{\sigma_{\Pi}}\int_{\Pi} d\Omega\frac{d\sigma}{d\Omega}\mathcal{O}(C_{ij}(\Omega)),
\end{equation}
which implies $\mathcal{O}(\overline{C}_{ij})=\overline{\mathcal{O}(\Omega)}$.

For non-linear functions, such as entropy and discord, the two averaging procedures are not equivalent and we use $\mathcal{O}(\overline{C}_{ij})$.  This form of averaging ensures that the results are interpretable through fictitious states and that the value matches the decay method.

Apart from linearity, another crucial property of quantities defined for fictitious states is convexity.  Separable states form a convex set which means that mixed states formed from separable substates are also separable.  This property is important at colliders because it ensures that observing entanglement in a fictitious state implies that there is genuine entanglement present for the corresponding quantum state.\footnote{Entanglement of the fictitious state, strictly speaking, implies that there exists an entangled quantum substate~\cite{Cheng:2023qmz,Cheng:2024btk}.  Appropriate cuts on phase space, however, can be used to define a genuine quantum state that is entangled.}  Bell local states also form a convex set so that observations of Bell nonlocality at colliders are authentic.

Zero discord states, on the other hand, do not form a convex set.  This means that the observation of quantum discord, using a fictitious state, does not guarantee that the underlying quantum state has non-zero quantum discord.  The canonical example is the state $\rho_D = a \ket{\uparrow\uparrow} \bra{\uparrow\uparrow} + b \ket{\leftarrow\leftarrow} \bra{\leftarrow\leftarrow}$, for $|a|^2 + |b|^2 = 1$ and both non-zero.  The quantum discord of $\ket{\uparrow\uparrow} \bra{\uparrow\uparrow}$ is zero as is the quantum discord of $\ket{\leftarrow\leftarrow} \bra{\leftarrow\leftarrow}$, but the quantum discord of $\rho_D$ is non-zero.

One approach to ensure that genuine quantum discord is measured at colliders is to restrict measurements to regions where the fictitious state is entangled.  This works for quantum discord because the set of entangled states is a subset of the set of non-zero discord states thus an entangled fictitious state implies there is an entangled substate which implies there is a non-zero discord substate.  This applies for quantum discord and would not necessarily apply to other quantum information quantities.

Ensuring that a collider measurement of quantum discord is meaningful generally requires additional knowledge of the quantum state being studied.  As discussed, a fictitious state with non-zero quantum discord but with substates that all have  zero quantum discord would not qualify as an observation of quantum discord.  Conservatively, the converse is not problematic since it would not lead us to erroneously claim observation of quantum discord.  With this in mind, we can guarantee a meaningful observation of quantum discord by restricting the phase space to regions where all substates have non-zero quantum discord.

In the $t\bar{t}$ state at the LHC, all signal regions we select in Sec.~\ref{sec:results} have this property of no substates with zero discord.  This property would need to be verified for other final states for which quantum discord is measured.

\subsection{Results at the LHC}
\label{sec:results}

To simulate $t\bar{t}$ events for a  measurement of quantum discord at the LHC, we generate the hard scattering process, $pp\rightarrow t\bar{t}$, at $\sqrt{s}=13$ TeV and at leading order with \texttt{Madgraph 5}~\cite{Alwall:2011uj}. We will focus on the dilepton final state resulting from $t\bar{t}\rightarrow W^{+}W^{-}+ b\bar{b}\rightarrow \ell^{+}\nu_{\ell}\ell^{-}\bar{\nu}_{\ell} + b\bar{b}$, including all possible combinations of $\ell^{\pm}= \{ \mu^{\pm},e^{\pm} \}$. The top quark decays are calculated with \texttt{Madspin}~\cite{Artoisenet:2012st}. Although the branching ratio of this final state is smaller than that of the semi-leptonic channel, the spin analyzing power is maximally correlated with the spin of the decaying top quarks.

After the hard scattering process and subsequent decays, hadronization is handled using \texttt{Pythia 8}~\cite{Bierlich:2022pfr}, and the full final state signature is passed to \texttt{Delphes 3}~\cite{deFavereau:2013fsa} to simulate the detector response. Thus, the relevant detector-level final state will be 
\begin{equation}
pp\rightarrow t\bar{t}\rightarrow \ell^{\pm} \ell^{\mp} + {\rm jets} + \slashed{E}_T.
\end{equation}
To reconstruct the $t\bar{t}$ density matrix our analysis proceeds in two steps. The first step is a basic event selection requiring that each event satisfy
\begin{itemize}
	\item At least two jets each of with $p_{T} > 25$ GeV and $|\eta|<2.5$.
	\item At least one $b$-tagged jet. If two $b$-jets are identified, we use these to reconstruct the event. If there is only one $b$-jet identified, we use the leading non $b$-tagged jet as the second candidate.
	\item Exactly two opposite-sign leptons with $p_{T}>25$ GeV and $|\eta|<2.5$. We consider the $ee$, $\mu\mu$, and $e\mu$ channels. Leptons must pass an isolation requirement of $I\leq 0.15$.\footnote{In \texttt{Delphes 3} the isolation variable is defined as $I = (\sum p_{T})/ (p_{T}^{\ell})$ where $p_{T}^{\ell}$ is the transverse momentum of the lepton under consideration and the sum runs over all particles, excluding the lepton itself, of all particles within a cone of radius $R=\sqrt{\Delta\eta^{2} + \Delta\phi^{2}}=0.3$.}
\end{itemize}
This basic event selection has an overall efficiency of $\mathcal{O}(10)\%$. 

The second, more involved, step of the analysis concerns the reconstruction of the top quark kinematics.  To do this, we assume that the total missing energy in the event is due to the two neutrinos accompanying the leptons.  Under this assumption, it is possible to algebraically solve for the neutrino momentum taking into account the kinematic constraint from the masses of the top quarks and $W$ bosons~\cite{Sonnenschein:2006ud}. To account for the detector resolution, the $b$-jet momenta and $\slashed{E}_{T}$ are randomly smeared around their reconstructed values. We apply the smearing 100 times, and solve for the neutrino momentum based on the kinematic constraints. For each iteration, the solution set has an up to four-fold degeneracy of the candidate neutrino momenta. We break this four-fold degeneracy by choosing the solution which leads to the smallest $M_{t\bar{t}}$, and this solution is assigned a weight based on the truth-level invariant mass distribution of $\ell^\pm$ and $b$ from the decay of a top~\cite{CMS:2019nrx}.

The $t$ and $\bar{t}$ momenta are then calculated as a weighted average over all collected solutions.  Furthermore, this procedure is completed twice per event, once for each possible lepton and $b$-jet pairing and the final reconstructed $t\bar{t}$ system is chosen as the pairing with the largest sum of weights. This procedure to reconstruct the top quark kinematics closely follows recent CMS and ATLAS studies on measurements of top quark polarization and $t\bar{t}$ spin correlations~\cite{CMS:2019nrx}, and entanglement~\cite{ATLAS:2023fsd}.

\begin{figure}
  \centering
  \includegraphics[scale=0.35]{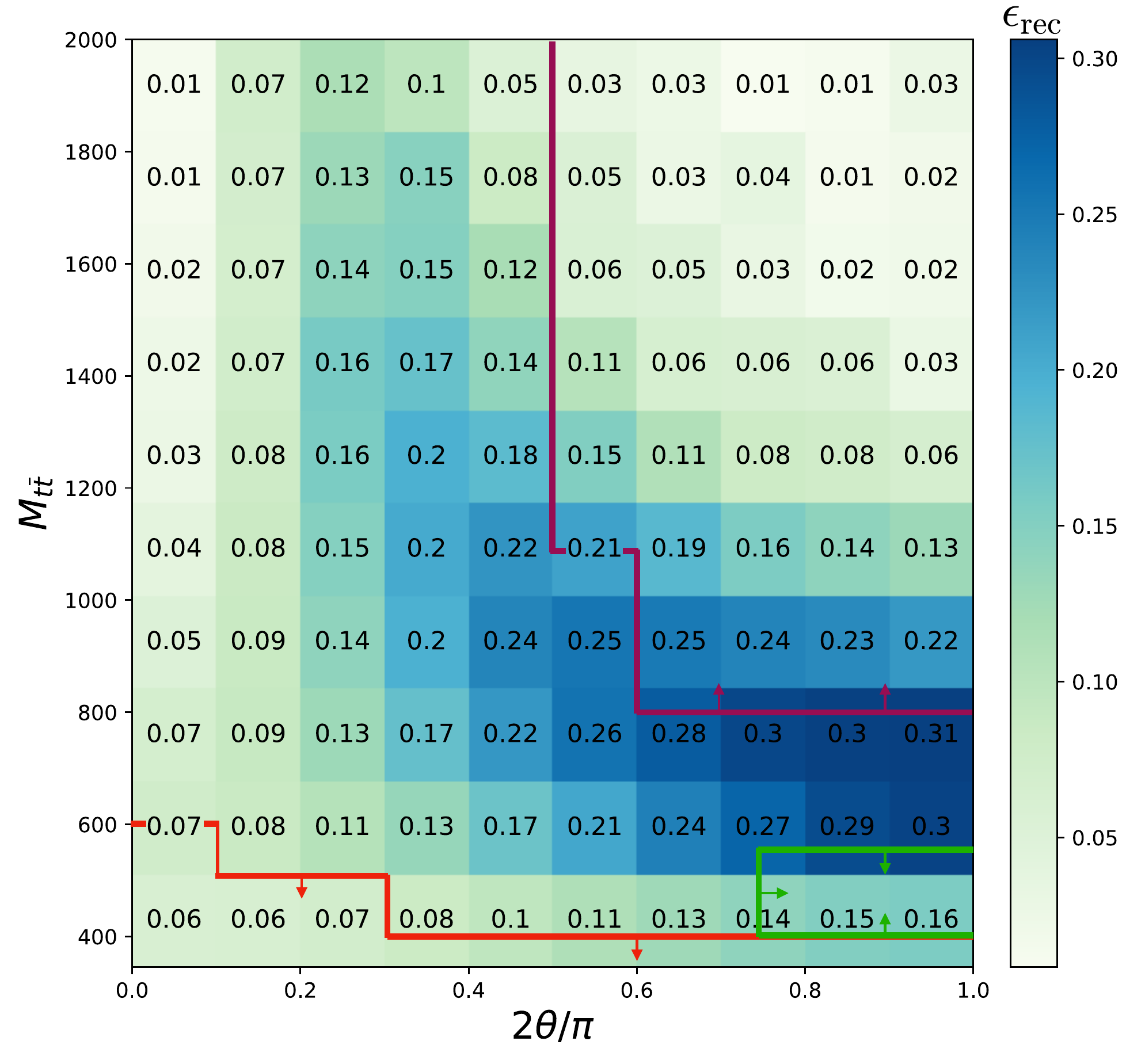}
  \caption{Reconstruction efficiency of the dilepton analysis with respect to the COM scattering angle and invariant mass of the top pair system. The efficiency per bin is defined as the ratio between the number of reconstructed events with the given kinematics and the number of events generated at parton-level at $\sqrt{s}=13$ TeV. The signal regions shown are the threshold (red), separable (green), and boosted (purple) regions in thin lines.}
\label{fig:rec_eff}
\end{figure}

In Fig.~\ref{fig:rec_eff}, we show the reconstruction efficiency of simulated events with respect to the $t\bar{t}$ invariant mass and the COM scattering angle of the top quark. The efficiency per bin is defined as the ratio between the number of reconstructed events with the given kinematics and the number of events generated at parton-level at $\sqrt{s}=13$ TeV.  We see that the reconstruction efficiency peaks for large scattering angle and moderate $t\bar{t}$ invariant mass. Previous studies on entanglement have focused on reconstructing events either at threshold or at high $p_T$ to maximize the sensitivity to entanglement. Expanding the focus of measuring quantum correlations beyond entanglement or Bell nonlocality, we will also consider the intermediate region of large scattering angles and $400~{\rm GeV} \leq M_{t\bar{t}} \leq 575$ GeV.  In this region, the underlying quantum substates are separable and in Fig.~\ref{fig:QI_quantumdiscord} we see that quantum discord is expected to appreciable in this region of phase space.

\begin{figure}
  \centering
  \includegraphics[scale=0.25]{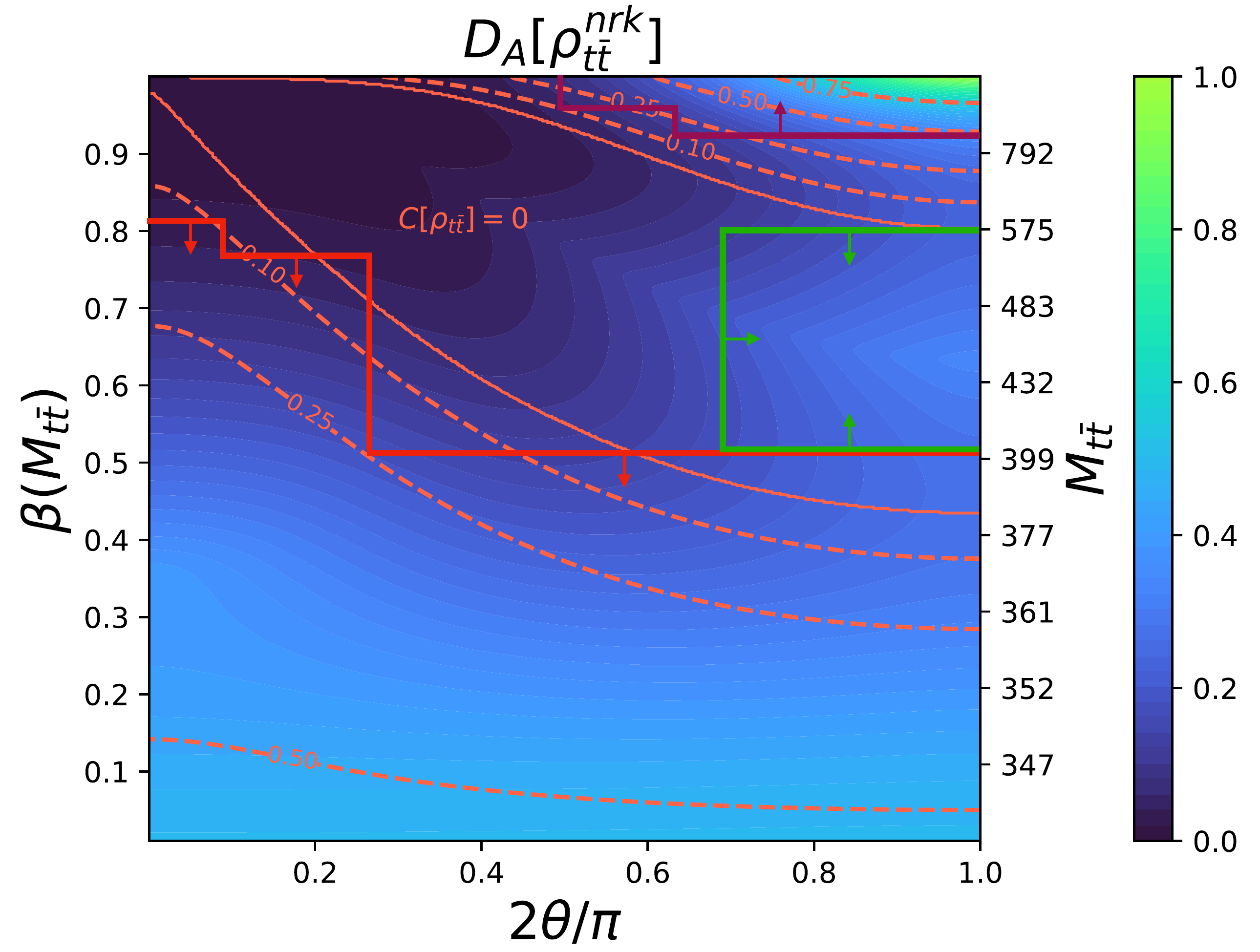}
  \caption{Quantum discord for $t\bar{t}$ in the helicity basis.  The signal regions shown are the threshold (red), separable (green), and boosted (purple) regions.  The light red dashed lines show contours of concurrence and the solid light red lines show the boundary between separable and entangled states. The right axis shows the corresponding value of $M_{t\bar{t}}$ for a given value of $\beta$, and the color bar denotes the contour values of $D_{A}(\rho_{t\bar{t}})$. }
  \label{fig:signal_regions}
\end{figure}

Thus, we define three signal regions of interest to measure quantum discord:
\begin{itemize}

\item \textbf{Threshold Region:}
\begin{align}
& M_{t\bar{t}} \leq 400\text{ GeV}, \nonumber \\
& M_{t\bar{t}} \leq 500 \text{ GeV}
\quad \text{and} \quad
\theta \leq 3\pi/20, \nonumber \\
& M_{t\bar{t}} \leq 600 \text{ GeV} 
\quad \text{and} \quad
\theta \leq \pi/20,
\end{align}
\item \textbf{Separable Region:}
\begin{equation}
400~\text{GeV} \leq M_{t\bar{t}} \leq 575~\text{GeV} 
\quad \text{and} \quad
\theta\geq\frac{3\pi}{8},
\end{equation}
\item \textbf{Boosted Region:}
\begin{align}
800\text{ GeV} \leq M_{t\bar{t}} 
\quad \text{and} \quad
\theta\geq\frac{3\pi}{10},\nonumber \\
1100\text{ GeV} \leq M_{t\bar{t}} 
\quad \text{and}\quad 
\theta\geq\frac{\pi}{4}.
\end{align}

\end{itemize}
The threshold and boosted regions are similar to the signal regions used in entanglement and Bell nonlocality studies.  The separable region, however, is completely unique to this study of quantum discord.  In fact, given the nature of quantum correlations, the events from this part of $t\bar{t}$ phase space are likely to be only useful, in the context of quantum information, for quantum discord.

In Fig.~\ref{fig:signal_regions}, we show these regions compared to contours of quantum discord with respect to $\beta(M_{t\bar{t}})$ and the COM scattering angle. The regions bounded by the red, green, and purple lines define the threshold, separable, and boosted regions, respectively.  The light red dashed lines show contours of concurrence for the $t\bar{t}$ density matrix, whereas the solid light red lines show the boundary of the region of separable states.  

After event selection, distributions of the spin correlation coefficients extracted from the angular asymmetry, Eq.~\eqref{eq:asymm}, are evaluated both at the parton and reconstructed level. The reconstructed distributions are unfolded using the \texttt{RooUnfold} package~\cite{Adye:2011gm}. Uncertainties for parton-level events are purely statistical whereas those of the reconstructed coefficients include both statistical uncertainties and the systematic effects of the detector simulation and unfolding. We use the Iterated Bayesian method~\cite{dagostini2010improved} and the uncertainty due to unfolding is evaluated as the variance of running pseudo-experiments; see Ref.~\cite{Han:2023fci} for an in-depth study of these effects. 

Uncertainties for both parton-level and reconstructed events using the kinematic method are statistical, where the average reconstruction efficiencies, from Fig.~\ref{fig:rec_eff} are used to calculate the number of events after the detector simulation. The fictitious state in a given signal region is then calculated as the average of the reconstructed quantum substates, defined by Eq.~\eqref{eq:fictitious_state} and with the substates evaluated using Eq.~\eqref{eq:ttbar_substate} with reconstructed kinematics. The uncertainty on a given spin correlation coefficient of the fictitious state is then given by the standard deviation of the corresponding coefficient of the substates from the sample of events that are reconstructed in a signal region. The errors on $\bar{C}_{ij}$ are then propagated to quantum discord by Monte Carlo pseudo-experiments.

\begin{table}
  \begin{center}
  \qquad\qquad\qquad Threshold Region\quad \;\;\; Separable Region\quad\quad \quad Boosted Region\\
    \begin{tabular}{|l|c|c|c} 
    \hline
       &$\braket{\epsilon_{rec}}$& $D_{A}(\rho_{t\bar{t}})$\\
      \hline
      Parton && $0.200 \pm 0.003$\\
      Reconstructed &$0.10$& $0.23 \pm 0.04$\\
      \hline
  
    \end{tabular}
        \begin{tabular}{|l|c|c} 
    \hline
      $\braket{\epsilon_{rec}}$& $D_{A}(\rho_{t\bar{t}})$\\
      \hline
        & $0.255\pm 0.008$\\
       $0.28$& $0.18\pm 0.05$\\
      \hline
    \end{tabular}
        \begin{tabular}{|l|c|c} 
    \hline
      $\braket{\epsilon_{rec}}$& $D_{A}(\rho_{t\bar{t}})$\\
      \hline
        & $0.197\pm 0.003$\\
       $0.08$& $0.20\pm 0.05$\\
      \hline
    \end{tabular}
  \end{center}
  \caption{Quantum discord evaluated using the decay method in the helicity basis. Events are simulated assuming the dilepton decay channel of $t\bar{t}$ with $\sqrt{s}=13$ TeV and $\mathcal{L}=139$ fb$^{-1}$. The uncertainties for parton-level events are purely statistical whereas uncertainties for reconstructed events are due to statistical and systematic uncertainties due to unfolding. For reconstructed events we include the average reconstructed efficiencies from Fig.~\ref{fig:rec_eff} for each signal region.}
  \label{table:Discord_table_decay}
\end{table}

\begin{table}[t]
  \begin{center}
  \qquad\qquad\qquad Threshold Region\quad \;\;\; Separable Region\quad\quad \quad Boosted Region\\
    \begin{tabular}{|l|c|c|c} 
    \hline
       &$\braket{\epsilon_{rec}}$& $D_{A}(\rho_{t\bar{t}})\times 10^{3}$\\
      \hline
      Parton && $173.42 \pm 0.07$\\
      Reconstructed & $0.10$ & $147.10 \pm 0.24$\\
      \hline
  
    \end{tabular}
        \begin{tabular}{|l|c|c}
    \hline
      $\braket{\epsilon_{rec}}$& $D_{A}(\rho_{t\bar{t}})\times 10^{3}$\\
      \hline
        & $249.72\pm 0.24$\\
       $0.28$& $232.54\pm 0.47$\\
      \hline
    \end{tabular}
        \begin{tabular}{|l|c|c}
    \hline
      $\braket{\epsilon_{rec}}$& $D_{A}(\rho_{t\bar{t}})\times 10^{3}$\\
      \hline
        & $200.81\pm 0.08$\\
       $0.08$ & $188.49\pm 0.25$\\
      \hline
    \end{tabular}
  \end{center}
  \caption{Quantum discord evaluated using the kinematic method in the helicity basis. Events are simulated assuming the dilepton decay channel of $t\bar{t}$ with $\sqrt{s}=13$ TeV and $\mathcal{L}=139$ fb$^{-1}$. The uncertainties are statistical as described in the main text. For reconstructed events we include the average reconstructed efficiencies for each signal region.}
  \label{table:Discord_table_kinematic}
\end{table}

In Table~\ref{table:Discord_table_decay} and Table~\ref{table:Discord_table_kinematic}, we compare expected values of quantum discord for the $t\bar{t}$ system at the LHC with $\sqrt{s}=13$ TeV and $\mathcal{L}=139$ fb$^{-1}$ in the threshold, separable, and boosted regions using the decay method and kinematic method, respectively.  We present results for the helicity basis. For comparison using the beam and lab frame bases see Appendix~\ref{sec:coeffs}. In the top row of each table we show results evaluated at parton-level and in the bottom row we show the expected reconstructed value after detector simulation. We see that although the reconstruction using the decay method performs fairly well, there are large uncertainties primarily from the unfolding procedure. 

We define the significance of the measured observable, $\mathcal{O}$, as
\begin{equation}
\text{significance}=\frac{\mathcal{O} - \mathcal{O}_{null}}{\delta \mathcal{O}},
\end{equation}
given by the ratio of the difference between the observed value and a null measurement,  and the uncertainty in the measurement, $\delta\mathcal{O}$.

Even with the large uncertainties inherent in the decay method, the significance for measuring quantum discord is $5.7\sigma$ in the threshold region, $3.6\sigma$ in the separable region, and $4.2\sigma$ in the boosted region. 

Since the kinematic method evaluates the spin correlation coefficients for each quantum substate event-by-event, the reconstructed uncertainties, being statistical, can reach the sub-percent level. Thus, measuring quantum discord using the kinematic method has a significance well over $5\sigma$ across the whole phase space.   In the threshold region the precision is $0.16\%$, in the separable region the precision is $0.20\%$, and in the boosted region the precision is $0.13\%$.

\begin{figure}
\centering
  \includegraphics[scale=0.55]{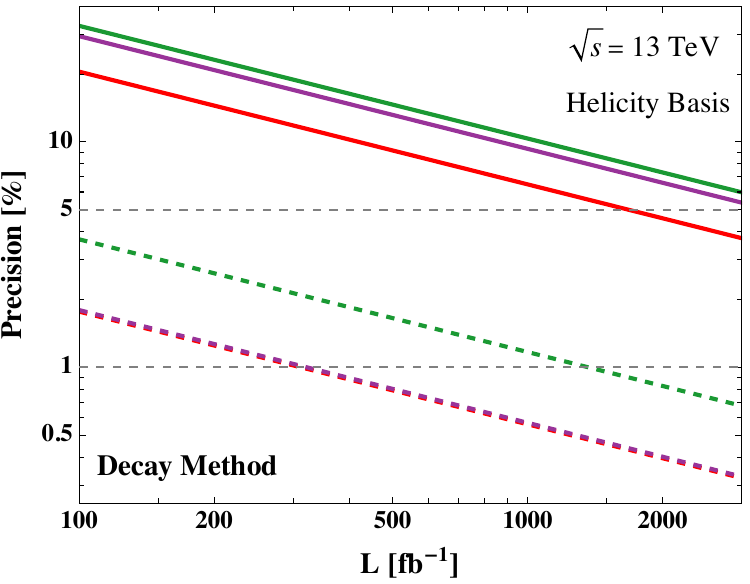}
  \qquad
  \includegraphics[scale=0.55]{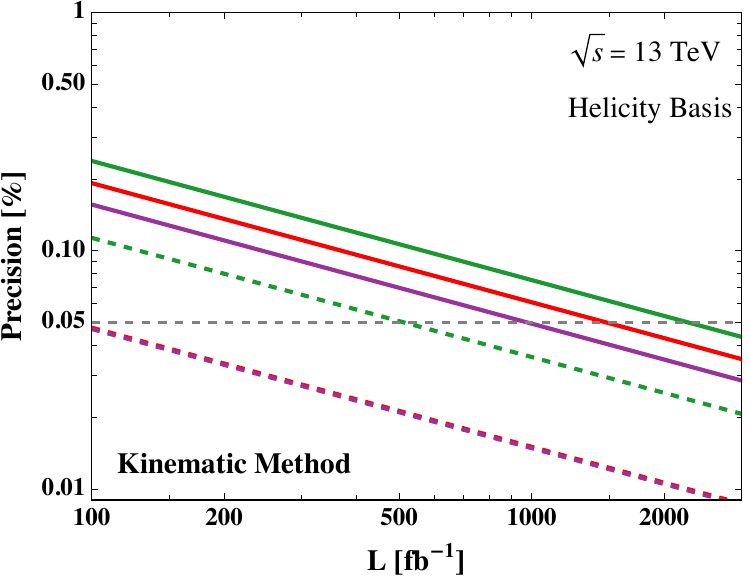}
  \caption{Expected precision for the measurement of quantum discord in top quarks at the LHC. Results at parton-level (dashed) and after detection simulation (solid) are shown.  The decay method is shown in the left panel and the kinematic method is shown in the right panel.  Red, green, and purple lines indicate results from the threshold, separable, and boosted regions, respectively.  All results are presented in the helicity basis.}
\label{fig:exp_precision}
\end{figure}

In Fig.~\ref{fig:exp_precision}, we show the expected precision,  defined as $\delta\mathcal{O}/\mathcal{O}$, of measuring quantum discord in top quarks for the LHC.  We project our estimates up to integrated luminosities of the HL-LHC, 3000 $\text{fb}^{-1}$. Results are presented using the helicity basis. The precision of measuring quantum discord using other bases are comparable or worse. In both panels, we show results from parton-level events (dashed) and after detection simulation (solid). In the left panel, we show results using the decay method where the red, green, and purple lines represent the results from event selections from the threshold, separable, and boosted regions, respectively. Our simulation shows that with present data, reconstructed values of discord could be measured with a precision close to 10\% using the decay method. This could improve to $\approx 3-5\%$ precision by the end of the HL-LHC, $L \approx 3000~\text{fb}^{-1}$.  Furthermore, each signal region leads to a roughly similar level of precision.

In the right panel of Fig.~\ref{fig:exp_precision}, we show the expected precision on a measurement of quantum discord using the kinematic method. The colors and line styles follow the same conventions as in the left panel. At the detector-level there is a dramatic difference between the decay and kinematic methods. The kinematic methods performs better by an order of magnitude. In particular, the precision on a measurement of quantum discord with current data, $\mathcal{L}=139$ fb$^{-1}$, is predicted to be better than 0.5\% across phase space using the kinematic method. 

One additional advantage that the kinematic method offers is that unfolding is not necessary whereas for the decay method it is crucial.  Spin correlations are typically not detectable without unfolding while the kinematics are only minimally affected by detector effects.  As a result, the difference between the parton-level and full results in Fig.~\ref{fig:exp_precision} for the decay method is due to the reduced number of events due to the reconstruction efficiency and from an increase in uncertainty due to unfolding.  For the kinematic method, the difference is smaller because only the reconstruction efficiency has an impact.

\section{Conclusions}
\label{sec:conclusions}

In recent years, there has been increasing interest in measuring quantities associated with quantum information at colliders. Experiments at the LHC accumulate a large data sample, and thus offer great opportunities for this exploration. In this paper, we study the quantum information for the $t\bar t$ system at the LHC, focusing on the quantum discord.

We presented the first fine-grained analysis of quantum information measures relevant to the $t\bar{t}$ system, including Von Neumann entropy, mutual information, conditional entropy, and quantum discord. These measures quantify different aspects of quantum correlations, with quantum discord uniquely isolating the total non-classical correlations among both separable and entangled states. We derived compact closed-form formulae for quantum discord in the helicity and beam bases, making it possible to compute discord efficiently in key regions of phase space. 

We explored the quantum discord of $t\bar t$ throughout the phase space using two complementary approaches: the decay method and the kinematic method. The decay method relies on angular distributions of $t\bar{t}$ decay products and follows standard experimental practices, while the kinematic method reconstructs spin density matrices directly from event kinematics, achieving superior precision. 

Our results show that the current analysis strategies of quantum correlations for the $t\bar{t}$ system can probe non-classical correlations detectable as non-zero discord at a significant level $\gtrsim 3\sigma$, offering a previously unexplored perspective on collider-produced quantum states. We identified three signal regions in phase space—threshold, separable, and boosted—where quantum discord can be probed in  $t\bar{t}$.  We demonstrated that with the HL-LHC dataset discord is projected to be measured with a precision of approximately $5\%$ using the decay method and sub-percent levels using the kinematic method. Even with current LHC datasets, discord can be measured with better than 0.5\% precision across phase space  with the kinematic method, showcasing the feasibility of exploring this novel aspect of quantum information at colliders.

This study establishes quantum discord as a powerful observable for probing quantum correlations in collider systems, including those present in separable states. We addressed critical theoretical challenges in measuring quantum discord, particularly the implications of non-convexity and the role of fictitious states. Unlike the set of separable states, the set of zero discord states is not a convex set, meaning that mixtures of states with zero discord can produce a state with non-zero discord.  This property underscores the importance of careful event selection to ensure meaningful measurements.

Our work presented in this article clears the way for future investigations of quantum discord. In particular, any differences between the discord of $t$ and the discord of $\bar{t}$ would imply a direct sensitivity to CP-violation~\cite{Afik:2022dgh}.  While we have focused on the leading order cross section for $t\bar t$ which does not exhibit these effects, we have outlined the necessary ingredients needed for robust studies in this direction. The study of this aspect of discord and its implications for electroweak cross sections or possible new physics is left for future work. Furthermore, we discussed various information measures that quantify even stronger correlations than entanglement or Bell nonlocality, such as negative conditional entropy, motivating the broad exploration of quantum information measures at the LHC for $t\bar{t}$ and other systems.

By systematically investigating quantum discord for the first time through a detailed collider analysis for the LHC, this work expands the toolkit for quantum information studies in particle physics and lays the groundwork for deeper insights into the quantum properties of high-energy collisions. 

\section*{Acknowledgements}

The authors thank Kun Cheng, Juan de Nova, and Arthur Wu for valuable discussions. 
TH and ML were supported in part by the U.S.~Department of Energy under grant No.~DE-SC0007914 and in part by Pitt PACC.  ML is also supported by the National Science Foundation under grant No.~PHY-2112829. NM and SS were supported in part by the U.S.~Department of Energy under grant No. DEFG02-13ER41976/DE-SC0009913. We have benefited from use of the \texttt{QuTip} package~\cite{Johansson:2011jer,Johansson:2012qtx}.

\appendix
\section{Simplified Formulae for the Quantum Discord of $t\bar{t}$}
\label{sec:App_A}

The general bipartite density matrix for a spin state of $t\bar{t}$ is given by
\begin{equation}
\rho = \frac{1}{4}\left(
\mathbb{I}_4 + \sum_i B^+_i \sigma_i \otimes \mathbb{I}_2 
+ \sum_i B^-_i \mathbb{I}_2 \otimes \sigma_i
+ \sum_{ij} C_{ij} \sigma_i \otimes \sigma_j
\right).
\end{equation}
Due to the symmetries of the leading order cross section for $t\bar{t}$ pair production, we may ignore the polarization factors
\begin{equation}
B^+_i = (0, 0, 0),
\quad\quad\quad
B^-_i = (0, 0, 0).
\end{equation}
In this limit, we note that the original density matrix is unitarily equivalent to
\begin{equation}
\rho^{\prime}_{AB}=\frac{1}{4}\left(
\mathbb{I}_4 + \sum_{ij} C_{ij} (U\sigma_i U^{\dagger}) \otimes (V\sigma_i V^{\dagger})\right),
\end{equation}
for any unitary matrices $U$ and $V$. This local unitary transformation on the $t\bar{t}$ system can be represented as a change of basis among the Pauli matrices as $(U\sigma^{\prime}_i U^{\dagger}) = A_{il}\sigma_{l}$ and $(V\sigma^{\prime}_j V^{\dagger}) = B_{jk}\sigma_{k}$. Thus, we have 
\begin{equation}
\rho^{\prime}_{AB}=\frac{1}{4}\left(
\mathbb{I}_4 + \sum_{ij} A_{il}C_{ij}B_{jk} \sigma_{l} \otimes \sigma_{k}
\right).
\end{equation}
Without loss of generality, we may choose $U$ and $V$ such that the matrices $A$ and $B$ result in the singular value decomposition of $C_{ij}$, i.e.  $\Lambda_{i}\delta_{ij}=A_{ki}C_{kl}B_{lj}$, where the $\Lambda_{i}$ are the singular values of $C_{ij}$. In cases where $C_{ij}$ is real and symmetric, $U$ and $V$ may instead be chosen to result in the diagonalization of $C_{ij}$. It follows that
\begin{equation}
\rho^{\prime}_{AB}=\frac{1}{4}\left(
\mathbb{I}_4 + \sum_{i} \Lambda_{i} \sigma_{i} \otimes \sigma_{i}
\right).
\end{equation}
Quantum discord is unchanged under the action of local, unitary operations, $D_{A}(\rho_{AB})=D_{A}(\rho^{\prime}_{AB})$.  We compute the quantum discord for $\rho^{\prime}_{AB}$.  Starting with the Von Neumann entropy
\begin{equation}
S(\rho^{\prime}_{AB})=-\sum_{i}e_{i}\log_{2}\left(e_{i}\right),
\end{equation}
where the $e_{i}$ are the eigenvalues of $\rho^{\prime}_{AB}$ (since eigenvalues are preserved under unitary transformation, these are simply the eigenvalues of the original state $\rho_{AB}$, i.e. $S(\rho^{\prime}_{AB})=S(\rho_{AB})$), and the reduced state entropy is
\begin{equation}
S(\rho^{\prime}_{B})=1.
\end{equation}
The $\rho^{\prime}_{\pm\hat{n}}$ state in the rotated basis is given by
\begin{equation}
\rho^{\prime}_{\pm\hat{n}}=\frac{1}{2}\left( \mathbb{I}_2 \pm (\Lambda \cdot \hat{n}) \vec{\sigma} \right),
\end{equation}
with a Von Neumann entropy of
\begin{equation}
S(\rho^{\prime}_{\pm\hat{n}})=-\frac{1}{2}(1+\bar{\Lambda})\log_{2}\left(\frac{1+\bar{\Lambda}}{2}\right)-\frac{1}{2}(1-\bar{\Lambda})\log_{2}\left(\frac{1-\bar{\Lambda}}{2}\right),
\end{equation}
where $\bar{\Lambda}=|\vec{\Lambda} \cdot \hat{n}|$.  The combination $p_{+n}S(\rho^{\prime}_{+\hat{n}}) + p_{-n}S(\rho^{\prime}_{-\hat{n}})$ is concave down at $\bar{\Lambda}=0$. Thus, the contribution of these terms to $D_{A}(\rho^{\prime}_{AB})$ is minimized by maximizing $\bar{\Lambda}$. We see that $\hat{n}$ must be chosen to point along the largest singular value (or largest eigenvalue in absolute value, if the diagonalization has been calculated instead). Let $\lambda = \text{max}(\Lambda_{i})$ be the maximal singular value. Then the quantum discord is
\begin{equation}
D_{A}(\rho^{\prime}_{AB})=1 + \sum_{i}e_{i}\log_{2}\left(e_{i}\right)-\frac{1}{2}(1+\lambda)\log_{2}\left(\frac{1+\lambda}{2}\right)-\frac{1}{2}(1-\lambda)\log_{2}\left(\frac{1-\lambda}{2}\right).
\end{equation}
In this form, the leading order value of quantum discord for the $t\bar{t}$ system may be calculated in a closed form in any basis once the values of $C_{ij}$ have been measured.

\subsection{The Beam Basis}

For the $t\bar{t}$ density matrix, we have
\begin{equation}
B^+_i = (0, 0, 0),
\quad\quad\quad
B^-_i = (0, 0, 0),
\quad\quad\quad
\text{(threshold, beam basis),}
\end{equation}
and a spin correlation matrix of
\begin{equation} \label{eq:c_beam_threshold}
C_{ij} = \left(\begin{array}{ccc}
C_\perp & 0 & 0 \\
0 & C_\perp & 0  \\
0 & 0 & C_z 
\end{array}\right),
\quad\quad\quad
\text{(threshold, beam basis).}
\end{equation}
The quantum discord is 
\begin{equation}\begin{aligned}
D_A =
& 1 + \frac{1}{2}(1+C_z) \log_2\left(\frac{1+C_z}{4}\right)
+ \frac{1}{4}(1+2C_\perp-C_z) \log_2\left(\frac{1+2C_\perp-C_z}{4}\right) \\
& + \frac{1}{4}(1-2C_\perp-C_z) \log_2\left(\frac{1-2C_\perp-C_z}{4}\right) \\
& - \frac{1}{2}(1+C_{\rm max}) \log_2 \left(\frac{1+C_{\rm max}}{2}\right) - \frac{1}{2}(1-C_{\rm max}) \log_2 \left(\frac{1-C_{\rm max}}{2}\right),
\end{aligned}\end{equation}
where $C_{\rm max} =\text{max}\{|C_{\perp}|, |C_{z}|\}$.

\subsection{The Helicity Basis}

At high-$p_{T}$, spin correlations in $t\bar{t}$ production are best characterized in the helicity basis where
\begin{equation}
B^+_i = (0, 0, 0),
\quad\quad\quad
B^-_i = (0, 0, 0),
\quad\quad\quad
\text{(high-$p_T$, helicity basis),}
\end{equation}
and a spin correlation matrix is
\begin{equation} \label{eq:c_hel_boosted}
C_{ij} = \left(\begin{array}{ccc}
C_k & C_{kr} & 0 \\
C_{kr} & C_r & 0  \\
0 & 0 & C_n 
\end{array}\right),
\quad\quad\quad
\text{(high-$p_T$, helicity basis).}
\end{equation}
In this basis, the quantum discord is
\begin{flalign}\nonumber
D_{A}(\rho_{AB})=1 &+ \frac{1}{4}(1-C_{k}-C_{n}-C_{r})\log_{2}\left(\frac{(1-C_{k}-C_{n}-C_{r})}{4}\right)\\\nonumber
&+ \frac{1}{4}(1+C_{k}-C_{n}+C_{r})\log_{2}\left(\frac{(1+C_{k}-C_{n}+C_{r})}{4}\right)\\\nonumber
&+ \frac{1}{4}(1+C_{n} - \Delta)\log_{2}\left(\frac{(1+C_{n} - \Delta)}{4}\right) + \frac{1}{4}(1+C_{n} + \Delta)\log_{2}\left(\frac{(1+C_{n} + \Delta)}{4}\right)\\
&-\frac{1}{2}(1+\lambda)\log_{2}\left(\frac{1+\lambda}{2}\right)-\frac{1}{2}(1-\lambda)\log_{2}\left(\frac{1-\lambda}{2}\right)
\end{flalign}
where $\Delta = \sqrt{C_{k}^{2} + 4C_{kr}^{2} +C_{r}^{2} - 2C_{k}C_{r}}$, and $\lambda = \text{max}\{|C_{n}|,\frac{1}{2}\left|C_{k}+C_{r}-\Delta\right|,\frac{1}{2}\left|C_{k}+C_{r}+\Delta\right|  \}$.

\section{Separable States with Non-Zero Quantum Discord}
\label{app:separable}

One of the notable features of quantum discord is that it can be non-zero, indicate the presence of quantum correlations, even for separable states.  In this appendix, we write down the separable state in the beam basis.  Recall in the beam basis the $t\bar{t}$ quantum state is described by $C_\perp$ and $C_Z$ via the spin correlation matrix in Eq.~\eqref{eq:c_beam_threshold}.

Consider the following density matrices
\begin{align}
\rho_{X+} &= 
\frac{1}{2} \ket{+} \bra{+} \otimes \ket{+} \bra{+} 
+ \frac{1}{2} \ket{-} \bra{-} \otimes \ket{-} \bra{-} , \\
\rho_{X-} &= 
\frac{1}{2} \ket{-} \bra{-} \otimes \ket{+} \bra{+}
+ \frac{1}{2} \ket{+} \bra{+} \otimes \ket{-} \bra{-}, \\
\rho_{Y+} &= \frac{1}{2} 
\ket{\leftarrow} \bra{\leftarrow} \otimes \ket{\leftarrow} \bra{\leftarrow}
+ \frac{1}{2}\ket{\rightarrow} \bra{\rightarrow} \otimes \ket{\rightarrow} \bra{\rightarrow} , \\
\rho_{Y-} &= 
\frac{1}{2}\ket{\leftarrow} \bra{\leftarrow} \otimes \ket{\rightarrow} \bra{\rightarrow}
+ \frac{1}{2} \ket{\rightarrow} \bra{\rightarrow} \otimes \ket{\leftarrow} \bra{\leftarrow}, \\
\rho_{Z+} &=
\frac{1}{2} \ket{\downarrow} \bra{\downarrow} \otimes \ket{\downarrow} \bra{\downarrow} 
+ \frac{1}{2} \ket{\uparrow} \bra{\uparrow}  \otimes \ket{\uparrow} \bra{\uparrow}, \\
\rho_{Z-} &= 
\frac{1}{2} \ket{\downarrow} \bra{\downarrow}  \otimes \ket{\uparrow} \bra{\uparrow}
+ \frac{1}{2} \ket{\uparrow} \bra{\uparrow} \otimes \ket{\downarrow} \bra{\downarrow}  .
\end{align}
These density matrices are all separable and any mixed states comprised of them will also be separable.  The most general density matrix from these states, respecting the symmetries of Eq.~\eqref{eq:c_beam_threshold}, is
\begin{equation}
\rho = p_1 \rho_{X+} + p_2 \rho_{X-}
+ p_1 \rho_{Y+} + p_2 \rho_{Y-}
+ p_3 \rho_{Z+} + (1-2p_1-2p_2-p_3) \rho_{Z-},
\end{equation}
with the simultaneous constraints
\begin{equation}
0 \leq p_1 \leq 1,
\quad
0 \leq p_2 \leq 1,
\quad
0 \leq p_3 \leq 1,
\quad
0 \leq 1-2p_1-2p_2-p_3 \leq 1.
\end{equation}
In terms of $C_\perp$ and $C_z$ we have
\begin{equation}
C_\perp = p_1 - p_2,
\qquad\qquad
C_z = -1 + 2p_1 + 2p_2 + 2p_3.
\end{equation}
When $C_\perp = 0$ only $\rho_{Z+}$ and $\rho_{Z-}$ contribute to the state which spans the set of zero discord states.  When $C_\perp \neq 0$ one cannot write the state using a single basis and quantum discord is non-zero.

Bipartite separable states can be classified by whether each subsystem is invariant under measurements, called classical, or not, called quantum~\cite{PhysRevA.77.022301}.  This leads to the classification of states as quantum-quantum, quantum-classical, classical-quantum, or classical-classical.  Since the $t\bar{t}$ system is CP-invariant, asymmetric classifications are not possible and states are either quantum-quantum or classical-classical.  Thus, all separable $t\bar{t}$ states with non-zero discord are quantum-quantum.

\section{Spin Correlation Coefficients}
\label{sec:coeffs}

In this appendix we collect detailed results for the reconstruction of the $t\bar{t}$ spin correlation coefficients. Events are simulated with the dilepton decay channel of $t\bar{t}$ with $\sqrt{s}=13$ TeV and $\mathcal{L}=139$ fb$^{-1}$. Detector-level events are simulated according to the description in the main text.  We present results for the threshold, separable, and boosted regions with respect to the helicity, beam, and lab frame bases. We also present results evaluated using both the decay and kinematic methods.

The threshold region spin correlation coefficients are given in Table~\ref{tab:Cij_threshold} using the decay method and in Table~\ref{tab:Cij_threshold_kinematic} using the kinematic method.  The separable region spin correlation coefficients are given in Table~\ref{tab:Cij_separable} using the decay method and in Table~\ref{tab:Cij_separable_kinematic} using the kinematic method.  The boosted region spin correlation coefficients are given in Table~\ref{tab:Cij_boosted} using the decay method and in Table~\ref{tab:Cij_boosted_kinematic} using the kinematic method.

%
%

\begin{table}
  \begin{center}
    Threshold Region\\
    \begin{tabular}{l|c|c|c|c} 
    \hline
      Helicity & $n$ & $r$ & $k$ & $D_{A}(\rho)$\\
      \hline
      $n$ & $-0.493\pm0.004$ & $0.002\pm0.004$ & $0.002\pm0.004$\\
      $r$ & $0.002\pm0.004$ & $-0.324\pm0.004$ & $0.079 \pm0.004$& $0.200\pm0.003$\\
      $k$ & $0.000\pm0.004$ & $0.079\pm0.004$ & $- 0.615\pm0.004$\\
       \hline
      $n$ & $-0.530\pm0.023$ & $0.004\pm0.006$ & $0.005\pm0.006$\\
      $r$ & $0.001\pm0.021$ & $-0.338\pm0.042$ & $0.092 \pm0.019$& $0.230\pm0.035$\\
      $k$ & $-0.001\pm0.004$ & $0.091\pm0.020$ & $- 0.653\pm0.089$\\
      \hline
      \hline
      Beam & $x$ & $y$ & $z$\\
      \hline
      $x$ & $-0.523\pm0.004$ & $-0.003\pm0.004$ & $0.001\pm0.004$\\
      $y$ & $0.000\pm0.004$ & $-0.526\pm0.004$ & $0.000\pm0.004$ & $0.221\pm0.003$\\
      $z$ & $0.000\pm0.004$ & $0.001\pm0.004$ & $- 0.386\pm0.004$\\
      \hline
      $x$ & $-0.567\pm0.039$ & $-0.004\pm0.006$ & $-0.003\pm0.006$\\
      $y$ & $0.018\pm0.12$ & $-0.564\pm0.014$ & $0.004\pm0.006$ & $0.209\pm0.027$\\
      $z$ & $-0.006\pm0.044$ & $0.001\pm0.005$ & $- 0.378\pm0.028$\\
      \hline
      \hline
      Lab & $x$ & $y$ & $z$\\
      \hline
      $x$ & $-0.517\pm0.004$ & $-0.002\pm0.004$ & $-0.001\pm0.004$\\
      $y$ & $0.000\pm0.004$ & $-0.520\pm0.004$ &$ 0.000\pm0.004$ & $0.214\pm0.003$\\
      $z$ & $0.001\pm0.004$ & $0.001\pm0.004$ & $- 0.371\pm0.004$\\
      \hline
      $x$ & $-0.559\pm0.029$ & $-0.009\pm0.006$ & $-0.013\pm0.038$\\
      $y$ & $-0.011\pm0.020$ & $-0.553\pm0.010$ &$ -0.006\pm0.036$ & $0.237\pm0.020$\\
      $z$ & $-0.021\pm0.065$ & $0.009\pm0.042$ & $- 0.378\pm0.083$\\
    \end{tabular}
  \end{center}
  \caption{Spin correlation matrices $C_{ij}$ evaluated using the decay method using the helicity (top), beam (middle), and lab frame (bottom) bases in the threshold region. For each basis, the top panel shows the parton-level results and the bottom panel shows the results after detector simulation and unfolding. The uncertainties for parton-level events are purely statistical whereas uncertainties for reconstructed events are due to statistical and systematic uncertainties due to unfolding.}
  \label{tab:Cij_threshold}
\end{table}

\begin{table}
  \begin{center}
    Threshold Region\\
    \begin{tabular}{l|c|c|c|c} 
    \hline
      Helicity $(\times 10^3)$& $n$ & $r$ & $k$ & $D_{A}(\rho)\times 10^3$\\
      \hline
      $n$ & $-479.62\pm0.09$ & $0.0\pm0.0$ & $0.0\pm0.0$\\
      $r$ & $0.0\pm0.0$ & $-300.49\pm0.14$ & $84.61\pm0.04$& $173.42\pm0.07$\\
      $k$ & $0.0\pm0.0$ & $84.61\pm0.04$ & $- 592.18\pm0.07$\\
      \hline
      $n$ & $-439.87\pm0.31$ & $0.0\pm0.0$ & $0.0\pm0.0$\\
      $r$ & $0.0\pm0.0$ & $-239.59\pm0.56$ & $84.70 \pm0.16$& $147.10\pm0.24$\\
      $k$ & $0.0\pm0.0$ & $84.70\pm0.16$ & $- 575.92\pm0.28$\\
      \hline
      \hline
      Beam $(\times 10^3)$& $x$ & $y$ & $z$\\
      \hline
      $x$ & $-509.36\pm0.09$ & $0.01\pm0.02$ & $-0.01\pm0.05$\\
      $y$ & $0.01\pm0.02$ & $-509.40\pm0.09$ & $0.00\pm0.05$ & $207.29\pm0.07$\\
      $z$ & $-0.01\pm0.05$ & $0.00\pm0.05$ & $-353.53\pm0.19$\\
      \hline
      $x$ & $-473.78\pm0.33$ & $-0.02\pm0.08$ & $-0.02\pm0.19$\\
      $y$ & $-0.02\pm0.08$ & $-472.85\pm0.33$ & $-0.03\pm0.19$ & $176.21\pm0.24$\\
      $z$ & $0.02\pm0.19$ & $-0.25\pm0.19$ & $-308.76\pm0.83$\\
      \hline
      \hline
      Lab $(\times 10^3)$& $x$ & $y$ & $z$\\
      \hline
      $x$ & $-502.31\pm0.09$ & $0.01\pm0.03$ & $0.01\pm0.07$\\
      $y$ & $0.01\pm0.03$ & $-502.35\pm0.09$ &$ 0.00\pm0.07$ & $199.90\pm0.06$\\
      $z$ & $-0.02\pm0.07$ & $0.00\pm0.07$ & $-339.42\pm0.19$\\
      \hline
      $x$ & $-469.73\pm0.33$ & $-0.10\pm0.11$ & $-0.02\pm0.25$\\
      $y$ & $0.06\pm0.11$ & $-468.71\pm0.32$ &$ -0.43\pm0.25$ & $172.40\pm0.26$\\
      $z$ & $0.07\pm0.25$ & $-0.07\pm0.25$ & $-301.30\pm0.83$\\
    \end{tabular}
  \end{center}
  \caption{Spin correlation matrices $C_{ij}$ evaluated using the kinematic method using the helicity (top), beam (middle), and lab frame (bottom) bases in the threshold region. For each basis, the top panel shows the parton-level results and the bottom panel shows the results after detector simulation. The uncertainties are statistical as described in the main text. All entries of the table have been scaled by a factor of $\times 10^3$.}
  \label{tab:Cij_threshold_kinematic}
\end{table}

\begin{table}
  \begin{center}
    Separable Region\\
    \begin{tabular}{l|c|c|c|c} 
    \hline
      Helicity & $n$ & $r$ & $k$ & $D_{A}(\rho)$\\
      \hline
      $n$ & $-0.346\pm0.004$ & $0.002\pm0.004$ & $0.003\pm0.004$\\
      $r$ & $0.002\pm0.004$ & $0.344\pm0.004$ & $0.099 \pm0.004$& $0.255\pm0.008$\\
      $k$ & $0.000\pm0.004$ & $0.099\pm0.004$ & $- 0.235\pm0.004$\\
      \hline
      $n$ & $-0.300\pm0.059$ & $0.002\pm0.013$ & $0.004\pm0.005$\\
      $r$ & $0.005\pm0.015$ & $0.310\pm0.010$ & $0.068 \pm0.005$& $0.175\pm0.048$\\
      $k$ & $0.000\pm0.004$ & $0.064\pm0.004$ & $- 0.230\pm0.034$\\
      \hline
      \hline
      Beam & $x$ & $y$ & $z$\\
      \hline
      $x$ & $-0.303\pm0.004$ & $-0.004\pm0.004$ & $0.000\pm0.004$\\
      $y$ & $0.000\pm0.004$ & $-0.302\pm0.004$ & $0.003\pm0.004$ & $0.265\pm0.008$\\
      $z$ & $0.001\pm0.004$ & $0.000\pm0.004$ & $0.364\pm0.004$\\
      \hline
      $x$ & $-0.255\pm0.028$ & $-0.005\pm0.004$ & $0.001\pm0.004$\\
      $y$ & $-0.004\pm0.018$ & $-0.260\pm0.006$ & $0.004\pm0.004$ & $0.147\pm0.016$\\
      $z$ & $0.000\pm0.023$ & $0.000\pm0.004$ & $0.284\pm0.007$\\
      \hline
      \hline
      Lab & $x$ & $y$ & $z$\\
      \hline
      $x$ & $-0.319\pm0.004$ & $0.003\pm0.004$ & $0.002\pm0.004$\\
      $y$ & $0.000\pm0.004$ & $-0.317\pm0.004$ & $0.001\pm0.004$ & $0.279\pm0.008$\\
      $z$ & $0.002\pm0.004$ & $0.000\pm0.004$ & $0.332\pm0.004$\\
       \hline
      $x$ & $-0.280\pm0.018$ & $0.004\pm0.012$ & $0.003\pm0.010$\\
      $y$ & $-0.003\pm0.005$ & $-0.252\pm0.048$ & $0.010\pm0.046$ & $0.137\pm0.029$\\
      $z$ & $-0.004\pm0.027$ & $-0.014\pm0.036$ & $0.251\pm0.027$\\
      
    \end{tabular}
  \end{center}
  \caption{Spin correlation matrices $C_{ij}$ evaluated using the decay method using the helicity (top), beam (middle), and lab frame (bottom) bases in the separable region. For each basis, the top panel shows the parton-level results and the bottom panel shows the results after detector simulation and unfolding. The uncertainties for parton-level events are purely statistical whereas uncertainties for reconstructed events are due to statistical and systematic uncertainties due to unfolding.}
  \label{tab:Cij_separable}
\end{table}

\begin{table}
  \begin{center}
    Separable Region\\
    \begin{tabular}{l|c|c|c|c} 
    \hline
      Helicity $(\times 10^3)$& $n$ & $r$ & $k$ & $D_{A}(\rho)\times 10^3$\\
      \hline
      $n$ & $-348.50\pm0.03$ & $0.00\pm0.00$ & $0.00\pm0.00$\\
      $r$ & $0.00\pm0.00$ & $357.17\pm0.16$ & $96.47 \pm0.05$& $249.72\pm0.24$\\
      $k$ & $0.000\pm0.000$ & $96.47 \pm0.05$ & $- 220.26\pm0.18$\\
      \hline
      $n$ & $-350.34\pm0.08$ & $0.00\pm0.00$ & $0.00\pm0.00$\\
      $r$ & $0.00\pm0.00$ & $395.75\pm0.33$ & $96.83 \pm0.11$& $232.54\pm0.47$\\
      $k$ & $0.00\pm0.00$ &$96.83 \pm0.11$ & $-177.95\pm0.39$\\
      \hline
      \hline
      Beam $(\times 10^3)$& $x$ & $y$ & $z$\\
      \hline
      $x$ & $-294.26\pm0.12$ & $0.00\pm0.08$ & $-0.01\pm0.02$\\
      $y$ & $0.00\pm0.08$ & $-294.31\pm0.12$ & $0.01\pm0.02$ & $254.66\pm0.25$\\
      $z$ & $-0.01\pm0.02$ & $0.01\pm0.02$ & $376.96\pm0.16$\\
      \hline
      $x$ & $-271.37\pm0.26$ & $-0.12\pm0.19$ & $-0.02\pm0.04$\\
      $y$ & $-0.12\pm0.19$ & $-277.45\pm0.25$ & $-0.08\pm0.04$ & $235.29\pm0.52$\\
      $z$ & $-0.02\pm0.04$ & $-0.08\pm0.04$ & $416.28\pm0.34$\\
      \hline
      \hline
      Lab $(\times 10^3)$& $x$ & $y$ & $z$\\
      \hline
      $x$ & $-298.38\pm0.11$ & $0.01\pm0.09$ & $-0.03\pm0.04$\\
      $y$ & $0.00\pm0.09$ & $-298.44\pm0.11$ & $0.03\pm0.04$ & $258.10\pm0.24$\\
      $z$ & $0.00\pm0.04$ & $-0.01\pm0.04$ & $368.06\pm0.15$\\
      \hline
      $x$ & $-274.14\pm0.26$ & $-0.08\pm0.21$ & $-0.02\pm0.07$\\
      $y$ & $-0.16\pm0.20$ & $-280.56\pm0.25$ & $-0.11\pm0.08$ & $235.58\pm0.50$\\
      $z$ & $-0.02\pm0.08$ & $-0.04\pm0.08$ & $407.09\pm0.31$\\
    \end{tabular}
  \end{center}
  \caption{Spin correlation matrices $C_{ij}$ evaluated using the kinematic method using the helicity (top), beam (middle), and lab frame (bottom) bases in the separable region. For each basis, the top panel shows the parton-level results and the bottom panel shows the results after detector simulation. The uncertainties are statistical as described in the main text. All entries of the table have been scaled by a factor of $\times 10^3$.}
    \label{tab:Cij_separable_kinematic}
\end{table}

\begin{table}
  \begin{center}
    Boosted Region\\
    \begin{tabular}{l|c|c|c|c} 
    \hline
      Helicity & $n$ & $r$ & $k$ & $D_{A}(\rho)$\\
      \hline
      $n$ & $-0.501\pm0.004$ & $-0.002\pm0.004$ & $0.002\pm0.004$\\
      $r$ & $-0.001\pm0.004$ & $0.620\pm0.004$ & $0.142 \pm0.004$& $0.197\pm0.003$\\
      $k$ & $-0.003\pm0.004$ & $0.137\pm0.004$ & $ 0.523\pm0.004$\\
      \hline
      $n$ & $-0.484\pm0.082$ & $0.010\pm0.064$ & $0.002\pm0.004$\\
      $r$ & $0.008\pm0.057$ & $0.620\pm0.023$ & $0.133 \pm0.037$& $0.203\pm0.048$\\
      $k$ & $-0.004\pm0.004$ & $0.128\pm0.017$ & $ 0.528\pm0.032$\\
      \hline
      \hline
      Beam & $x$ & $y$ & $z$\\
      \hline
      $x$ & $-0.047\pm0.004$ & $-0.002\pm0.004$ & $0.002\pm0.004$\\
      $y$ & $-0.003\pm0.004$ & $-0.043\pm0.004$ & $0.0\pm0.004$ & $0.011\pm0.002$\\
      $z$ & $-0.001\pm0.004$ & $0.0\pm0.004$ & $0.730\pm0.004$\\
       \hline
      $x$ & $-0.039\pm0.036$ & $-0.002\pm0.004$ & $0.001\pm0.004$\\
      $y$ & $0.013\pm0.022$ & $-0.036\pm0.008$ & $-0.001\pm0.004$ & $0.011\pm0.008$\\
      $z$ & $0.017\pm0.013$ & $-0.001\pm0.004$ & $0.722\pm0.006$\\
      \hline
      \hline
      Lab & $x$ & $y$ & $z$\\
      \hline
      $x$ & $-0.151\pm0.004$ & $-0.001\pm0.004$ & $-0.001\pm0.004$\\
      $y$ & $-0.001\pm0.004$ & $-0.151\pm0.004$ & $ -0.002\pm0.004$ & $0.071\pm0.003$\\
      $z$ & $0.0\pm0.004$ & $0.001\pm0.004$ & $0.503\pm0.004$\\
      \hline
      $x$ & $-0.148\pm0.031$ & $0.022\pm0.033$ & $0.001\pm0.062$\\
      $y$ & $0.015\pm0.023$ & $-0.136\pm0.009$ & $ 0.003\pm0.012$ & $0.065\pm0.017$\\
      $z$ & $0.023\pm0.051$ & $-0.003\pm0.01$ & $0.499\pm0.037$\\
    \end{tabular}
  \end{center}
  \caption{Spin correlation matrices $C_{ij}$ evaluated using the decay method using the helicity (top), beam (middle), and lab frame (bottom) bases in the boosted region. For each basis, the top panel shows the parton-level results and the bottom panel shows the results after detector simulation and unfolding. The uncertainties for parton-level events are purely statistical whereas uncertainties for reconstructed events are due to statistical and systematic uncertainties due to unfolding.}
  \label{tab:Cij_boosted}
\end{table}

\begin{table}
  \begin{center}
    Boosted Region\\
    \begin{tabular}{l|c|c|c|c} 
    \hline
      Helicity $(\times 10^3)$& $n$ & $r$ & $k$ & $D_{A}(\rho)\times 10^3$\\
      \hline
      $n$ & $-504.84\pm0.12$ & $0.00\pm0.00$ & $0.00\pm0.00$\\
      $r$ & $0.00\pm0.00$ & $628.15\pm0.14$ & $140.68 \pm0.05$& $200.81\pm0.08$\\
      $k$ & $0.00\pm0.00$ & $140.68 \pm0.05$& $ 535.88\pm0.12$\\
      \hline
      $n$ & $-488.41\pm0.40$ & $0.00\pm0.00$ & $0.00\pm0.00$\\
      $r$ & $0.00\pm0.00$ & $607.73\pm0.52$ & $146.74 \pm0.18$& $188.49\pm0.25$\\
      $k$ & $0.00\pm0.00$ & $146.74 \pm0.18$ & $ 538.91\pm0.35$\\
      \hline
      \hline
      Beam $(\times 10^3)$& $x$ & $y$ & $z$\\
      \hline
      $x$ & $-40.66\pm0.31$ & $0.03\pm0.31$ & $0.00\pm0.06$\\
      $y$ & $0.03\pm0.31$ & $-40.83\pm0.31$ & $-0.01\pm0.06$ & $9.39\pm0.10$\\
      $z$ & $0.00\pm0.06$ & $-0.01\pm0.06$ & $740.68\pm0.07$\\
      \hline
      $x$ & $-13.91\pm1.12$ & $0.53\pm1.02$ & $0.05\pm0.24$\\
      $y$ & $0.53\pm1.02$ & $-60.47\pm1.11$ & $0.55\pm0.22$ & $8.02\pm0.33$\\
      $z$ & $0.05\pm0.24$ & $0.55\pm0.22$ & $732.26\pm0.24$\\
      \hline
      \hline
      Lab $(\times 10^3)$& $x$ & $y$ & $z$\\
      \hline
      $x$ & $-60.06\pm0.32$ & $0.03\pm0.32$ & $0.02\pm0.09$\\
      $y$ & $0.02\pm0.32$ & $-60.25\pm0.32$ & $-0.02\pm0.09$ & $17.53\pm0.14$\\
      $z$ & $-0.05\pm0.09$ & $0.00\pm0.09$ & $694.14\pm0.09$\\
      \hline
      $x$ & $-28.69\pm1.16$ & $0.58\pm1.06$ & $0.25\pm0.31$\\
      $y$ & $0.45\pm1.06$ & $-76.43\pm1.15$ & $0.59\pm0.31$ & $13.89\pm0.43$\\
      $z$ & $-0.05\pm0.31$ & $0.47\pm0.31$ & $696.26\pm0.28$\\
    \end{tabular}
  \end{center}
  \caption{Spin correlation matrices $C_{ij}$ evaluated using the kinematic method using the helicity (top), beam (middle), and lab frame (bottom) bases in the boosted region. For each basis, the top panel shows the parton-level results and the bottom panel shows the results after detector simulation. The uncertainties are statistical as described in the main text.  All entries of the table have been scaled by a factor of $\times 10^3$.}
  \label{tab:Cij_boosted_kinematic}
\end{table}

\cleardoublepage
\bibliographystyle{jhep}
\bibliography{refs}

\providecommand{\href}[2]{#2}\begingroup\raggedright\begin{thebibliography}{100}

\bibitem{Afik:2020onf}
Y.~Afik and J.R.M.n.~de~Nova, \emph{{Entanglement and quantum tomography with top quarks at the LHC}}, \href{https://doi.org/10.1140/epjp/s13360-021-01902-1}{\emph{Eur. Phys. J. Plus} {\bfseries 136} (2021) 907} [\href{https://arxiv.org/abs/2003.02280}{{\ttfamily 2003.02280}}].

\bibitem{Djouadi:2024lyv}
A.~Djouadi, J.~Ellis and J.~Quevillon, \emph{{Discriminating between Pseudoscalar Higgs and Toponium States at the LHC and Beyond}},  \href{https://arxiv.org/abs/2412.15138}{{\ttfamily 2412.15138}}.

\bibitem{White:2024bjp}
C.D.~White and M.J.~White, \emph{{The magic of top quarks}},  in \emph{{17th International Workshop on Top Quark Physics}}, 12, 2024 [\href{https://arxiv.org/abs/2412.07479}{{\ttfamily 2412.07479}}].

\bibitem{Wildridge:2024yeg}
A.J.~Wildridge, J.P.~Rodgers, E.M.~Colbert, Y.~yao, A.W.~Jung and M.~Liu, \emph{{Bumblebee: Foundation Model for Particle Physics Discovery}},  in \emph{{38th conference on Neural Information Processing Systems}}, 12, 2024 [\href{https://arxiv.org/abs/2412.07867}{{\ttfamily 2412.07867}}].

\bibitem{Altomonte:2024upf}
C.~Altomonte, A.J.~Barr, M.~Eckstein, P.~Horodecki and K.~Sakurai, \emph{{Prospects for quantum process tomography at high energies}},  \href{https://arxiv.org/abs/2412.01892}{{\ttfamily 2412.01892}}.

\bibitem{Gao:2024leu}
L.~Gao, A.~Ruzi, Q.~Li, C.~Zhou, L.~Chen, X.~Zhang et~al., \emph{{Quantum state tomography with muons}},  \href{https://arxiv.org/abs/2411.12518}{{\ttfamily 2411.12518}}.

\bibitem{Fang:2024ple}
Y.~Fang, C.~Gao, Y.-Y.~Li, J.~Shu, Y.~Wu, H.~Xing et~al., \emph{{Quantum Frontiers in High Energy Physics}},  \href{https://arxiv.org/abs/2411.11294}{{\ttfamily 2411.11294}}.

\bibitem{Wu:2024ovc}
Y.~Wu, R.~Jiang, A.~Ruzi, Y.~Ban and Q.~Li, \emph{{Testing Bell inequalities and probing quantum entanglement at CEPC}},  \href{https://arxiv.org/abs/2410.17025}{{\ttfamily 2410.17025}}.

\bibitem{Sullivan:2024wzl}
M.~Sullivan, \emph{{Constraining New Physics with $h\rightarrow VV$ Tomography}},  \href{https://arxiv.org/abs/2410.10980}{{\ttfamily 2410.10980}}.

\bibitem{Cheng:2024rxi}
K.~Cheng, T.~Han and M.~Low, \emph{{Quantum Tomography at Colliders: With or Without Decays}},  \href{https://arxiv.org/abs/2410.08303}{{\ttfamily 2410.08303}}.

\bibitem{Ravina:2024ard}
B.~Ravina, \emph{{Observation of quantum entanglement in top quark pairs at the ATLAS experiment}},  in \emph{{12th Large Hadron Collider Physics Conference}}, 10, 2024 [\href{https://arxiv.org/abs/2410.04590}{{\ttfamily 2410.04590}}].

\bibitem{Du:2024sly}
Y.~Du, X.-G.~He, C.-W.~Liu and J.-P.~Ma, \emph{{Impact of parity violation on quantum entanglement and Bell nonlocality}},  \href{https://arxiv.org/abs/2409.15418}{{\ttfamily 2409.15418}}.

\bibitem{CMS:2024zkc}
{\scshape CMS} collaboration, \emph{{Measurements of polarization and spin correlation and observation of entanglement in top quark pairs using lepton+jets events from proton-proton collisions at $\sqrt{s}$ = 13 TeV}},  \href{https://arxiv.org/abs/2409.11067}{{\ttfamily 2409.11067}}.

\bibitem{Ye:2024dqx}
B.-L.~Ye, L.-Y.~Xue, Z.-Q.~Zhu, D.-D.~Shi and S.-M.~Fei, \emph{{Entropic uncertainty relations and quantum Fisher information of top quarks in a large hadron collider}}, \href{https://doi.org/10.1103/PhysRevD.110.055025}{\emph{Phys. Rev. D} {\bfseries 110} (2024) 055025} [\href{https://arxiv.org/abs/2409.08918}{{\ttfamily 2409.08918}}].

\bibitem{Ruzi:2024cbt}
A.~Ruzi, Y.~Wu, R.~Ding, S.~Qian, A.M.~Levin and Q.~Li, \emph{{Testing Bell inequalities and probing quantum entanglement at a muon collider}}, \href{https://doi.org/10.1007/JHEP10(2024)211}{\emph{JHEP} {\bfseries 10} (2024) 211} [\href{https://arxiv.org/abs/2408.05429}{{\ttfamily 2408.05429}}].

\bibitem{Gabrielli:2024kbz}
E.~Gabrielli and L.~Marzola, \emph{{Entanglement and Bell Inequality Violation in B \textrightarrow{} \ensuremath{\phi}\ensuremath{\phi} Decays}}, \href{https://doi.org/10.3390/sym16081036}{\emph{Symmetry} {\bfseries 16} (2024) 1036} [\href{https://arxiv.org/abs/2408.05010}{{\ttfamily 2408.05010}}].

\bibitem{LoChiatto:2024dmx}
P.~Lo~Chiatto, \emph{{Interference Resurrection of the $\tau$ Dipole through Quantum Tomography}},  \href{https://arxiv.org/abs/2408.04553}{{\ttfamily 2408.04553}}.

\bibitem{ParticleDataGroup:2024cfk}
{\scshape Particle Data Group} collaboration, \emph{{Review of particle physics}}, \href{https://doi.org/10.1103/PhysRevD.110.030001}{\emph{Phys. Rev. D} {\bfseries 110} (2024) 030001}.

\bibitem{Dong:2024xsb}
Z.~Dong, D.~Gon\c{c}alves, K.~Kong, A.J.~Larkoski and A.~Navarro, \emph{{Analytical Insights on Hadronic Top Quark Polarimetry}},  \href{https://arxiv.org/abs/2407.07147}{{\ttfamily 2407.07147}}.

\bibitem{Larkoski:2024uoc}
A.J.~Larkoski, \emph{{QCD masterclass lectures on jet physics and machine learning}}, \href{https://doi.org/10.1140/epjc/s10052-024-13341-0}{\emph{Eur. Phys. J. C} {\bfseries 84} (2024) 1117} [\href{https://arxiv.org/abs/2407.04897}{{\ttfamily 2407.04897}}].

\bibitem{Dong:2024xsg}
Z.~Dong, D.~Gon\c{c}alves, K.~Kong, A.J.~Larkoski and A.~Navarro, \emph{{Hadronic Top Quark Polarimetry with ParticleNet}},  \href{https://arxiv.org/abs/2407.01663}{{\ttfamily 2407.01663}}.

\bibitem{Cheng:2024btk}
K.~Cheng, T.~Han and M.~Low, \emph{{Optimizing Entanglement and Bell Inequality Violation in Top Anti-Top Events}},  \href{https://arxiv.org/abs/2407.01672}{{\ttfamily 2407.01672}}.

\bibitem{Wu:2024asu}
S.~Wu, C.~Qian, Q.~Wang and X.-R.~Zhou, \emph{{Bell nonlocality and entanglement in e+e-\textrightarrow{}YY\textasciimacron{} at BESIII}}, \href{https://doi.org/10.1103/PhysRevD.110.054012}{\emph{Phys. Rev. D} {\bfseries 110} (2024) 054012} [\href{https://arxiv.org/abs/2406.16298}{{\ttfamily 2406.16298}}].

\bibitem{deNova:2024qeo}
J.R.M.n.~de~Nova, P.F.~Palacios, P.A.~Guerrero, I.~Zapata and F.~Sols, \emph{{Resonant analogue configurations in atomic condensates}},  \href{https://arxiv.org/abs/2406.10027}{{\ttfamily 2406.10027}}.

\bibitem{White:2024nuc}
C.D.~White and M.J.~White, \emph{{The magic of entangled top quarks}},  \href{https://arxiv.org/abs/2406.07321}{{\ttfamily 2406.07321}}.

\bibitem{Maltoni:2024wyh}
F.~Maltoni, D.~Pagani and S.~Tentori, \emph{{Top-quark pair production as a probe of light top-philic scalars and anomalous Higgs interactions}}, \href{https://doi.org/10.1007/JHEP09(2024)098}{\emph{JHEP} {\bfseries 09} (2024) 098} [\href{https://arxiv.org/abs/2406.06694}{{\ttfamily 2406.06694}}].

\bibitem{CMS:2024pts}
{\scshape CMS} collaboration, \emph{{Observation of quantum entanglement in top quark pair production in proton\textendash{}proton collisions at $\sqrt{s} = 13$ TeV}}, \href{https://doi.org/10.1088/1361-6633/ad7e4d}{\emph{Rept. Prog. Phys.} {\bfseries 87} (2024) 117801} [\href{https://arxiv.org/abs/2406.03976}{{\ttfamily 2406.03976}}].

\bibitem{Afik:2024uif}
Y.~Afik, Y.~Kats, J.R.M.n.~de~Nova, A.~Soffer and D.~Uzan, \emph{{Entanglement and Bell nonlocality with bottom-quark pairs at hadron colliders}},  \href{https://arxiv.org/abs/2406.04402}{{\ttfamily 2406.04402}}.

\bibitem{Kowalska:2024kbs}
K.~Kowalska and E.M.~Sessolo, \emph{{Entanglement in flavored scalar scattering}}, \href{https://doi.org/10.1007/JHEP07(2024)156}{\emph{JHEP} {\bfseries 07} (2024) 156} [\href{https://arxiv.org/abs/2404.13743}{{\ttfamily 2404.13743}}].

\bibitem{ATLAS:2024kxj}
{\scshape ATLAS} collaboration, \emph{{Climbing to the Top of the ATLAS 13 TeV data}},  \href{https://arxiv.org/abs/2404.10674}{{\ttfamily 2404.10674}}.

\bibitem{Maltoni:2024csn}
F.~Maltoni, C.~Severi, S.~Tentori and E.~Vryonidou, \emph{{Quantum tops at circular lepton colliders}}, \href{https://doi.org/10.1007/JHEP09(2024)001}{\emph{JHEP} {\bfseries 09} (2024) 001} [\href{https://arxiv.org/abs/2404.08049}{{\ttfamily 2404.08049}}].

\bibitem{Subba:2024mnl}
A.~Subba and R.~Rahaman, \emph{{On bipartite and tripartite entanglement at present and future particle colliders}},  \href{https://arxiv.org/abs/2404.03292}{{\ttfamily 2404.03292}}.

\bibitem{Morales:2024jhj}
R.A.~Morales, \emph{{Tripartite entanglement and Bell non-locality in loop-induced Higgs boson decays}}, \href{https://doi.org/10.1140/epjc/s10052-024-12921-4}{\emph{Eur. Phys. J. C} {\bfseries 84} (2024) 581} [\href{https://arxiv.org/abs/2403.18023}{{\ttfamily 2403.18023}}].

\bibitem{Duch:2024pwm}
M.~Duch, A.~Strumia and A.~Titov, \emph{{New physics in spin entanglement}},  \href{https://arxiv.org/abs/2403.14757}{{\ttfamily 2403.14757}}.

\bibitem{Aguilar-Saavedra:2024whi}
J.A.~Aguilar-Saavedra, \emph{{Tripartite entanglement in H\textrightarrow{}ZZ,WW decays}}, \href{https://doi.org/10.1103/PhysRevD.109.113004}{\emph{Phys. Rev. D} {\bfseries 109} (2024) 113004} [\href{https://arxiv.org/abs/2403.13942}{{\ttfamily 2403.13942}}].

\bibitem{Aoude:2024xpx}
R.~Aoude, G.~Elor, G.N.~Remmen and O.~Sumensari, \emph{{Positivity in Amplitudes from Quantum Entanglement}},  \href{https://arxiv.org/abs/2402.16956}{{\ttfamily 2402.16956}}.

\bibitem{Wu:2024mtj}
S.~Wu, C.~Qian, Y.-G.~Yang and Q.~Wang, \emph{{Generalized Quantum Measurement in Spin-Correlated Hyperon-Antihyperon Decays}}, \href{https://doi.org/10.1088/0256-307X/41/11/110301}{\emph{Chin. Phys. Lett.} {\bfseries 41} (2024) 110301} [\href{https://arxiv.org/abs/2402.16574}{{\ttfamily 2402.16574}}].

\bibitem{Jung:2024ans}
A.~Jung, \emph{{Properties of the Top Quark}}, \href{https://doi.org/10.3390/universe10030106}{\emph{Universe} {\bfseries 10} (2024) 106}.

\bibitem{Aguilar-Saavedra:2024vpd}
J.A.~Aguilar-Saavedra, \emph{{Full quantum tomography of top quark decays}}, \href{https://doi.org/10.1016/j.physletb.2024.138849}{\emph{Phys. Lett. B} {\bfseries 855} (2024) 138849} [\href{https://arxiv.org/abs/2402.14725}{{\ttfamily 2402.14725}}].

\bibitem{Barr:2024djo}
A.J.~Barr, M.~Fabbrichesi, R.~Floreanini, E.~Gabrielli and L.~Marzola, \emph{{Quantum entanglement and Bell inequality violation at colliders}}, \href{https://doi.org/10.1016/j.ppnp.2024.104134}{\emph{Prog. Part. Nucl. Phys.} {\bfseries 139} (2024) 104134} [\href{https://arxiv.org/abs/2402.07972}{{\ttfamily 2402.07972}}].

\bibitem{Sarkar:2024dxe}
S.~Sarkar, \emph{{Model-independent inference of quantum interaction from statistics}}, \href{https://doi.org/10.1103/PhysRevA.110.L020402}{\emph{Phys. Rev. A} {\bfseries 110} (2024) L020402} [\href{https://arxiv.org/abs/2402.08003}{{\ttfamily 2402.08003}}].

\bibitem{Belhaj:2024bqk}
A.~Belhaj, S.E.~Ennadifi and L.~Jebli, \emph{{Probing quantum entanglement from quantum correction to newtonian potential energy}}, \href{https://doi.org/10.1088/1402-4896/ad241c}{\emph{Phys. Scripta} {\bfseries 99} (2024) 035217} [\href{https://arxiv.org/abs/2401.14342}{{\ttfamily 2401.14342}}].

\bibitem{Blasone:2024dud}
M.~Blasone, G.~Lambiase and B.~Micciola, \emph{{Entanglement distribution in Bhabha scattering with an entangled spectator particle}}, \href{https://doi.org/10.1103/PhysRevD.109.096022}{\emph{Phys. Rev. D} {\bfseries 109} (2024) 096022} [\href{https://arxiv.org/abs/2401.10715}{{\ttfamily 2401.10715}}].

\bibitem{Aguilar-Saavedra:2024hwd}
J.A.~Aguilar-Saavedra, \emph{{A closer look at post-decay $t \bar t$ entanglement}}, \href{https://doi.org/10.1103/PhysRevD.109.096027}{\emph{Phys. Rev. D} {\bfseries 109} (2024) 096027} [\href{https://arxiv.org/abs/2401.10988}{{\ttfamily 2401.10988}}].

\bibitem{Maltoni:2024tul}
F.~Maltoni, C.~Severi, S.~Tentori and E.~Vryonidou, \emph{{Quantum detection of new physics in top-quark pair production at the LHC}}, \href{https://doi.org/10.1007/JHEP03(2024)099}{\emph{JHEP} {\bfseries 03} (2024) 099} [\href{https://arxiv.org/abs/2401.08751}{{\ttfamily 2401.08751}}].

\bibitem{Aguilar-Saavedra:2024fig}
J.A.~Aguilar-Saavedra and J.A.~Casas, \emph{{Entanglement Autodistillation from Particle Decays}}, \href{https://doi.org/10.1103/PhysRevLett.133.111801}{\emph{Phys. Rev. Lett.} {\bfseries 133} (2024) 111801} [\href{https://arxiv.org/abs/2401.06854}{{\ttfamily 2401.06854}}].

\bibitem{Li:2024luk}
S.~Li, W.~Shen and J.M.~Yang, \emph{{Can Bell inequalities be tested via scattering cross-section at colliders ?}}, \href{https://doi.org/10.1140/epjc/s10052-024-13584-x}{\emph{Eur. Phys. J. C} {\bfseries 84} (2024) 1195} [\href{https://arxiv.org/abs/2401.01162}{{\ttfamily 2401.01162}}].

\bibitem{CMS:2024hgo}
{\scshape CMS} collaboration, \emph{{Probing entanglement in top quark production with the CMS detector}}, .

\bibitem{Simpson:2024hbr}
E.L.~Simpson, \emph{{A new spin on top-quark physics: using angular distributions to probe top-quark properties, and make the first observation of entanglement between quarks}}, Ph.D. thesis, Glasgow U., 2024.
\newblock 10.5525/gla.thesis.84277.

\bibitem{CMS:2024vqh}
{\scshape CMS} collaboration, \emph{{Measurements of polarization, spin correlations, and entanglement in top quark pairs using lepton+jets events from pp collisions at $\sqrt{s}=13~\mathrm{TeV}$}}, .

\bibitem{Blasone:2024bmo}
M.~Blasone, S.~De~Siena and C.~Matrella, \emph{{Realism-based nonlocality in neutrino oscillations}}, \href{https://doi.org/10.1088/1742-6596/2883/1/012001}{\emph{J. Phys. Conf. Ser.} {\bfseries 2883} (2024) 012001}.

\bibitem{Khor:2023xar}
B.J.J.~Khor, D.M.~K\"urk\c{c}\"uoglu, T.J.~Hobbs, G.N.~Perdue and I.~Klich, \emph{{Confinement and Kink Entanglement Asymmetry on a Quantum Ising Chain}}, \href{https://doi.org/10.22331/q-2024-09-06-1462}{\emph{Quantum} {\bfseries 8} (2024) 1462} [\href{https://arxiv.org/abs/2312.08601}{{\ttfamily 2312.08601}}].

\bibitem{DeFabritiis:2023llu}
P.~De~Fabritiis, F.M.~Guedes, M.S.~Guimaraes, I.~Roditi and S.P.~Sorella, \emph{{Using Weyl operators to study Mermin\textquoteright{}s inequalities in quantum field theory}}, \href{https://doi.org/10.1103/PhysRevD.109.045020}{\emph{Phys. Rev. D} {\bfseries 109} (2024) 045020} [\href{https://arxiv.org/abs/2312.06918}{{\ttfamily 2312.06918}}].

\bibitem{Liu:2023bnr}
Q.~Liu and I.~Low, \emph{{Hints of entanglement suppression in hyperon-nucleon scattering}}, \href{https://doi.org/10.1016/j.physletb.2024.138899}{\emph{Phys. Lett. B} {\bfseries 856} (2024) 138899} [\href{https://arxiv.org/abs/2312.02289}{{\ttfamily 2312.02289}}].

\bibitem{Ehataht:2023zzt}
K.~Ehat\"aht, M.~Fabbrichesi, L.~Marzola and C.~Veelken, \emph{{Probing entanglement and testing Bell inequality violation with e+e-\textrightarrow{}\ensuremath{\tau}+\ensuremath{\tau}- at Belle II}}, \href{https://doi.org/10.1103/PhysRevD.109.032005}{\emph{Phys. Rev. D} {\bfseries 109} (2024) 032005} [\href{https://arxiv.org/abs/2311.17555}{{\ttfamily 2311.17555}}].

\bibitem{Cheng:2023qmz}
K.~Cheng, T.~Han and M.~Low, \emph{{Optimizing fictitious states for Bell inequality violation in bipartite qubit systems with applications to the tt\textasciimacron{} system}}, \href{https://doi.org/10.1103/PhysRevD.109.116005}{\emph{Phys. Rev. D} {\bfseries 109} (2024) 116005} [\href{https://arxiv.org/abs/2311.09166}{{\ttfamily 2311.09166}}].

\bibitem{Kats:2023zxb}
Y.~Kats and D.~Uzan, \emph{{Prospects for measuring quark polarization and spin correlations in $b\overline{b }$ and $c\overline{c }$ samples at the LHC}}, \href{https://doi.org/10.1007/JHEP03(2024)063}{\emph{JHEP} {\bfseries 03} (2024) 063} [\href{https://arxiv.org/abs/2311.08226}{{\ttfamily 2311.08226}}].

\bibitem{ATLAS:2023fsd}
{\scshape ATLAS} collaboration, \emph{{Observation of quantum entanglement with top quarks at the ATLAS detector}}, \href{https://doi.org/10.1038/s41586-024-07824-z}{\emph{Nature} {\bfseries 633} (2024) 542} [\href{https://arxiv.org/abs/2311.07288}{{\ttfamily 2311.07288}}].

\bibitem{Altomonte:2023mug}
C.~Altomonte and A.J.~Barr, \emph{{Quantum state-channel duality for the calculation of Standard Model scattering amplitudes}}, \href{https://doi.org/10.1016/j.physletb.2023.138303}{\emph{Phys. Lett. B} {\bfseries 847} (2023) 138303} [\href{https://arxiv.org/abs/2312.02242}{{\ttfamily 2312.02242}}].

\bibitem{Han:2023fci}
T.~Han, M.~Low and T.A.~Wu, \emph{{Quantum entanglement and Bell inequality violation in semi-leptonic top decays}}, \href{https://doi.org/10.1007/JHEP07(2024)192}{\emph{JHEP} {\bfseries 07} (2024) 192} [\href{https://arxiv.org/abs/2310.17696}{{\ttfamily 2310.17696}}].

\bibitem{Bernal:2023jba}
A.~Bernal, \emph{{Quantum tomography of helicity states for general scattering processes}}, \href{https://doi.org/10.1103/PhysRevD.109.116007}{\emph{Phys. Rev. D} {\bfseries 109} (2024) 116007} [\href{https://arxiv.org/abs/2310.10838}{{\ttfamily 2310.10838}}].

\bibitem{Jung:2023fpv}
A.~Jung and J.~Kieseler, \emph{{Top Quarks from Tevatron to the LHC}}, \href{https://doi.org/10.3390/sym15101915}{\emph{Symmetry} {\bfseries 15} (2023) 1915}.

\bibitem{Sakurai:2023nsc}
K.~Sakurai and M.~Spannowsky, \emph{{Three-Body Entanglement in Particle Decays}}, \href{https://doi.org/10.1103/PhysRevLett.132.151602}{\emph{Phys. Rev. Lett.} {\bfseries 132} (2024) 151602} [\href{https://arxiv.org/abs/2310.01477}{{\ttfamily 2310.01477}}].

\bibitem{FerreiradaSilva:2023mhf}
P.~Ferreira~da Silva, \emph{{Physics of the Top Quark at the LHC: An Appraisal and Outlook of the Road Ahead}}, \href{https://doi.org/10.1146/annurev-nucl-102419-052854}{\emph{Ann. Rev. Nucl. Part. Sci.} {\bfseries 73} (2023) 255}.

\bibitem{Ma:2023yvd}
K.~Ma and T.~Li, \emph{{Testing Bell inequality through ${\boldsymbol h{\bf\rightarrow}\boldsymbol\tau\boldsymbol\tau }$ at CEPC*}}, \href{https://doi.org/10.1088/1674-1137/ad62d8}{\emph{Chin. Phys. C} {\bfseries 48} (2024) 103105} [\href{https://arxiv.org/abs/2309.08103}{{\ttfamily 2309.08103}}].

\bibitem{DeFabritiis:2023tkh}
P.~De~Fabritiis, F.M.~Guedes, M.S.~Guimaraes, G.~Peruzzo, I.~Roditi and S.P.~Sorella, \emph{{Weyl operators, Tomita-Takesaki theory, and Bell-Clauser-Horne-Shimony-Holt inequality violations}}, \href{https://doi.org/10.1103/PhysRevD.108.085026}{\emph{Phys. Rev. D} {\bfseries 108} (2023) 085026} [\href{https://arxiv.org/abs/2309.02941}{{\ttfamily 2309.02941}}].

\bibitem{Bi:2023uop}
Q.~Bi, Q.-H.~Cao, K.~Cheng and H.~Zhang, \emph{{New observables for testing Bell inequalities in W boson pair production}}, \href{https://doi.org/10.1103/PhysRevD.109.036022}{\emph{Phys. Rev. D} {\bfseries 109} (2024) 036022} [\href{https://arxiv.org/abs/2307.14895}{{\ttfamily 2307.14895}}].

\bibitem{Bernal:2023ruk}
A.~Bernal, P.~Caban and J.~Rembieli\'nski, \emph{{Entanglement and Bell inequalities violation in $H\rightarrow ZZ$ with anomalous coupling}}, \href{https://doi.org/10.1140/epjc/s10052-023-12216-0}{\emph{Eur. Phys. J. C} {\bfseries 83} (2023) 1050} [\href{https://arxiv.org/abs/2307.13496}{{\ttfamily 2307.13496}}].

\bibitem{Aoude:2023hxv}
R.~Aoude, E.~Madge, F.~Maltoni and L.~Mantani, \emph{{Probing new physics through entanglement in diboson production}}, \href{https://doi.org/10.1007/JHEP12(2023)017}{\emph{JHEP} {\bfseries 12} (2023) 017} [\href{https://arxiv.org/abs/2307.09675}{{\ttfamily 2307.09675}}].

\bibitem{Aguilar-Saavedra:2023hss}
J.A.~Aguilar-Saavedra, \emph{{Postdecay quantum entanglement in top pair production}}, \href{https://doi.org/10.1103/PhysRevD.108.076025}{\emph{Phys. Rev. D} {\bfseries 108} (2023) 076025} [\href{https://arxiv.org/abs/2307.06991}{{\ttfamily 2307.06991}}].

\bibitem{Dudal:2023mij}
D.~Dudal, P.~De~Fabritiis, M.S.~Guimaraes, I.~Roditi and S.P.~Sorella, \emph{{Maximal violation of the Bell-Clauser-Horne-Shimony-Holt inequality via bumpified Haar wavelets}}, \href{https://doi.org/10.1103/PhysRevD.108.L081701}{\emph{Phys. Rev. D} {\bfseries 108} (2023) L081701} [\href{https://arxiv.org/abs/2307.04611}{{\ttfamily 2307.04611}}].

\bibitem{Morales:2023gow}
R.A.~Morales, \emph{{Exploring Bell inequalities and quantum entanglement in vector boson scattering}}, \href{https://doi.org/10.1140/epjp/s13360-023-04784-7}{\emph{Eur. Phys. J. Plus} {\bfseries 138} (2023) 1157} [\href{https://arxiv.org/abs/2306.17247}{{\ttfamily 2306.17247}}].

\bibitem{Dong:2023xiw}
Z.~Dong, D.~Gon\c{c}alves, K.~Kong and A.~Navarro, \emph{{Entanglement and Bell inequalities with boosted tt\textasciimacron{}}}, \href{https://doi.org/10.1103/PhysRevD.109.115023}{\emph{Phys. Rev. D} {\bfseries 109} (2024) 115023} [\href{https://arxiv.org/abs/2305.07075}{{\ttfamily 2305.07075}}].

\bibitem{Bittencourt:2023asd}
V.A.S.V.~Bittencourt, M.~Blasone, S.~De~Siena and C.~Matrella, \emph{{Quantifying quantumness in three-flavor neutrino oscillations}}, \href{https://doi.org/10.1140/epjc/s10052-024-12631-x}{\emph{Eur. Phys. J. C} {\bfseries 84} (2024) 301} [\href{https://arxiv.org/abs/2305.06095}{{\ttfamily 2305.06095}}].

\bibitem{Dudal:2023pbc}
D.~Dudal, P.~De~Fabritiis, M.S.~Guimaraes, G.~Peruzzo and S.P.~Sorella, \emph{{BRST invariant formulation of the Bell-CHSH inequality in gauge field theories}}, \href{https://doi.org/10.21468/SciPostPhys.15.5.201}{\emph{SciPost Phys.} {\bfseries 15} (2023) 201} [\href{https://arxiv.org/abs/2304.01028}{{\ttfamily 2304.01028}}].

\bibitem{DeFabritiis:2023ubj}
P.~De~Fabritiis, I.~Roditi and S.P.~Sorella, \emph{{Mermin's inequalities in Quantum Field Theory}}, \href{https://doi.org/10.1016/j.physletb.2023.138198}{\emph{Phys. Lett. B} {\bfseries 846} (2023) 138198} [\href{https://arxiv.org/abs/2303.12195}{{\ttfamily 2303.12195}}].

\bibitem{Ghosh:2023rpj}
D.~Ghosh and R.~Sharma, \emph{{Bell violation in 2 \textrightarrow{} 2 scattering in photon, gluon and graviton EFTs}}, \href{https://doi.org/10.1007/JHEP08(2023)146}{\emph{JHEP} {\bfseries 08} (2023) 146} [\href{https://arxiv.org/abs/2303.03375}{{\ttfamily 2303.03375}}].

\bibitem{Fabbrichesi:2023cev}
M.~Fabbrichesi, R.~Floreanini, E.~Gabrielli and L.~Marzola, \emph{{Bell inequalities and quantum entanglement in weak gauge boson production at the LHC and future colliders}}, \href{https://doi.org/10.1140/epjc/s10052-023-11935-8}{\emph{Eur. Phys. J. C} {\bfseries 83} (2023) 823} [\href{https://arxiv.org/abs/2302.00683}{{\ttfamily 2302.00683}}].

\bibitem{ATLAS:2023jzs}
{\scshape ATLAS} collaboration, \emph{{Observation of quantum entanglement in top-quark pair production using $pp$ collisions of $\sqrt{s} = 13$\textasciitilde{}\textbackslash{}TeV\textbackslash{} with the ATLAS detector}}, .

\bibitem{Gross:2022hyw}
F.~Gross et~al., \emph{{50 Years of Quantum Chromodynamics}}, \href{https://doi.org/10.1140/epjc/s10052-023-11949-2}{\emph{Eur. Phys. J. C} {\bfseries 83} (2023) 1125} [\href{https://arxiv.org/abs/2212.11107}{{\ttfamily 2212.11107}}].

\bibitem{Altakach:2022ywa}
M.M.~Altakach, P.~Lamba, F.~Maltoni, K.~Mawatari and K.~Sakurai, \emph{{Quantum information and CP measurement in H\textrightarrow{}\ensuremath{\tau}+\ensuremath{\tau}- at future lepton colliders}}, \href{https://doi.org/10.1103/PhysRevD.107.093002}{\emph{Phys. Rev. D} {\bfseries 107} (2023) 093002} [\href{https://arxiv.org/abs/2211.10513}{{\ttfamily 2211.10513}}].

\bibitem{Mantani:2022dao}
L.~Mantani, \emph{{Quantum SMEFT tomography: top quark pair}}, \href{https://doi.org/10.22323/1.414.0858}{\emph{PoS} {\bfseries ICHEP2022} (2022) 858} [\href{https://arxiv.org/abs/2211.03428}{{\ttfamily 2211.03428}}].

\bibitem{Severi:2022qjy}
C.~Severi and E.~Vryonidou, \emph{{Quantum entanglement and top spin correlations in SMEFT at higher orders}}, \href{https://doi.org/10.1007/JHEP01(2023)148}{\emph{JHEP} {\bfseries 01} (2023) 148} [\href{https://arxiv.org/abs/2210.09330}{{\ttfamily 2210.09330}}].

\bibitem{Aguilar-Saavedra:2022mpg}
J.A.~Aguilar-Saavedra, \emph{{Laboratory-frame tests of quantum entanglement in H\textrightarrow{}WW}}, \href{https://doi.org/10.1103/PhysRevD.107.076016}{\emph{Phys. Rev. D} {\bfseries 107} (2023) 076016} [\href{https://arxiv.org/abs/2209.14033}{{\ttfamily 2209.14033}}].

\bibitem{Ashby-Pickering:2022umy}
R.~Ashby-Pickering, A.J.~Barr and A.~Wierzchucka, \emph{{Quantum state tomography, entanglement detection and Bell violation prospects in weak decays of massive particles}}, \href{https://doi.org/10.1007/JHEP05(2023)020}{\emph{JHEP} {\bfseries 05} (2023) 020} [\href{https://arxiv.org/abs/2209.13990}{{\ttfamily 2209.13990}}].

\bibitem{Aguilar-Saavedra:2022wam}
J.A.~Aguilar-Saavedra, A.~Bernal, J.A.~Casas and J.M.~Moreno, \emph{{Testing entanglement and Bell inequalities in H\textrightarrow{}ZZ}}, \href{https://doi.org/10.1103/PhysRevD.107.016012}{\emph{Phys. Rev. D} {\bfseries 107} (2023) 016012} [\href{https://arxiv.org/abs/2209.13441}{{\ttfamily 2209.13441}}].

\bibitem{Ehlers:2022oke}
P.J.~Ehlers, \emph{{Entanglement between Quarks in Hadrons}}, Ph.D. thesis, Washington U., Seattle, 2022.

\bibitem{Ehlers:2022oal}
P.J.~Ehlers, \emph{{Entanglement between valence and sea quarks in hadrons of 1+1 dimensional QCD}}, \href{https://doi.org/10.1016/j.aop.2023.169290}{\emph{Annals Phys.} {\bfseries 452} (2023) 169290} [\href{https://arxiv.org/abs/2209.09867}{{\ttfamily 2209.09867}}].

\bibitem{Afik:2022dgh}
Y.~Afik and J.R.M.n.~de~Nova, \emph{{Quantum Discord and Steering in Top Quarks at the LHC}}, \href{https://doi.org/10.1103/PhysRevLett.130.221801}{\emph{Phys. Rev. Lett.} {\bfseries 130} (2023) 221801} [\href{https://arxiv.org/abs/2209.03969}{{\ttfamily 2209.03969}}].

\bibitem{Hung:2022azf}
L.-Y.~Hung, K.~Ji and T.~Wang, \emph{{Scrambling and entangling spinning particles}}, \href{https://doi.org/10.1007/JHEP02(2023)197}{\emph{JHEP} {\bfseries 02} (2023) 197} [\href{https://arxiv.org/abs/2208.12128}{{\ttfamily 2208.12128}}].

\bibitem{Fabbrichesi:2022ovb}
M.~Fabbrichesi, R.~Floreanini and E.~Gabrielli, \emph{{Constraining new physics in entangled two-qubit systems: top-quark, tau-lepton and photon pairs}}, \href{https://doi.org/10.1140/epjc/s10052-023-11307-2}{\emph{Eur. Phys. J. C} {\bfseries 83} (2023) 162} [\href{https://arxiv.org/abs/2208.11723}{{\ttfamily 2208.11723}}].

\bibitem{Kurashvili:2022ybg}
P.~Kurashvili and L.~Chotorlishvili, \emph{{Quantum discord and entropic measures of two relativistic fermions}}, \href{https://doi.org/10.1088/1751-8121/aca7a0}{\emph{J. Phys. A} {\bfseries 55} (2022) 495303} [\href{https://arxiv.org/abs/2207.12963}{{\ttfamily 2207.12963}}].

\bibitem{Hentschinski:2022rsa}
M.~Hentschinski, K.~Kutak and R.~Straka, \emph{{Maximally entangled proton and charged hadron multiplicity in Deep Inelastic Scattering}}, \href{https://doi.org/10.1140/epjc/s10052-022-11122-1}{\emph{Eur. Phys. J. C} {\bfseries 82} (2022) 1147} [\href{https://arxiv.org/abs/2207.09430}{{\ttfamily 2207.09430}}].

\bibitem{Aguilar-Saavedra:2022uye}
J.A.~Aguilar-Saavedra and J.A.~Casas, \emph{{Improved tests of entanglement and Bell inequalities with LHC tops}}, \href{https://doi.org/10.1140/epjc/s10052-022-10630-4}{\emph{Eur. Phys. J. C} {\bfseries 82} (2022) 666} [\href{https://arxiv.org/abs/2205.00542}{{\ttfamily 2205.00542}}].

\bibitem{Ramos:2022gia}
G.S.~Ramos and M.V.T.~Machado, \emph{{Investigating the QCD dynamical entropy in high-energy hadronic collisions}}, \href{https://doi.org/10.1103/PhysRevD.105.094009}{\emph{Phys. Rev. D} {\bfseries 105} (2022) 094009} [\href{https://arxiv.org/abs/2203.10986}{{\ttfamily 2203.10986}}].

\bibitem{Aoude:2022imd}
R.~Aoude, E.~Madge, F.~Maltoni and L.~Mantani, \emph{{Quantum SMEFT tomography: Top quark pair production at the LHC}}, \href{https://doi.org/10.1103/PhysRevD.106.055007}{\emph{Phys. Rev. D} {\bfseries 106} (2022) 055007} [\href{https://arxiv.org/abs/2203.05619}{{\ttfamily 2203.05619}}].

\bibitem{Afik:2022kwm}
Y.~Afik and J.R.M.n.~de~Nova, \emph{{Quantum information with top quarks in QCD}}, \href{https://doi.org/10.22331/q-2022-09-29-820}{\emph{Quantum} {\bfseries 6} (2022) 820} [\href{https://arxiv.org/abs/2203.05582}{{\ttfamily 2203.05582}}].

\bibitem{Larkoski:2022lmv}
A.J.~Larkoski, \emph{{General analysis for observing quantum interference at colliders}}, \href{https://doi.org/10.1103/PhysRevD.105.096012}{\emph{Phys. Rev. D} {\bfseries 105} (2022) 096012} [\href{https://arxiv.org/abs/2201.03159}{{\ttfamily 2201.03159}}].

\bibitem{Severi:2021cnj}
C.~Severi, C.D.E.~Boschi, F.~Maltoni and M.~Sioli, \emph{{Quantum tops at the LHC: from entanglement to Bell inequalities}}, \href{https://doi.org/10.1140/epjc/s10052-022-10245-9}{\emph{Eur. Phys. J. C} {\bfseries 82} (2022) 285} [\href{https://arxiv.org/abs/2110.10112}{{\ttfamily 2110.10112}}].

\bibitem{Bittencourt:2021xcx}
V.A.S.V.~Bittencourt, A.E.~Bernardini and M.~Blasone, \emph{{Lepton-Antineutrino Entanglement and Chiral Oscillations}}, \href{https://doi.org/10.3390/universe7080293}{\emph{Universe} {\bfseries 7} (2021) 293}.

\bibitem{Karlberg:2021kwr}
A.~Karlberg, G.P.~Salam, L.~Scyboz and R.~Verheyen, \emph{{Spin correlations in final-state parton showers and jet observables}}, \href{https://doi.org/10.1140/epjc/s10052-021-09378-0}{\emph{Eur. Phys. J. C} {\bfseries 81} (2021) 681} [\href{https://arxiv.org/abs/2103.16526}{{\ttfamily 2103.16526}}].

\bibitem{Eckstein:2021pgm}
M.~Eckstein and P.~Horodecki, \emph{{Probing the limits of quantum theory with quantum information at subnuclear scales}}, \href{https://doi.org/10.1098/rspa.2021.0806}{\emph{Proc. Roy. Soc. Lond. A} {\bfseries 478} (2022) 20210806} [\href{https://arxiv.org/abs/2103.12000}{{\ttfamily 2103.12000}}].

\bibitem{Fabbrichesi:2021npl}
M.~Fabbrichesi, R.~Floreanini and G.~Panizzo, \emph{{Testing Bell Inequalities at the LHC with Top-Quark Pairs}}, \href{https://doi.org/10.1103/PhysRevLett.127.161801}{\emph{Phys. Rev. Lett.} {\bfseries 127} (2021) 161801} [\href{https://arxiv.org/abs/2102.11883}{{\ttfamily 2102.11883}}].

\bibitem{Vollmann:2021moq}
M.~Vollmann, \emph{{Resummation of large electroweak terms for indirect dark matter detection}},  in \emph{{Strong interactions from QCD to new strong dynamics at LHC and Future Colliders}}, pp.~79--86, 1, 2021.

\bibitem{Feal:2020myr}
X.~Feal, C.~Pajares and R.~Vazquez, \emph{{Thermal and hard scales in transverse momentum distributions, fluctuations, and entanglement}}, \href{https://doi.org/10.1103/PhysRevC.104.044904}{\emph{Phys. Rev. C} {\bfseries 104} (2021) 044904} [\href{https://arxiv.org/abs/2012.02894}{{\ttfamily 2012.02894}}].

\bibitem{Iskander:2020rkb}
G.~Iskander, J.~Pan, M.~Tyler, C.~Weber and O.K.~Baker, \emph{{Quantum Entanglement and Thermal Behavior in Charged-Current Weak Interactions}}, \href{https://doi.org/10.1016/j.physletb.2020.135948}{\emph{Phys. Lett. B} {\bfseries 811} (2020) 135948} [\href{https://arxiv.org/abs/2010.00709}{{\ttfamily 2010.00709}}].

\bibitem{Aoude:2020mlg}
R.~Aoude, M.-Z.~Chung, Y.-t.~Huang, C.S.~Machado and M.-K.~Tam, \emph{{Silence of Binary Kerr Black Holes}}, \href{https://doi.org/10.1103/PhysRevLett.125.181602}{\emph{Phys. Rev. Lett.} {\bfseries 125} (2020) 181602} [\href{https://arxiv.org/abs/2007.09486}{{\ttfamily 2007.09486}}].

\bibitem{Ramos:2020kaj}
G.S.~Ramos and M.V.T.~Machado, \emph{{Investigating entanglement entropy at small-$x$ in DIS off protons and nuclei}}, \href{https://doi.org/10.1103/PhysRevD.101.074040}{\emph{Phys. Rev. D} {\bfseries 101} (2020) 074040} [\href{https://arxiv.org/abs/2003.05008}{{\ttfamily 2003.05008}}].

\bibitem{Grossi:2024jae}
M.~Grossi, G.~Pelliccioli and A.~Vicini, \emph{{From angular coefficients to quantum observables: a phenomenological appraisal in di-boson systems}}, \href{https://doi.org/10.1007/JHEP12(2024)120}{\emph{JHEP} {\bfseries 12} (2024) 120} [\href{https://arxiv.org/abs/2409.16731}{{\ttfamily 2409.16731}}].

\bibitem{Han:2025ewp}
T.~Han, M.~Low and Y.~Su, \emph{{Entanglement and Bell Nonlocality in $\tau^+ \tau^-$ at the BEPC}},  \href{https://arxiv.org/abs/2501.04801}{{\ttfamily 2501.04801}}.

\bibitem{Fabbrichesi:2025ywl}
M.~Fabbrichesi, M.~Low and L.~Marzola, \emph{{The trace distance between density matrices, a nifty tool in new-physics searches}},  \href{https://arxiv.org/abs/2501.03311}{{\ttfamily 2501.03311}}.

\bibitem{Fu:2024bki}
J.-H.~Fu, Y.-J.~Li, H.-M.~Yang, Y.-B.~Li, Y.-J.~Zhang and C.-P.~Shen, \emph{{Toponium: the smallest bound state and simplest hadron in quantum mechanics}},  \href{https://arxiv.org/abs/2412.11254}{{\ttfamily 2412.11254}}.

\bibitem{Zurek_2000}
W.~Zurek, \emph{Einselection and decoherence from an information theory perspective}, \href{https://doi.org/10.1002/andp.200051211-1204}{\emph{Annalen der Physik} {\bfseries 512} (2000) 855–864}.

\bibitem{Ollivier:2001fdq}
H.~Ollivier and W.H.~Zurek, \emph{{Introducing Quantum Discord}}, \href{https://doi.org/10.1103/PhysRevLett.88.017901}{\emph{Phys. Rev. Lett.} {\bfseries 88} (2001) 017901} [\href{https://arxiv.org/abs/quant-ph/0105072}{{\ttfamily quant-ph/0105072}}].

\bibitem{Bera_2017}
A.~Bera, T.~Das, D.~Sadhukhan, S.~Singha~Roy, A.~Sen(De) and U.~Sen, \emph{Quantum discord and its allies: a review of recent progress}, \href{https://doi.org/10.1088/1361-6633/aa872f}{\emph{Reports on Progress in Physics} {\bfseries 81} (2017) 024001}.

\bibitem{Fano:1983zz}
U.~Fano, \emph{{Pairs of two-level systems}}, \href{https://doi.org/10.1103/RevModPhys.55.855}{\emph{Rev. Mod. Phys.} {\bfseries 55} (1983) 855}.

\bibitem{Fano:1957zz}
U.~Fano, \emph{{Description of States in Quantum Mechanics by Density Matrix and Operator Techniques}}, \href{https://doi.org/10.1103/RevModPhys.29.74}{\emph{Rev. Mod. Phys.} {\bfseries 29} (1957) 74}.

\bibitem{Araki:1970ba}
H.~Araki and E.H.~Lieb, \emph{{Entropy inequalities}}, \href{https://doi.org/10.1007/BF01646092}{\emph{Commun. Math. Phys.} {\bfseries 18} (1970) 160}.

\bibitem{PhysRevA.84.042124}
N.~Li and S.~Luo, \emph{Classical and quantum correlative capacities of quantum systems}, \href{https://doi.org/10.1103/PhysRevA.84.042124}{\emph{Phys. Rev. A} {\bfseries 84} (2011) 042124}.

\bibitem{Zurek:2003zz}
W.H.~Zurek, \emph{{Decoherence, einselection, and the quantum origins of the classical}}, \href{https://doi.org/10.1103/RevModPhys.75.715}{\emph{Rev. Mod. Phys.} {\bfseries 75} (2003) 715} [\href{https://arxiv.org/abs/quant-ph/0105127}{{\ttfamily quant-ph/0105127}}].

\bibitem{PhysRevA.82.052122}
S.~Luo, S.~Fu and N.~Li, \emph{Decorrelating capabilities of operations with application to decoherence}, \href{https://doi.org/10.1103/PhysRevA.82.052122}{\emph{Phys. Rev. A} {\bfseries 82} (2010) 052122}.

\bibitem{Xi_2011}
Z.~Xi, X.-M.~Lu, X.~Wang and Y.~Li, \emph{The upper bound and continuity of quantum discord}, \href{https://doi.org/10.1088/1751-8113/44/37/375301}{\emph{Journal of Physics A: Mathematical and Theoretical} {\bfseries 44} (2011) 375301}.

\bibitem{Luo:2008ecu}
S.~Luo, \emph{{Quantum discord for two-qubit systems}}, \href{https://doi.org/10.1103/PhysRevA.77.042303}{\emph{Phys. Rev. A} {\bfseries 77} (2008) 042303}.

\bibitem{Horodecki_2005}
M.~Horodecki, J.~Oppenheim and A.~Winter, \emph{Partial quantum information}, \href{https://doi.org/10.1038/nature03909}{\emph{Nature} {\bfseries 436} (2005) 673–676}.

\bibitem{Friis_2017}
N.~Friis, S.~Bulusu and R.A.~Bertlmann, \emph{Geometry of two-qubit states with negative conditional entropy}, \href{https://doi.org/10.1088/1751-8121/aa5dfd}{\emph{Journal of Physics A: Mathematical and Theoretical} {\bfseries 50} (2017) 125301}.

\bibitem{Barger:1988jj}
V.D.~Barger, J.~Ohnemus and R.J.N.~Phillips, \emph{{Spin Correlation Effects in the Hadroproduction and Decay of Very Heavy Top Quark Pairs}}, \href{https://doi.org/10.1142/S0217751X89000297}{\emph{Int. J. Mod. Phys. A} {\bfseries 4} (1989) 617}.

\bibitem{Brandenburg:2002xr}
A.~Brandenburg, Z.G.~Si and P.~Uwer, \emph{{QCD corrected spin analyzing power of jets in decays of polarized top quarks}}, \href{https://doi.org/10.1016/S0370-2693(02)02098-1}{\emph{Phys. Lett. B} {\bfseries 539} (2002) 235} [\href{https://arxiv.org/abs/hep-ph/0205023}{{\ttfamily hep-ph/0205023}}].

\bibitem{Baumgart:2012ay}
M.~Baumgart and B.~Tweedie, \emph{{A New Twist on Top Quark Spin Correlations}}, \href{https://doi.org/10.1007/JHEP03(2013)117}{\emph{JHEP} {\bfseries 03} (2013) 117} [\href{https://arxiv.org/abs/1212.4888}{{\ttfamily 1212.4888}}].

\bibitem{NNPDF:2021njg}
{\scshape NNPDF} collaboration, \emph{{The path to proton structure at 1\% accuracy}}, \href{https://doi.org/10.1140/epjc/s10052-022-10328-7}{\emph{Eur. Phys. J. C} {\bfseries 82} (2022) 428} [\href{https://arxiv.org/abs/2109.02653}{{\ttfamily 2109.02653}}].

\bibitem{Buckley:2014ana}
A.~Buckley, J.~Ferrando, S.~Lloyd, K.~Nordstr\"om, B.~Page, M.~R\"ufenacht et~al., \emph{{LHAPDF6: parton density access in the LHC precision era}}, \href{https://doi.org/10.1140/epjc/s10052-015-3318-8}{\emph{Eur. Phys. J. C} {\bfseries 75} (2015) 132} [\href{https://arxiv.org/abs/1412.7420}{{\ttfamily 1412.7420}}].

\bibitem{Alwall:2011uj}
J.~Alwall, M.~Herquet, F.~Maltoni, O.~Mattelaer and T.~Stelzer, \emph{{MadGraph 5 : Going Beyond}}, \href{https://doi.org/10.1007/JHEP06(2011)128}{\emph{JHEP} {\bfseries 06} (2011) 128} [\href{https://arxiv.org/abs/1106.0522}{{\ttfamily 1106.0522}}].

\bibitem{Artoisenet:2012st}
P.~Artoisenet, R.~Frederix, O.~Mattelaer and R.~Rietkerk, \emph{{Automatic spin-entangled decays of heavy resonances in Monte Carlo simulations}}, \href{https://doi.org/10.1007/JHEP03(2013)015}{\emph{JHEP} {\bfseries 03} (2013) 015} [\href{https://arxiv.org/abs/1212.3460}{{\ttfamily 1212.3460}}].

\bibitem{Bierlich:2022pfr}
C.~Bierlich et~al., \emph{{A comprehensive guide to the physics and usage of PYTHIA 8.3}}, \href{https://doi.org/10.21468/SciPostPhysCodeb.8}{\emph{SciPost Phys. Codeb.} {\bfseries 2022} (2022) 8} [\href{https://arxiv.org/abs/2203.11601}{{\ttfamily 2203.11601}}].

\bibitem{deFavereau:2013fsa}
{\scshape DELPHES 3} collaboration, \emph{{DELPHES 3, A modular framework for fast simulation of a generic collider experiment}}, \href{https://doi.org/10.1007/JHEP02(2014)057}{\emph{JHEP} {\bfseries 02} (2014) 057} [\href{https://arxiv.org/abs/1307.6346}{{\ttfamily 1307.6346}}].

\bibitem{Sonnenschein:2006ud}
L.~Sonnenschein, \emph{{Analytical solution of ttbar dilepton equations}}, \href{https://doi.org/10.1103/PhysRevD.78.079902}{\emph{Phys. Rev. D} {\bfseries 73} (2006) 054015} [\href{https://arxiv.org/abs/hep-ph/0603011}{{\ttfamily hep-ph/0603011}}].

\bibitem{CMS:2019nrx}
{\scshape CMS} collaboration, \emph{{Measurement of the top quark polarization and $\mathrm{t\bar{t}}$ spin correlations using dilepton final states in proton-proton collisions at $\sqrt{s} =$ 13 TeV}}, \href{https://doi.org/10.1103/PhysRevD.100.072002}{\emph{Phys. Rev. D} {\bfseries 100} (2019) 072002} [\href{https://arxiv.org/abs/1907.03729}{{\ttfamily 1907.03729}}].

\bibitem{Adye:2011gm}
T.~Adye, \emph{{Unfolding algorithms and tests using RooUnfold}},  in \emph{{PHYSTAT 2011}}, (Geneva), pp.~313--318, CERN, 2011, \href{https://doi.org/10.5170/CERN-2011-006.313}{DOI} [\href{https://arxiv.org/abs/1105.1160}{{\ttfamily 1105.1160}}].

\bibitem{dagostini2010improved}
G.~D'Agostini, \emph{Improved iterative bayesian unfolding},  2010.

\bibitem{Johansson:2011jer}
J.R.~Johansson, P.D.~Nation and F.~Nori, \emph{{QuTiP: An open-source Python framework for the dynamics of open quantum systems}}, \href{https://doi.org/10.1016/j.cpc.2012.02.021}{\emph{Comput. Phys. Commun.} {\bfseries 183} (2012) 1760} [\href{https://arxiv.org/abs/1110.0573}{{\ttfamily 1110.0573}}].

\bibitem{Johansson:2012qtx}
J.R.~Johansson, P.D.~Nation and F.~Nori, \emph{{QuTiP 2: A Python framework for the dynamics of open quantum systems}}, \href{https://doi.org/10.1016/j.cpc.2012.11.019}{\emph{Comput. Phys. Commun.} {\bfseries 184} (2013) 1234} [\href{https://arxiv.org/abs/1211.6518}{{\ttfamily 1211.6518}}].

\bibitem{PhysRevA.77.022301}
S.~Luo, \emph{Using measurement-induced disturbance to characterize correlations as classical or quantum}, \href{https://doi.org/10.1103/PhysRevA.77.022301}{\emph{Phys. Rev. A} {\bfseries 77} (2008) 022301}.

\end{thebibliography}\endgroup
\end{document}